\documentclass[twoside,twocolumn,9pt]{article}
\usepackage{extsizes}
\usepackage[super,sort&compress,comma]{natbib} 
\usepackage[version=3]{mhchem}
\usepackage[left=1.5cm, right=1.5cm, top=1.785cm, bottom=2.0cm]{geometry}
\usepackage{balance}
\usepackage{mathptmx}
\usepackage{sectsty}
\usepackage{graphicx} 
\usepackage{lastpage}
\usepackage[format=plain,justification=justified,singlelinecheck=false,font={stretch=1.125,small,sf},labelfont=bf,labelsep=space]{caption}
\usepackage{float}
\usepackage{fancyhdr}
\usepackage{fnpos}
\usepackage[english]{babel}
\addto{\captionsenglish}{%
  
}
\usepackage{array}
\usepackage{droidsans}
\usepackage{charter}
\usepackage[T1]{fontenc}
\usepackage[usenames,dvipsnames]{xcolor}
\usepackage{setspace}
\usepackage[compact]{titlesec}
\usepackage{hyperref}

\usepackage{epstopdf}

\definecolor{cream}{RGB}{222,217,201}

\begin{document}

\pagestyle{fancy}
\thispagestyle{plain}
\fancypagestyle{plain}{
\renewcommand{\headrulewidth}{0pt}
}

\makeFNbottom
\makeatletter
\renewcommand\LARGE{\@setfontsize\LARGE{15pt}{17}}
\renewcommand\Large{\@setfontsize\Large{12pt}{14}}
\renewcommand\large{\@setfontsize\large{10pt}{12}}
\renewcommand\footnotesize{\@setfontsize\footnotesize{7pt}{10}}
\makeatother

\renewcommand{\thefootnote}{\fnsymbol{footnote}}
\renewcommand\footnoterule{\vspace*{1pt}%
\color{cream}\hrule width 3.5in height 0.4pt \color{black}\vspace*{5pt}} 
\setcounter{secnumdepth}{5}

\makeatletter 
\renewcommand\@biblabel[1]{#1}            
\renewcommand\@makefntext[1]%
{\noindent\makebox[0pt][r]{\@thefnmark\,}#1}
\makeatother 
\renewcommand{\figurename}{\small{Fig.}~}
\sectionfont{\sffamily\Large}
\subsectionfont{\normalsize}
\subsubsectionfont{\bf}
\setstretch{1.125} 
\setlength{\skip\footins}{0.8cm}
\setlength{\footnotesep}{0.25cm}
\setlength{\jot}{10pt}
\titlespacing*{\section}{0pt}{4pt}{4pt}
\titlespacing*{\subsection}{0pt}{15pt}{1pt}

\fancyfoot{}
\fancyfoot[LO,RE]{\vspace{-7.1pt}\includegraphics[height=9pt]{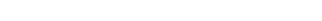}}
\fancyfoot[CO]{\vspace{-7.1pt}\hspace{13.2cm}\includegraphics{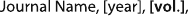}}
\fancyfoot[CE]{\vspace{-7.2pt}\hspace{-14.2cm}\includegraphics{RF}}
\fancyfoot[RO]{\footnotesize{\sffamily{1--\pageref{LastPage} ~\textbar  \hspace{2pt}\thepage}}}
\fancyfoot[LE]{\footnotesize{\sffamily{\thepage~\textbar\hspace{3.45cm} 1--\pageref{LastPage}}}}
\fancyhead{}
\renewcommand{\headrulewidth}{0pt} 
\renewcommand{\footrulewidth}{0pt}
\setlength{\arrayrulewidth}{1pt}
\setlength{\columnsep}{6.5mm}
\setlength\bibsep{1pt}

\makeatletter 
\newlength{\figrulesep} 
\setlength{\figrulesep}{0.5\textfloatsep} 

\newcommand{\topfigrule}{\vspace*{-1pt}%
\noindent{\color{cream}\rule[-\figrulesep]{\columnwidth}{1.5pt}} }

\newcommand{\botfigrule}{\vspace*{-2pt}%
\noindent{\color{cream}\rule[\figrulesep]{\columnwidth}{1.5pt}} }

\newcommand{\dblfigrule}{\vspace*{-1pt}%
\noindent{\color{cream}\rule[-\figrulesep]{\textwidth}{1.5pt}} }

\newcommand{\thib}[1]{\textcolor{blue}{\textbf{#1}}}

\newcommand{\onlinecite}[1]{[\!\!\!\citenum{#1}]} 

\makeatother

\twocolumn[
  \begin{@twocolumnfalse}
{\includegraphics[height=30pt]{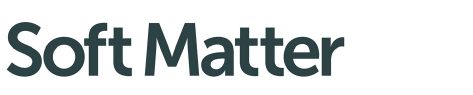}\hfill\raisebox{0pt}[0pt][0pt]{\includegraphics[height=55pt]{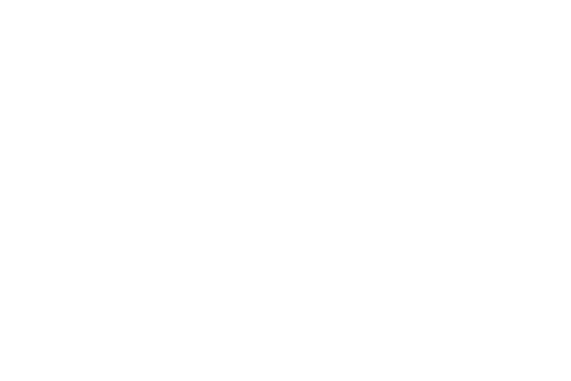}}\\[1ex]
\includegraphics[width=18.5cm]{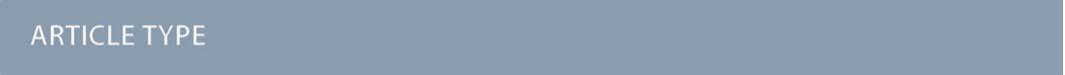}}\par
\vspace{1em}
\sffamily
\begin{tabular}{m{4.5cm} p{13.5cm} }

\includegraphics{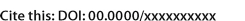} & \noindent\LARGE{\textbf{Ductile-to-brittle transition and yielding in soft amorphous materials: perspectives and open questions}} \\
\vspace{0.3cm} & \vspace{0.3cm} \\

 & \noindent\large{Thibaut Divoux,\textit{$^{o}$} Elisabeth Agoritsas,\textit{$^k$} Stefano Aime,\textit{$^{a}$}  Catherine Barentin,\textit{$^{b}$} Jean-Louis Barrat,\textit{$^{f}$} Roberto Benzi,\textit{$^{m}$} Ludovic Berthier,\textit{$^{d}$} Dapeng Bi,\textit{$^{z}$}  Giulio Biroli,\textit{$^{g}$} Daniel Bonn,\textit{$^{c}$} Philippe Bourrianne,\textit{$^{u}$} Mehdi Bouzid,\textit{$^{t}$} Emanuela Del Gado,\textit{$^{q}$}  Hélène Delano\"e-Ayari,\textit{$^{b}$} Kasra Farain,\textit{$^{c}$} Suzanne Fielding,\textit{$^{ag}$}  Matthias Fuchs,\textit{$^v$}  Jasper van der Gucht,\textit{$^s$} Silke Henkes,\textit{$^{ah}$}  Maziyar Jalaal,\textit{$^{aa}$} Yogesh M. Joshi,\textit{$^{x}$} Ana\"el Lema\^itre,\textit{$^{j}$} Robert L. Leheny,\textit{$^l$} Sébastien Manneville,\textit{$^{o}$} Kirsten Martens,\textit{$^{f}$} Wilson C.K. Poon,\textit{$^{e}$} Marko Popovi\'c,\textit{$^{p}$} Itamar Procaccia,\textit{$^{i,af}$} Laurence Ramos,\textit{$^d$} James A. Richards,\textit{$^{e}$} Simon Rogers,\textit{$^{y}$} Saverio Rossi,\textit{$^h$} Mauro Sbragaglia,\textit{$^{m}$} Gilles Tarjus,\textit{$^h$} Federico Toschi,\textit{$^{n,ae}$} Véronique Trappe,\textit{$^w$}  Jan Vermant,\textit{$^{r}$} Matthieu Wyart,\textit{$^k$} Francesco Zamponi,\textit{$^{ad,g}$} Davoud Zare\textit{$^{ab}$} } \\

\includegraphics{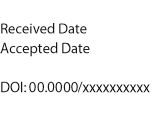} & \noindent\normalsize{Soft amorphous materials are viscoelastic solids ubiquitously found around us, from clays and cementitious pastes to emulsions and physical gels encountered in food or biomedical engineering. Under an external deformation, these materials undergo a noteworthy transition from a solid to a liquid state that reshapes the material microstructure. This yielding transition was the main theme of a workshop held from January 9 to 13, 2023 at the Lorentz Center in Leiden. The manuscript presented here offers a critical perspective on the subject, synthesizing insights from the various brainstorming sessions and informal discussions that unfolded during this week of vibrant exchange of ideas. The result of these exchanges takes the form of a series of open questions that represent outstanding experimental, numerical, and theoretical challenges to be tackled in the near future.} 

\end{tabular}

 \end{@twocolumnfalse} \vspace{0.6cm}

  ]

\renewcommand*\rmdefault{bch}\normalfont\upshape
\rmfamily
\section*{}
\vspace{-1cm}


\footnotetext{\textit{$^{a}$~ Molecular, Macromolecular Chemistry, and Materials, ESPCI Paris, Paris, France.}}
\footnotetext{\textit{$^{b}$~Univ. de Lyon, Université Claude Bernard Lyon 1, CNRS, Institut Lumière Matière - F-69622 Villeurbanne, France.}}
\footnotetext{\textit{$^{c}$~Soft Matter Group, Van der Waals-Zeeman Institute, University of Amsterdam, Science Park 904, 1098XH Amsterdam, Netherlands. }}
\footnotetext{\textit{$^{d}$~Laboratoire Charles Coulomb (L2C), Université Montpellier, CNRS, Montpellier, France.}}
\footnotetext{\textit{$^{e}$~SUPA and the School of Physics and Astronomy, The University of Edinburgh, Peter Guthrie Tait Road, Edinburgh EH9 3FD, United Kingdom.}}
\footnotetext{\textit{$^{f}$~Université Grenoble Alpes, CNRS, LIPhy, 38000 Grenoble, France.}}
\footnotetext{\textit{$^{g}$~Laboratoire de Physique de l'Ecole Normale Supérieure, ENS, Université PSL, CNRS, Sorbonne Université, Université de Paris, F-75005 Paris, France.}}
\footnotetext{\textit{$^{h}$~LPTMC, CNRS-UMR 7600, Sorbonne Université, 4 Pl. Jussieu, F-75005 Paris, France.}}
\footnotetext{\textit{$^{ag}$~Department of Physics, Durham University, South Road, Durham DH1 3LE, UK.}}
\footnotetext{\textit{$^{j}$~Navier, \'Ecole des Ponts, Univ Gustave Eiffel, CNRS, Marne-la-Vallée, France.}}
\footnotetext{\textit{$^{k}$~Department of Quantum Matter Physics (DQMP), University of Geneva, Quai Ernest-Ansermet 24, CH-1211 Geneva, Switzerland.}}
\footnotetext{\textit{$^{l}$~Department of Physics and Astronomy, Johns Hopkins University, Baltimore, Maryland 21218, USA.}}
\footnotetext{\textit{$^{m}$~Dipartimento di Fisica and INFN, Università di Roma ``Tor Vergata" - Via della Ricerca Scientifica, I-00133 Rome, Italy.}}
\footnotetext{\textit{$^{n}$~Department of Applied Physics and Science Education, Eindhoven University of Technology - P.O. Box 513, 9 5600 MB Eindhoven, The Netherlands.}}
\footnotetext{\textit{$^{ae}$~ CNR-IAC, Via dei Taurini 19, 00185 Rome, Italy.}}
\footnotetext{\textit{$^{o}$~ENSL, CNRS, Laboratoire de physique - F-69342 Lyon, France.}}
\footnotetext{\textit{$^{p}$~Max Planck Institute for the Physics of Complex Systems, N\"othnitzer Str.38, 01187 Dresden, Germany.}}
\footnotetext{\textit{$^{q}$~Georgetown University, Department of Physics, Institute for Soft Matter Synthesis and Metrology, Washington DC, USA..}}
\footnotetext{\textit{$^{r}$~Department of Materials, ETH Z\"urich, Vladimir Prelog Weg 5, 8032, Z\"urich, Switzerland.}}
\footnotetext{\textit{$^{s}$~Physical Chemistry and Soft Matter, Wageningen University \& Research, Stippeneng 4, 6708WE Wageningen, Netherlands.}}
\footnotetext{\textit{$^{t}$~Univ. Grenoble Alpes, CNRS, Grenoble INP, 3SR, F-38000, Grenoble, France.}}
\footnotetext{\textit{$^{u}$~PMMH, ESPCI Paris, France.}}
\footnotetext{\textit{$^{v}$~Fachbereich Physik, Universit\"at Konstanz, 78457 Konstanz, Germany}}
\footnotetext{\textit{$^{w}$~Department of Physics, University of Fribourg, Chemin du Musée 3, Fribourg, 1700, Switzerland.}}
\footnotetext{\textit{$^{x}$~Department of Chemical Engineering, Indian Institute of Technology, Kanpur, 208016, Uttar Pradesh, India.}}
\footnotetext{\textit{$^{y}$~Department of Chemical and Biomolecular Engineering, University of Illinois at Urbana-Champaign, Urbana, Illinois 61801, USA.}}
\footnotetext{\textit{$^{z}$~Department of Physics, Northeastern University, Boston, MA 02115, USA.}}
\footnotetext{\textit{$^{aa}$~Institute of Physics, University of Amsterdam; Science Park 904, Amsterdam, The Netherlands.}}
\footnotetext{\textit{$^{ab}$~Fonterra Research and Development Centre, Dairy Farm Road, Fitzherbert, Palmerston North 4442, New Zealand.}}
\footnotetext{\textit{$^{ac}$~Univ. Grenoble Alpes, CNRS, LIPhy, 38000 Grenoble, France.}}
\footnotetext{\textit{$^{ad}$~Dipartimento di Fisica, Sapienza Università di Roma, Piazzale Aldo Moro 5, 00185 Rome, Italy .}}
\footnotetext{\textit{$^{i}$~Dept. of Chemical Physics, The Weizmann Institute of Science, Rehovot 76100, Israel.}}
\footnotetext{\textit{$^{af}$~Sino-Europe Complex Science Center, School of Mathematics, North University of China, Shanxi, Taiyuan 030051, China.}}
\footnotetext{\textit{$^{ah}$~Lorentz Institute, Leiden University, 2300 RA Leiden, The Netherlands}}

Emulsions, polymer and colloidal gels, microgels, or concentrated particulate suspensions such as cement pastes constitute various examples of so-called soft amorphous materials. At the microscopic level, their constituents form disordered, possibly hierarchical, structures that span over a broad range of mesoscopic length scales. They often develop interactions that are comparable to or larger than thermal fluctuations, which hampers, without fully stifling, relaxation processes: these materials may relax towards low-energy states, but often without reaching equilibrium, leading to the emergence of time-dependent phenomena such as aging. Usually, their low-energy states have solid-like characteristics, with dynamic moduli showing weak frequency dependence, and an elastic modulus significantly larger than the viscous modulus. 

When subjected to external forces, soft amorphous solids undergo a solid-to-liquid transition known as the \textit{yielding transition}. The characteristics of this transition vary significantly depending on factors such as material properties, system preparation, the geometry confining the material, and boundary conditions. Similarly, following flow cessation, soft amorphous materials undergo a liquid-to-solid transition in diverse ways. Notably, memory effects from previous flow history can significantly impact the mechanical properties of the sample. This phenomenon is observed in various materials, including gels, soft glasses used in food products, cementitious items, and biological materials, potentially with important consequences on their practical use.

A general discussion of both yielding and memory phenomena took place during a one-week workshop at the Lorentz Center in Leiden from January 9-13, 2023. The workshop aimed at reviewing current knowledge in the field and at identifying upcoming challenges.  This manuscript summarizes the enriching discussions from the workshop. It is organized as follows: Section~\ref{sec:SIY} delves into the shear-induced yielding transition in soft amorphous materials, discussing its ductile or brittle nature, and the associated local scenarios. Section~\ref{sec:memory} focuses on memory effects in soft amorphous solids imprinted through shear history, exploring their impact on sample properties and relevant applications. Finally, Sections~\ref{sec:Bio} and \ref{sec:MatDe} provide insights into two specific examples: biological materials as soft glasses and material design through the utilization of shear history and memory effects. The manuscript concludes with a list of pressing questions that should guide the community's research agenda in the coming years. 

\section{Shear-induced yielding of soft amorphous materials}
\label{sec:SIY}

Traditionally, in structural and mechanical engineering, ductility and brittleness have been considered to be material properties, and different structural materials have been characterized as brittle (e.g., glasses) or ductile (e.g., metals) based on the way they macroscopically fail under deformation. Nanometer-scale AFM analyses, however, have revealed ductile fracture modes even in vitreous materials at temperatures much lower than their glass transition, highlighting that macroscopic brittleness or ductility are instead simply the result of similar plastic processes and damage accumulation occurring at microscopic length scales.\cite{Celarie2003Glass}

In soft matter, where the range and hierarchy of microstructural motifs and time scales can be extreme and become easily accessible, material failure (under an imposed deformation or stress) is a complex process that may manifests itself via a rich phenomenology. Here, we focus on the behavior of soft amorphous materials under deformation: they yield and eventually flow, with yielding being often preceded by, and associated with, mechanical and flow instabilities, whose prominence and persistence depend on their mechanical or rheological history, as well as on the imposed stress or deformation rate.\cite{Bonn:2017} At low deformation rates, the yielding of soft amorphous solids emerges from avalanches of localized plastic events,\cite{MaloneyLemaitre2004a,MaloneyLemaitre2006,Schall:2007} which are activated by the externally applied stress or strain,\cite{ArgonKuo1979,ChattorajCaroliLemaitre2010,Nicolas:2018} and is often associated with extended and pronounced shear localization or banding, where part of the material remains ``stuck'' while the rest is already flowing. \cite{Fielding:2014,Divoux:2016}

At first sight, yielding seems obviously a manifestation of the ductility of soft materials, and may be associated with dramatically sharp drops of the stress as fluidization takes place~\cite{Lu:2003,Dimitriou:2014,Keshavarz:2017}. Ultimately, the very same soft materials that yield may fracture,\cite{Gibaud:2010,Lindstrom:2012,Colombo:2014,Perge:2014b} therefore raising the question of what factors and processes control, over the relevant range of length scales and time scales, the emergence of brittle versus ductile behavior in yield stress fluids and soft materials. 

\begin{figure*}[t!]
\centering
\includegraphics[width=0.9\linewidth]{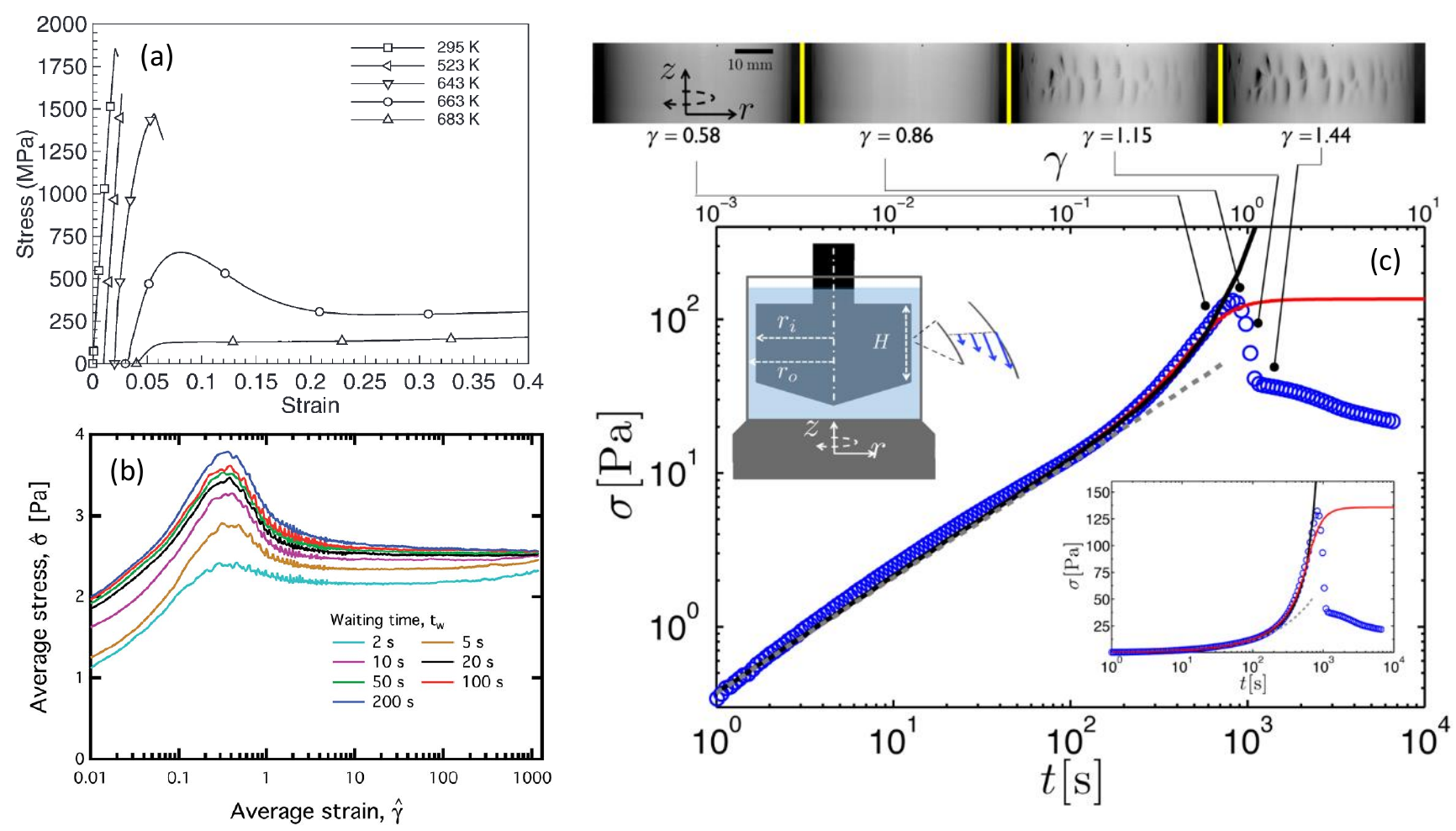}
\caption{Typical stress-strain (or time) curves measured under shear (a)~in a zirconium-based metallic glass alloy (Vitreloy~1) at different temperatures (extracted from Ref.~\onlinecite{Lu:2003}), (b)~in a model waxy crude oil system at a fixed shear rate ($\dot \gamma=2$~s$^{-1}$) for various waiting times elapsed since the rejuvenation of the sample (extracted from Ref.~\onlinecite{Dimitriou:2014}), and (c)~in an acid-induced casein gel (extracted from Ref.~\onlinecite{Keshavarz:2017}). In (c), the inset shows the same data in semilogarithmic scale, and the Taylor-Couette cell in which the experiments are performed. The upper panel shows the images of the side view of the Couette cell at different strains recorded simultaneously to the experiment reported in the main graph; macroscopic cracks are visible beyond the stress maximum.}
\label{fig:DBT_experimental}
\end{figure*}

\subsection{\large Ductile \textit{vs.} brittle yielding}

The ductile-to-brittle transition (DBT) refers to a clear qualitative change of material behavior when subject to an external deformation. This transition is typically observed by monitoring the stress dynamics under imposed deformation and is linked to a change in the material {\it rigidity}. For instance, the DBT can be achieved by changing the density or the packing fraction of the material, the temperature, the pressure, or simply the initial preparation of the sample. The DBT is well known in many areas of material sciences, ranging from metallic glasses\cite{Lu:2003,Schuh:2007,Gu:2009} to nanofibers,\cite{Luo:2016} fat crystals,\cite{Macias:2018,Macias:2018b} and biological materials.\cite{Meyers:2006,Peterlik:2006,Barthelat:2016}  
At the ductile end of the spectrum, the initial, linear increase of the stress with applied strain (i.e., the elastic response) is followed by a continuous crossover towards a stress plateau or by a smooth stress overshoot over a large strain range (viscoplastic response). At the brittle end of the spectrum, the short-time elastic response is followed by an abrupt stress drop at low strain.\cite{Ozawa:2022,Vasisht:2020,Leocmach:2014,Keshavarz:2017,Aime:2018,Colombo:2014} 
However, it is not yet clear whether the DBT occurs at some critical value of the parameter controlling the system rigidity, i.e., in the language of non-equilibrium statistical mechanics, whether there exists a (dynamical) phase transition underlying DBT. \cite{Barlow:2020,Ozawa:2018,Rossi:2022} Another key issue is whether the DBT in soft amorphous materials can be described along with that in hard glasses. In particular, soft glassy materials generically show a dual solid--liquid behavior at rest, i.e., viscoelasticity. Therefore, they may display elastic and/or viscous responses depending on the applied strain rate, which strongly impacts the stress response and leads to a key distinction between quasi-static approaches and finite-rate descriptions of yielding.\cite{Barlow:2020,Kamani:2021,Ozawa:2022,Singh:2020}

 \subsubsection{Experimental observations of ductile and brittle yielding in soft materials}
   \label{sec:BDT_experimental}

Figure~\ref{fig:DBT_experimental}(a) shows an example of ductile-to-brittle transition occurring by decreasing the temperature in a metallic glass. In practice, brittle yielding in such a hard material involves \textit{fractures} or \textit{cracks} within the material, i.e., interfaces are created, and the material eventually separates into two or more pieces. 
A very similar phenomenology is observed for the stress dynamics, for instance, in waxy crude oil, upon changing the sample age, i.e., the yielding process becomes more brittle for increasing time elapsed since the latest rejuvenation [see Fig.~\ref{fig:DBT_experimental}(b)]. Here, the sudden stress drop characteristic of brittle yielding is usually associated with the formation of {\it shear bands}, i.e., shear gets localized in small subregions of the sample. Note that the stress drop observed in experiments can also be concomitant with the growth of macroscopic cracks as illustrated in Fig.~\ref{fig:DBT_experimental}(c) in the case of acid-induced casein gels.  
Subsequent discussions in this paper will be mostly restricted to the situation of Fig.~\ref{fig:DBT_experimental}(b), where continuity of the material structure is warranted, and no new interface or fracture is nucleated.

\begin{figure}
    \centering
    \includegraphics{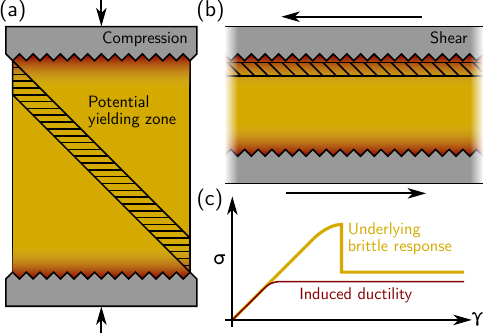}
    \caption{Stress inhomogeneity and yielding in experimental geometries. (a)~Compressive testing. Rough or uneven boundaries cause regions of inhomogeneous stress (red shading) near boundaries compared to the idealized stress field in the sample (yellow shading). Possible yielding region (hatched) lies within idealized stress region. (b)~Shear rheometer testing. Possible region of yielding lies within the zone of stress inhomogeneity. (c)~Resulting stress-strain schematic for a brittle material. Underlying brittle response predicted for compressive testing \textit{vs} induced ductility expected for experimental shear geometries.}
    \label{fig:UTM}
\end{figure}

Regarding experimental protocols, laboratory mechanical testing of amorphous materials in the soft matter community most often takes place in a rheometer, where a simple shear is applied in either parallel-plate or cone-and-plate geometry. However, stress-strain curves akin to the one shown in Fig.~\ref{fig:DBT_experimental}(b) may also be generated under compression or extension using a ``universal testing machine'' (UTM), or biaxial and triaxial apparatuses. A crucial issue is, therefore, whether there are differences regarding the response of a given material between different geometries. If the specimen is brittle, using compression in UTM, we expect the bulk compressive stress to rise until shear bands (or fractures) occur, usually at $45^\circ$ from the compression direction as sketched in Fig.~\ref{fig:UTM}(a). Strain localization along a $45^\circ$ plane closer to the compressing surfaces (where there are stress inhomogeneities) is suppressed by the rigid boundary itself. On the other hand, in a rheometer, the thickness of the sample may be such that stress heterogeneities induced by, e.g., irregularities in the rheometer plates, propagate throughout the sample [see Fig.~\ref{fig:UTM}(b)]. Furthermore, the geometry does not prevent shear deformation near the boundaries (as in a compressive test). Under these circumstances, even a sample that is {\it structurally} homogeneous may display {\it stress} inhomogeneities, and, therefore, yield in a way that may appear to be ductile in a rheometer while showing brittle yielding in a UTM [see Fig.~\ref{fig:UTM}(c)]. Therefore, caution must be taken when interpreting the macroscopic mechanical response obtained from rheological tests, whether in a UTM or in a rheometer.

Finally, as discussed in the next section, stress-strain curves can be predicted in computer simulations and/or phenomenological models, which, however, have mostly pertained (at least to date) to a simple shear geometry. In simulations, shear is mostly implemented using the Lees-Edwards periodic boundary condition, i.e., without boundary walls, and therefore focus on \textit{bulk} properties. Extending numerical approaches to encompass compression, including boundaries, and thus model the case of a UTM, would certainly prove very useful not only for practical applications, but also for a deeper understanding of the influence of boundaries (see also Sect.~\ref{sec:BC} below).

 \subsubsection{Numerical insights on the ductile-to-brittle transition}
 
\begin{figure*}[tb]
\centering
\includegraphics[width=1\textwidth]{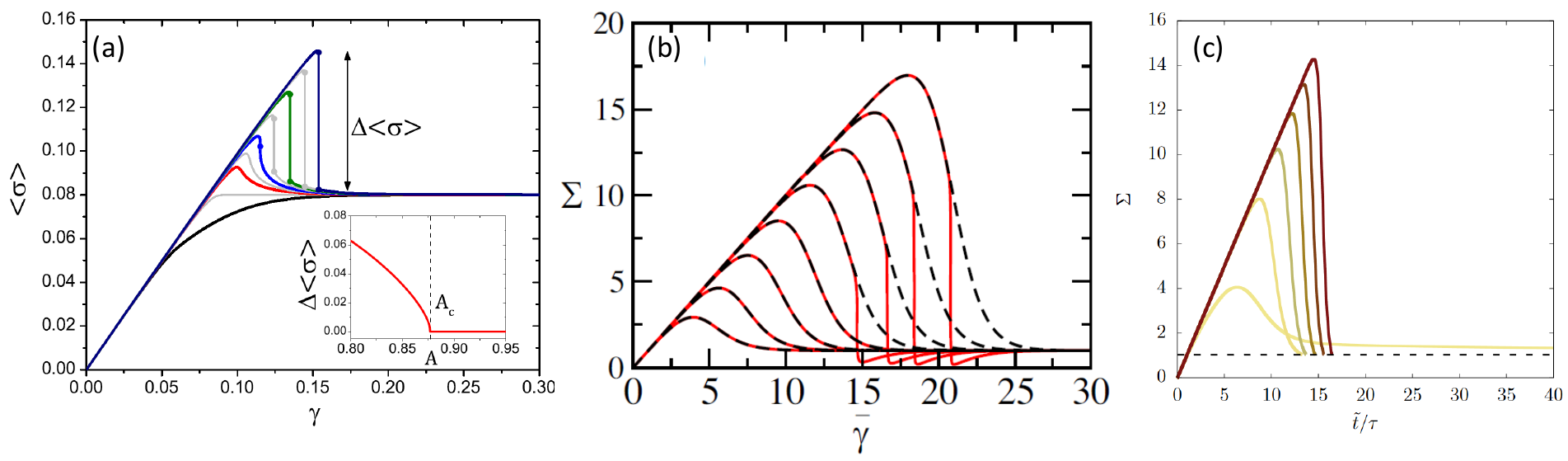}
\caption{Stress-strain curves obtained from numerical computations of (a)~a mean-field elasto-plastic model for increasing degree of annealing from bottom to top; the inset shows the amplitude of the stress drop vs.~the degree of annealing (extracted from Ref.~\onlinecite{Ozawa:2018}), (b)~a fluidity model with homogeneous flow enforced (dashed lines) and shear banding allowed (solid lines) for increasing waiting time from left to right (extracted from Ref.~\onlinecite{Barlow:2020}), and (c)~a fluidity model with decreasing initial fluidity from left to right (extracted from Ref.~\onlinecite{Benzi:2021PRE}).}
\label{fig:DBT_numerical}
\end{figure*}

The DBT can be investigated numerically using either \textit{microscopic} models, aimed at reproducing the relevant dynamics at small scales, or by models where the macroscopic dynamics are prescribed, usually referred to as \textit{continuous} models. Figure~\ref{fig:DBT_numerical} gathers a few examples of stress-strain curves obtained through various recent models using both approaches. In the microscopic approach, which includes molecular dynamics simulations \cite{PatiNinarello:2017,Ozawa:2018,BarbotLerbingerLemaitreVandembroucqPatinet2020,Singh:2020} and numerical resolutions of elastoplastic models,\cite{Martens:2012,Nicolas:2018,Liu:2018,Ozawa:2018,Rossi:2022} one usually focuses on the so-called athermal quasi-static limit (AQS)~\cite{MaloneyLemaitre2004a,MaloneyLemaitre2006}, where a number of small strain steps are imposed successively, each step being followed by a relaxation period when the system is driven close to mechanical equilibrium using {\it ad hoc} dissipative forces. AQS provides a useful bridge between the system dynamics and the available microscopic configurations, which may be studied using statistical mechanics. In general, the link between the microstructure and the dynamics cannot be disentangled by continuous models, although the DBT can be qualitatively reproduced from simple fluidity models \cite{Barlow:2020,Benzi:2021PRE,Benzi:2023} as shown in Fig.~\ref{fig:DBT_numerical}(b,c). 

Still, the AQS limit can be difficult to achieve experimentally for colloidal systems. As a matter of fact, any small but finite shear rate introduces a time scale in the system. This time scale may not be relevant for the DBT to occur~\cite{Singh:2020}, but it becomes important if one is interested in studying the statistical properties of the stick-slip-like dynamics that occur after the yield point. This leads to the key physical question of determining the correct theoretical framework to describe the DBT: should the DBT be described based on critical-like phenomena in the framework of statistical mechanics of disordered systems, or based on dynamical approaches as proposed in continuous models? Then, one should focus on the {\it possible predictions}, if any, which can be obtained using the different tools, and on the material properties that one should consider relevant in order to test the various approaches.

In Ref.~\onlinecite{Ozawa:2018}, a unifying picture was proposed to understand ductile \textit{vs.}~brittle behavior, where the crucial information is the preparation of the system. In this framework, the fictive temperature (i.e., the temperature at which the system has fallen out of equilibrium during cooling) becomes a relevant external parameter: at low fictive temperature (corresponding to very deep, very slowly cooled glassy states), the system exhibits a brittle response, whereas, at large fictive temperature (corresponding to poorly annealed systems), the solid-to-liquid transition is ductile. Similar results can be achieved for athermal systems using multiple control parameters (e.g., packing fraction, system size, annealing conditions, quenched disorder, etc.) to tune the material response.\cite{Ozawa:2020,Ozawa:2022} Here again, comparing possible predictions and available experimental data will be critical to draw definitive conclusions. 
 
 \subsubsection{Mesoscopic insights on the DBT in the light of quadrupolar events}

At the microscopic scale, deformation in dense disordered materials proceeds via the accumulation of rearrangements (shear transformations)\cite{ArgonKuo1979,FalkLanger1998,SchuhLund2003} that take place in small regions a few particles wide and release the macroscopic stress. At low temperatures and moderate strain rates, each rearrangement is followed by a nearly complete mechanical equilibration of the surrounding system: this is a fast process, established by the propagation of acoustic waves, which introduces long-range (Eshelby) stresses and strains\cite{eshelby1957determination,Picard:2002} that display a quadrupolar pattern in 2D or in 3D cuts. Such events and the associated Eshelby fields have been observed in numerical simulations\cite{MaloneyLemaitre2004a,MaloneyLemaitre2004b,MaloneyLemaitre2006,TanguyLeonforteBarrat2006,LemaitreCaroli2009} and experiments in colloidal or granular materials.\cite{Schall:2007,le2014emergence}

When a soft amorphous material is subjected to AQS deformation, every rearrangement may only occur through the crossing of an instability threshold.\cite{MaloneyLemaitre2004b} As the external strain drives every local packing towards some unstability threshold, the material is progressively brought into a state comprising a significant population of marginally stable regions.\cite{10KLPb} In such a system, Eshelby fields may trigger secondary events and thus give rise to plastic avalanches.\cite{MaloneyLemaitre2004a,MaloneyLemaitre2006,BaileySchiotzLemaitreJacobsen2007,MaloneyRobbins2008,LernerProcaccia2009,11HKLP,SalernoRobbins2013} These avalanches appear to be quite robust: moderate thermal fluctuations may trigger Eshelby events before they actually reach their instability threshold without changing significantly the avalanche dynamics,\cite{ChattorajLemaitre2013} while finite strain rates introduce a cutoff on avalanche sizes.\cite{LemaitreCaroli2009,ChattorajCaroliLemaitre2010} In model supercooled liquids, this avalanche regime is relevant to a very broad domain of temperatures and strain rates, and only ceases to be relevant in the Newtonian regime.\cite{ChattorajLemaitre2013} In that context, a non-Newtonian behavior emerges from correlations between rearrangements that, like avalanches, result from the induction of secondary events via the Eshelby mechanism.

Within this picture, ductile behavior takes place when Eshelby events and plastic avalanches are broadly scattered throughout the material, while brittleness results from the localization of the plastic activity along shear bands. Avalanches appear as an intermediate between isolated events and shear bands, which can therefore be viewed as resulting from the emergence of spatial organization among avalanches,\cite{HoudouxNguyenAmonCrassous2018} and not just of isolated events. The connection between Eshelby events, avalanches, and shear banding, however, remains a matter of debate and may involve the combination of elastic and structural effects.

Thus, an important challenge in the research on plasticity consists in identifying local predictors for plastic instabilities.\cite{Richard:2020} In particular, it has become possible to characterize the proximity of mechanical instabilities via measurements of local yield stresses.\cite{PuosiOlivierMartens2015,Patinet:2016} These studies show that, in better relaxed systems, local packings present statistically larger values of local yield stresses. By identifying how the distribution of local barriers is affected by strain,\cite{BarbotLerbingerLemaitreVandembroucqPatinet2020} they have also evidenced a significant role of rejuvenation: in a well-relaxed system, the first events to occur cause a local softening of the surrounding matrix, which facilitates the occurrence of more plastic events at the softened locations, which may lead to shear localization. From this perspective, hence, the DTB transition is governed by a dynamical interplay between plastic activity (including possible avalanches) and rejuvenation. It is favored by a high softness contrast between the initial state and rejuvenated packings, which explains that it is increasingly likely to take place in more relaxed (stable) system.

The identification of these effects leaves open the question of the role of elastic interactions between Eshelby quadrupoles in the formation of shear bands. This issue was partially clarified in previous studies,\cite{12DHP,13DHP} which have demonstrated that, with increasing shear strain $\gamma$, there is a threshold value $\gamma^*$ at which the minimum of the energy functional promotes an alignment of the quadrupoles along a line in two dimensions and along a plane in three dimensions. This alignment organizes the displacement field around a shear band. However, this theory had its limitations: the intensity of the quadrupoles and their core sizes were measured from simulations rather than arising directly from the theory itself. 

Recent progress in understanding the role of quadrupolar plastic events in determining the mechanical response of amorphous solids\cite{21LMMPRS,23CMP} presents an opportunity for achieving a self-consistent theory of shear banding. A key observation suggests that, in general, the plastic quadrupoles and their contribution to the energy functional renormalize the elastic moduli, yet they do not change the nature of linear elasticity theory. Conversely, when the quadrupolar field is non-uniform, gradients of this field act as dipoles, resulting in screening effects, the emergence of characteristic scales, and a fundamental change in mechanical response compared to the predictions of elasticity theory. These results were demonstrated in both simulations and experiments.\cite{21LMMPRS,23CMP,22MMPRSZ,22BMP,23CMPR,23JPS} 
Expanding this theory to address instabilities and shear banding necessitates further development, specifically including nonlinear quadrupolar and dipolar interactions. This ongoing work holds the potential to allow a self-consistent determination of the intensity and density of quadrupoles that are involved in creating the shear band. An example of such a theory of mechanical instability in metamaterials is outlined in Ref.~\onlinecite{20BLBM}, and could be adapted for describing shear banding in amorphous solids. 

 \subsubsection{Outlook} 
 
One key outcome of the discussions that took place at the Lorentz Center is that the DBT needs to be defined more quantitatively. To this aim, one should first systematically look for any signature of the DBT at the microscopic or mesoscopic scale, and/or try to identify static or dynamical precursors for the DBT. In that context, the goal would be to answer the following questions: How can one define brittle \textit{vs.}~ductile behavior at the mesoscale and/or at the microscale? Can we connect the nature of the precursors to the DBT and the subsequent failure scenario?

Yet, a significant technical hurdle for future experiments is the detection of failure precursors. Although predictions of failure have been identified in the rheological signatures of soft materials in specific cases,\cite{Saintmichel:2017,Cho:2022} rheometry typically is not sensitive enough to detect such precursors due to the small fraction of particles involved in microstructural rearrangements, which neither significantly nor reliably impact the macroscopic mechanical response.\cite{Pommella:2020,Koivisto:2016} Detection should rely on multiprobe approaches that combine mechanical testing with methods that couple directly to precursors, among which X-ray and light scattering,\cite{Aime:2018,Pommella:2019,le2014emergence} electrical spectroscopy,\cite{Helal:2016} acoustic emissions,\cite{Garcimartin:1997} change in the speed of sound,\cite{Scuderi:2016} and fluorogenic mechanophores,\cite{Slootman:2020} to name a few, have been shown to unravel changes in the sample microstructure long before shear-induced or stress-induced failure. 

However, these tools might not be sufficient. Indeed, despite the wealth of data collected by the last-generation array of seismic sensors along the San Andreas fault, no precursory signal was detected before the earthquake that took place in Parkfield.~\cite{Bakun:2005} This episode illustrates that new strategies are required to detect precursors to failure accurately. A first strategy could be to design a setup combining different probes to monitor various length scales and time scales simultaneously. A more ambitious, second strategy entails designing innovative and more selective probes referred to as ``\textit{smart probes},'' capable, for instance, of detecting only non-affine motion. Such experimental probes could connect to recent theory and simulation advances that have identified microstructural features, like regions of low local yield stress or high ``softness" parameter, that correlate with sites of initial failure.\cite{Patinet:2016,Zhang:2021} These technical developments will be pivotal in making significant progress in failure prediction. Finally, it appears evident that both structural heterogeneities\cite{Bouzid2017network,tauber2020microscopic} as well the rigidity percolation transition\cite{berthier2019rigidity,javerzat2023evidences} play crucial roles in influencing the DBT in soft amorphous systems. 
Therefore, future experiments and simulations should examine more carefully how rigidity percolation and stress propagation, especially under well-controlled structural disorder, determine the fate of materials.

More generally, in soft amorphous materials, the DBT transition has been mainly considered when the system is forced at a constant shear rate in a simple, plane shear geometry. As noted above, it may be useful to understand the DBT in other geometries, e.g., under compression in the UTM. Moreover, it is important to understand whether or not the DBT also occurs under a constant external stress and/or in cyclic shear protocol. In order to compare the DBT in different geometries and under various testing protocols, both in experiments and in numerical computations, it is essential to introduce a {\it quantitative} measure of brittleness. Preliminary attempts focused on the energy release during brittle fracture and on stress drops in metallic glasses.\cite{Lewandowski:2005} Defining and investigating similar observables in soft glassy systems should help characterize the DBT and identify universal features, if any,  common to soft and hard materials. 

Finally, from a more general perspective, a key question remains about whether understanding the DBT provides any deeper insight into the shear-induced solid-to-liquid transition in amorphous systems, and into the physics of the yielding transition. Indeed, the solid-to-liquid transition is often considered as a dynamical phase transition as in other disordered systems.\cite{Bonn:2017,aime2023} In the case of brittle materials, this point of view is consistent with the existence of an abrupt drop of the macroscopic stress at the yield point. In ductile materials, if the yielding transition occurs {\it because} of the nucleation of a shear band, the macroscopic stress does not show any abrupt change {\it in time}. 
However,  the velocity gradient may exhibit a rather sharp interface  {\it in space}, corresponding to a sharp change in the local shear rate. In the latter case, yielding still falls in the framework of a first-order (non-equilibrium) dynamical phase transition. Thus, one may argue that the DBT and the shear-induced solid-to-liquid transition could be discussed in a more general theoretical framework. This point deserves further investigation in the near future.

\subsection{\large Long-lasting heterogeneous flows}

Whatever the brittleness or ductility of the material, the shear-induced yielding transition that brings the material under shear from a solid-like state to a liquid-like state is never instantaneous, but rather takes some significant amount of time (or strain) to develop and to lead to a steady state. In particular, under an applied shear rate, the fluidization involves a stress overshoot, which may be followed by a long-lasting stress relaxation associated with shear bands of finite duration.\cite{Divoux:2010,Divoux:2012,Martin:2012} Under applied shear stress, the strain rate initially displays a power-law decrease, followed by a rapid increase over several orders of magnitude before a steady state is reached.\cite{Divoux:2011b,Grenard:2014,Caton:2008,Siebenburger:2012a} 

In Ref.~\onlinecite{Moorcroft:2013}, criteria for the formation of shear bands were developed within a set of minimal assumptions, aimed at being independent of the particular constitutive model or material under consideration (and therefore applying not only to yield stress fluids, but complex fluids more widely). These criteria suggest that one might generically expect shear bands to form both (i) in shear startup, associated with the presence of a stress overshoot, and (ii) following creep under an imposed shear stress, when the shear rate signal curves upwards as the material yields. Both criteria are supported by observations in several experimental systems, particle based simulations and constitutive models, as reviewed in Ref.~\cite{Fielding:2016}.

However, the startup criterion  (i, above) was derived within some caveats~\onlinecite{Moorcroft:2013,Fielding:2016,Moorcroft:2014} -- most notably assuming only two interacting mechanical variables (the shear stress and one component of normal stress or one fluidity variable, for example) -- and Ref.~\cite{Moorcroft:2014} indeed confirmed it to hold in some models of polymeric fluids, but not others. It also showed that the magnitude of the banding effect depends on the solvent-to-solution viscosity ratio (in models of highly elastic polymeric fluids; the dependence is much weaker in models of yield stress fluids).  Recent theoretical and numerical works~\cite{Sharma:2021,Sharma:2023preprint} confirmed that the start-up dynamics are influenced by variations in the viscoelastic constitutive model, and in  the solvent-to-solution viscosity ratio;\cite{Sharma:2023preprint} and that the correlation between the stress decay and the growth of linearized perturbations is not universal.\cite{Sharma:2021}  The startup criterion has also been challenged in experiments on repulsive glasses.\cite{Briole:2021} However, wall slip plays a key role in these experiments, whereas the criterion was derived assuming zero slip.

Taken together, these works suggest that the evidence for shear banding associated with startup stress overshoot is quite widespread,\cite{Fielding:2016} but that further work is needed concerning its generality. The creep criterion (ii, above)  holds for arbitrarily many interacting mechanical variables and is likely to be more generic.

\subsubsection{Overview on (transient) shear banding}
    
Shear bands are known to occur in many different soft amorphous systems.\cite{Becu:2006,Schall:2007,Schall:2010,Besseling:2010,Fall:2010b,Coussot:2010,Chikkadi:2014,Divoux:2016} In fact, there exist different definitions of shear bands depending on the material properties. In ductile-like materials, shear bands are usually identified with a {\it flowing state}, which nucleates within a {\it rigid state} during the solid-to-liquid transition. However, the term ``shear band'' also refers to fracture-like failure processes, as observed in hard brittle-like materials.\cite{Shi:2005,Greer:2013} In the latter case, the material {\it failure time} $\tau_f$ is the time at which a sample-spanning fracture is observed, and eventually, the stress reaches its equilibrium value \cite{Vanel:2009}. On the other hand, in ductile materials, $\tau_f$ is the time after which no rigid state can be detected in the system, i.e., the system has achieved full fluidization, so that $\tau_f$ corresponds to a {\it fluidization time}.\cite{Divoux:2016} When the system evolves toward a state with permanent shear bands rather than toward a fully fluidized state, $\tau_f$ is defined as the time at which the size of the shear band no longer grows. $\tau_f$ depends on the forcing mechanism, and two main cases have been considered, i.e., constant applied shear rate or constant applied stress. In the latter case, an alternative definition of $\tau_f$ was proposed in the literature as the time at which the strain rate reaches a minimum.\cite{Fielding:2014,Popovic:2022} This time scale, which rather corresponds to the onset of heterogeneous flows, shows interesting yet different scalings from the (full) fluidization time,\cite{Caton:2008,Popovic:2022} and is not discussed in the following. 

A challenging question is to determine how many different theoretical frameworks are needed to describe the various types of shear bands and time scales associated with shear banding. For instance, in the absence of fracture, one could imagine building a consistent picture by generalizing known concepts from nucleation dynamics that occur in dynamical phase transitions (see, for instance, Ref.~\onlinecite{Bray:2002}). In all cases, a key quantity to predict theoretically is $\tau_f$ as a function of the external forcing mechanism and the material properties. 
The case of transient or long-lived shear bands is particularly interesting because modeling these systems is greatly simplified using a field theoretical approach, i.e., a continuous model, where, however, the choice of the {\it order parameter} is crucial. Then, in order to describe quantitatively the time evolution of the system, one needs to predict how the size of the shear band grows in time. By contrast, continuous models may not be required to describe the dynamics involved in the formation of fracture-like shear bands. Yet, in brittle materials, it is still unclear whether shear bands are the {\it cause} or the {\it effect} of the abrupt stress drop at the yielding transition. In the first case, shear band nucleation and dynamics should be described as for ductile materials,\cite{Barlow:2020} whereas in the second case, $\tau_f$ is dictated by a different mechanism, which may depend on the system state and preparation. For ductile-like materials, it was shown both in experiments and models that the nucleation of transient shear bands is the mechanism underlying the stress overshoot in start-up shear experiments.\cite{Divoux:2010,Moorcroft:2011,Moorcroft:2013,Benzi:2021PRL} Finally, the impact of aging, possibly combined with inertia, on the yielding scenario, and on heterogeneous flows should also be considered.\cite{Nicolas:2016,Jain:2018,Kushwaha:2022}

\begin{figure*}[t!]
\centering
\includegraphics[width=1\textwidth]{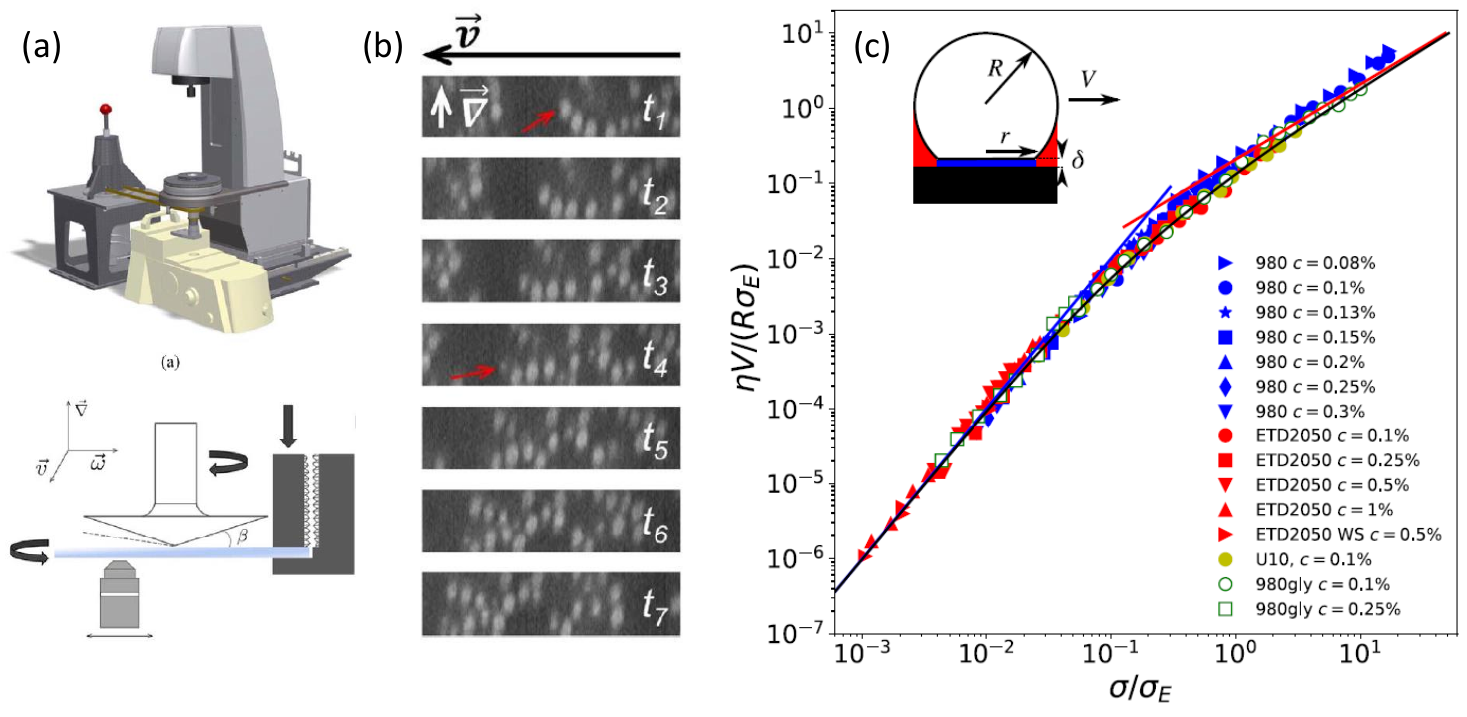}
\caption{(a)~Schematic of the rheo-confocal setup for high-speed imaging, and a sketch of the shear cell, in which a second motor is positioned on the side to counter-rotate the lower glass plate. (b)~Images taken during the shear-induced yielding of a depletion gel at 1~s$^{-1}$. Projections in the velocity-gradient plane of shear reconstructed from fast axial scanning with a focus-tunable lens. The red arrow indicates the same particle, which drifted out of the selected orthogonal planes. Extracted from Ref.~\onlinecite{Colombo:2019}. (c)~Dimensionless master curve of the slip velocity $V$ vs.~stress at the wall $\sigma$ as measured in microfluidic experiments performed on dense suspensions of microgels of various compositions. The parameters $\eta$ and $\sigma_E$ denote, respectively, the solvent viscosity and an elastic stress scale characterizing the nonlinear friction regime that is determined by fitting bulk velocity profiles. Different symbols stand for different Carbopol microgel samples. Inset: sketch of a microgel of radius $R$ in contact with the wall through a lubrication layer of thickness $\delta$. Extracted from Ref.~\onlinecite{Pemeja:2019}.}
\label{fig:wallslip}
\end{figure*}
On more general grounds, any theoretical approach should be consistent with known {\it robust}  features observed in most amorphous materials like, for instance, the aging properties of the system. This provides nontrivial constraints for possible theoretical frameworks. Moreover, such a theory should not only provide well-defined qualitative predictions to be checked but also be in quantitative agreement with existing experimental and numerical data. 
In particular, for ductile yielding, one should account for {\it non-local} rheological properties characterized by finite size effects and well-defined spatial scales. These features have been considered in recent developments, where the scaling properties of $\tau_f$ have been computed and successfully compared to experimental data.\cite{Benzi:2019,Benzi:2021PRL,Benzi:2021PRE} The presence of non-local effects naturally raises the problem of boundary conditions, which may play an important role, as discussed in more detail in Sect.~\ref{sec:BC} below.

 \subsubsection{Microscopic picture} 

From a microscopic point of view, long-lasting heterogeneous flows are related to the nature of the microstructure and of the interactions between the microscopic constituents composing the sample. To date, long-lasting transient shear bands were reported in both soft repulsive glasses\cite{Divoux:2010,Divoux:2011} and gels composed of colloids with short-range attractive interactions.\cite{Martin:2012,Grenard:2014} However, numerous gels with similar adhesive forces rather display \textit{steady-state} shear banding,\cite{Martin:2012} which makes it unclear at this stage whether some degree of attraction is required to observe long-lasting shear banding. Other important control parameters include the sample age, i.e., the time elapsed since the latest rejuvenation, as well as the shape and/or surface roughness of the particles. The latter was recently illustrated on gels made from rough or smooth silica particles, which display strongly different yielding scenarios.\cite{Muller:2023}

Because experiments remain challenging, faster progress might come by comparing with numerical simulations. Recent molecular dynamics simulations have demonstrated that long-lasting heterogeneous flows and shear banding can be observed in both attractive particle systems~\cite{chaudhuri2012inhomogeneous} and purely repulsive glasses.\cite{Vasisht:2020,Vasisht:2020b} In the latter case, long-lasting transients are related to the presence of overconstrained microscopic domains trapped in the bulk of the sample upon preparation. Their structural reorganization under shear controls the emergence and the persistence of the shear banding. Finally, mesoscale simulations have shed some light on the delayed yielding observed during creep tests, i.e., constant stress experiments. A mean-field version of the elastoplastic scenario captured the main features of delayed yielding in soft glassy materials, including the nonmonotonic response of the shear rate, and the power-law scaling of the fluidization time, whose exponent increases for decreasing sample age.\cite{Liu:2018}
    
 \subsubsection{Outlook: understanding yielding at the particle scale}
 \label{sec:outlookyielding}
    
Making a significant step toward rationalizing long-lasting heterogeneous flows will require designing well-controlled experimental systems, e.g., colloids with tunable interactions that can be easily controlled in situ.\cite{Siebenburger:2012a} A long discussion pointed out that we still need to understand the spatially-resolved scenario of the shear-induced yielding transition in gels at the particle scale. Unraveling the yielding scenario at the particle scale will require fast confocal microscopy [see Fig.~\ref{fig:wallslip}(a,b)] and/or rheo-tomography.\cite{Rajaram:2010,Rajaram:2011,Chan:2013,Colombo:2019} These experiments should primarily concentrate on identifying clearly defined microscale and mesoscale observables, which can be compared with the outcomes of numerical simulations. This comparative analysis aims to offer a consistent description of yielding across a broad range of length scales.

\subsection{\large Influence of boundary conditions: from wall slip to wall-induced plasticity}
\label{sec:BC}

As argued above in Sect.~\ref{sec:BDT_experimental}, the surface properties of the confining walls, as well as the geometry of the experiment, may affect the yielding scenario, and even provide a ductile-like response to a brittle material.
Apparent wall slip is another signature of the effect of boundary conditions on the deformation and flow of soft amorphous materials. Although it has been the topic of numerous reviews,\cite{hatzikiriakos:2012,Cloitre:2017,Malkin:2018} wall slip remains often considered as a mere issue that is disconnected from bulk rheology, if not as an experimental artifact. Key issues remain that wall slip is delicate to measure experimentally without any local probe, and that it often couples with the flow, leading to complex heterogeneous dynamics in both space and time.\cite{Gibaud:2008} In the following, we summarize discussions on experimental and numerical aspects raised by boundary conditions and their connection with the shear-induced solid-to-liquid transition.

\subsubsection{Experimental insights: soft glasses vs gels}
    
In the last decades, wall slip in soft glassy materials has received various levels of attention, mainly focused on steady shear. The case of dense suspensions has been so far the most studied.\cite{Kalyon:2005} In the case of hard, weakly deformable particles, apparent wall slip arise because of steric depletion, particle migration, and shear-induced layering in the vicinity of the wall.\cite{Ballesta:2012} In contrast, in the case of soft deformable particles, the slip phenomenology is controlled by both lubrication forces through elasto-hydrodynamic forces, and by the nature of the interactions between the particles and the wall.\cite{Seth:2008} In practice, wall slip is quantified by a friction law that relates the stress at the wall to the slip velocity, defined as the difference between the wall velocity and that of the sample extrapolated at the wall. This relationship, which was reported to be a power law whose exponent depends on the system under study,\cite{Meeker:2004a,Divoux:2015,Lemerrer:2015,Zhang:2017} was recently understood in terms of a balance of length scales between the particle size and the thickness of the lubrication layer [see Fig.~\ref{fig:wallslip}(c)].\cite{Pemeja:2019}

In more dilute dispersions, including gels, wall slip is also observed experimentally.\cite{Walls:2003,Grenard:2014} To date, there is, however, no generic framework to describe wall slip in dilute systems, and it remains unclear whether its description can be coarse-grained to a mere comparison of well-defined length scales. Indeed, gels naturally form clusters due to interparticle attractive forces, which makes it hard to identify a relevant length scale between the particle size and the cluster size, which is itself a complex function of shear. Therefore, the multiscale architecture of colloidal gels, as well as the possibility for adhesion, rolling, and sliding between (aggregates of) colloids and the walls, constitute a significant challenge for models to offer a consistent description of wall slip in such dilute systems. 

Finally, besides the steady state, wall slip plays a crucial role during transient flows, even in the presence of rough boundary conditions. \cite{Divoux:2012} For instance, in shear start-up flow, the sample may yield at the wall or in the bulk, which results in different dynamics, especially in systems that display strong aging dynamics.\cite{Gibaud:2009} In that case, comparing two characteristic time scales, e.g., the aging time and the inverse of the plastic event rate near the wall, could be more relevant than comparing two length scales. Moreover, this picture prompts us to better understand the link between wall slip and plastic events in dilute systems, for these two concepts could be deeply related. In this context, a promising avenue lies in non-local effects, which offer an elegant way of coupling the material behavior near the wall to its bulk dynamics.\cite{Seth:2012,Mansard:2014,Benzi:2019,Benzi:2021PRL}

Besides these fundamental considerations, addressing wall slip should help develop global strategies to suppress, reduce, and even tune wall slip for a broad variety of soft materials, from dense suspensions to colloidal gels, in steady-state flow as well as in transient regimes.
One classical strategy used to reduce slip is to roughen the solid walls: for soft pastes, it has been shown empirically that a roughness comparable to the particle size efficiently reduces wall slip, whereas, for colloidal gels, the required roughness is not clearly identified, again due to the multiscale nature of such systems. A second strategy consists in modifying the interactions between the particles and the walls. For instance, it was shown that attractive wall--particle interactions may allow one to suppress the solvent lubrication layer and, consequently, wall slip for both soft pastes \cite{Seth:2012} and colloidal gels.\cite{Walls:2003} Such strategies will have a crucial impact in various engineering areas, including additive manufacturing.\cite{jalaal:2015,mackay:2018,martouzet:2021,vander:2023}

\subsubsection{Recent numerical progress}

With the exception of dense suspensions of soft particles,\cite{Seth:2008} numerical simulations devoted to modeling the impact of wall slip in dispersed systems remain scarce. A noticeable recent contribution can be found in Ref.~\onlinecite{Jung:2021}, where wall slip is reproduced in two-dimensional elastohydrodynamic simulations that capture both the particle elasticity and fluid mechanics of dense thermal soft particle suspensions. These simulations capture some salient features of experimental flow curves, e.g., the so-called kink or reduction in stress observed at low shear rates toward stresses below the bulk yield stress, but predict a linear friction law. Moreover, continuous models based on fluidity, i.e., the local rate of plastic events, were shown to capture accurately the impact of apparent wall slip via non-local effects during shear start-up and creep flow in soft glassy materials. \cite{Benzi:2019,Benzi:2021PRL,Benzi:2021PRE} This versatile approach also allows taking into account elasto-hydrodynamic forces and, therefore, captures the impact of lubrication forces in both transient and steady-state flow.\cite{Benzi:2023} Yet, more work is needed, especially at the particle scale, to capture the non-linearity of the friction laws reported experimentally [see Fig.~\ref{fig:wallslip}(c)]. In particular, molecular dynamic simulations of dense suspensions, as well as dilute gels, which would help understand the multiscale physics at stake in wall slip, are still sorely lacking. 

\subsubsection{Outlook: wall slip as localization of plastic events near the wall?} 

The discussions that took place at the Lorentz Center call for a change of paradigm, in which wall slip would be treated on an equal footing with the bulk constitutive equation. In the context of steady flow, it was indeed shown within the framework of elastoplastic models that the introduction of plastic events at the wall had effects very similar to wall slip.\cite{Nicolas:2013} Future work should focus on identifying experimentally the differences and similarities between wall slip and plasticity near the wall, so as to clarify the contribution of wall slip to the overall yielding dynamics, especially in colloidal gels. This experimental challenge aligns closely with the ones emphasized in Sections~\ref{sec:BDT_experimental} and \ref{sec:outlookyielding}, i.e., it pertains to understanding the distinct characteristics of the yielding process under shear as opposed to compression, especially at the particle scale.

\section{Memory effects and processing history}
\label{sec:memory}

\subsection{\large Rheological memory in soft amorphous materials}

In the rheology literature, the term ``\textit{rheological memory}'' has been loosely used and commonly applied to discuss the transient, possibly long-lived, mechanical response of systems perturbed from some quiescent state, in both the linear and the nonlinear viscoelastic regimes. In this section, we first define more rigorously the concept of rheological memory before reviewing its signatures in soft amorphous materials. 

\subsubsection{Defining and encoding rheological memory}

\begin{figure*}[!th]
    \centering
    \includegraphics[width=0.9\linewidth]{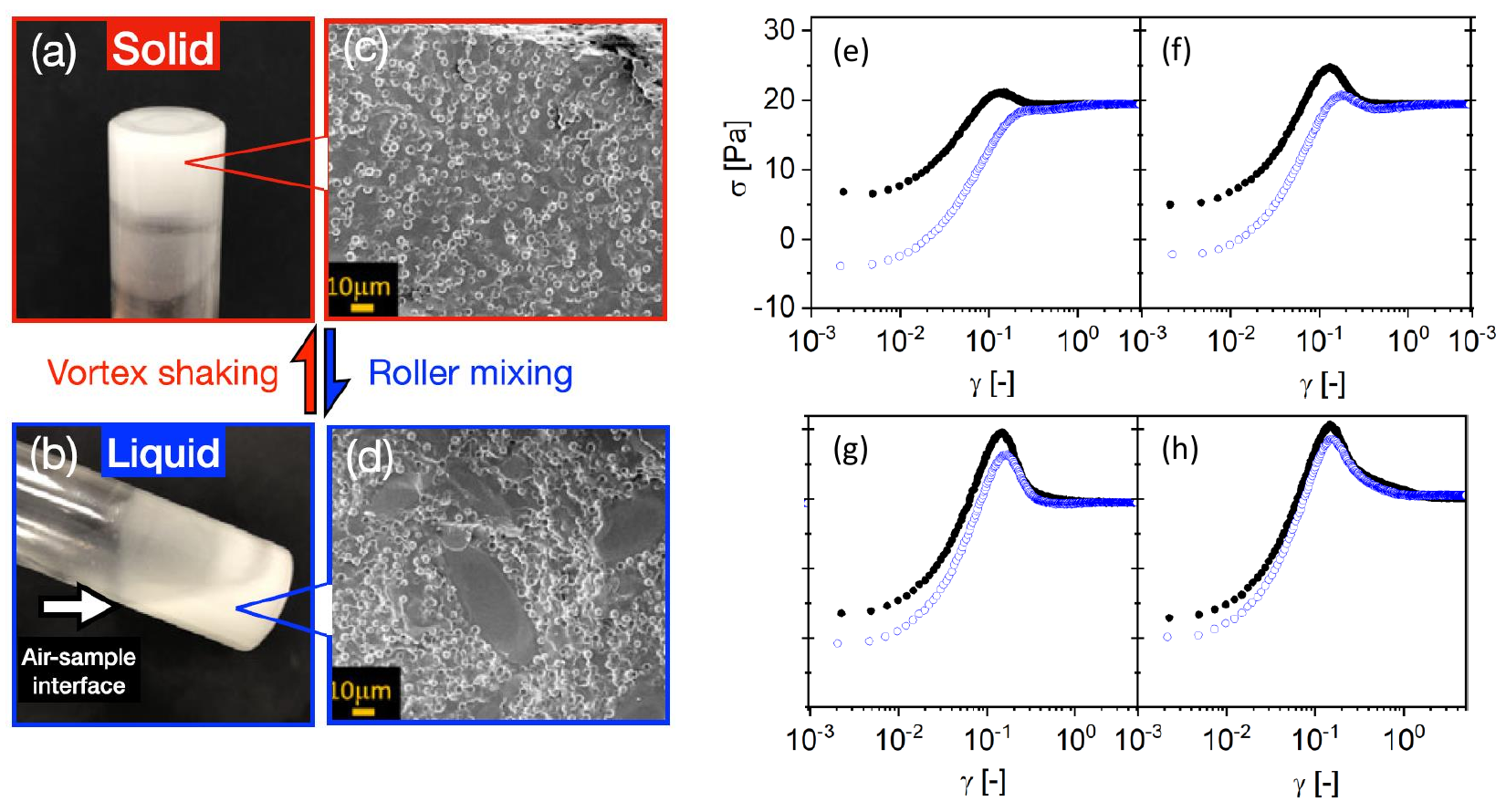}
    \caption{Examples of memory effects in Soft Glassy Materials. (a)–-(d) Images of a fumed-silica-based binary suspension
(small hydrophobic silica colloids $\phi_S=0.02$ and large silica microspheres $\phi_L=0.3$) in a glass vial, showing a liquid-solid transition under alternating high-shear vortex mixing and gentle roller mixing. Extracted from Ref.~\onlinecite{Jiang:2022} (e)--(h) Dependence of the stress response of a dense microgel suspension on the preshear direction. The shear rate $\dot \gamma=0.25$~s$^{-1}$ is applied in the positive direction in all experiments, whereas the preshear ($|\dot \gamma_0|=20$~s$^{-1}$) is applied in the positive (black circles) or negative (blue open circles) direction. The sample is left at rest ($\dot \gamma=0$) for a duration $t_w$ [$=1$, 10, 1000, and 10000~s from (e) to (h)] between the preshear and the shear start-up test. Extracted from Ref.~\onlinecite{DiDio:2022}.}
    \label{fig:Memory}
\end{figure*}

A typical rheological protocol employed to highlight memory effects consists in applying a step strain of the order of the yield strain for a fixed period of time. In that case, the ``memory'' is lost over some characteristic relaxation time, as illustrated in a pioneering work on dense aqueous foams.\cite{Hohler:1999} More recently, the same meaning of ``rheological memory'' has been used in various contexts, including plate tectonics.\cite{Fuchs:2022}

In control engineering, however, such systems, which relax over some characteristic time, would be considered as {\it monostable}, i.e., with no memory. Indeed, in a recent review \cite{Keim:2019} of ``memory formation in matter,'' the word ``memory'' has been clearly defined as ``the ability to encode, access, and erase signatures of past history in the state of a system.'' A practical realization has been obtained by applying continuous oscillatory shear with a fixed strain amplitude on particulate systems, e.g., dense suspensions or bubble rafts. After a transient period where the sample microstructure is reorganized by the oscillatory strain, a steady state is reached, where oscillations do not produce any further rearrangement. A signature of the applied strain amplitude is thus ``stored'' by the system, and can be read out by observing at what strain the response of the system starts to be irreversible or display an abrupt change in the displacement of its constituents.\cite{Keim:2011} In such a framework, a single system can store multiple memories by being submitting to more than one strain amplitude. These memories are transient and compete with each other, yet they can be stabilized if noise is added.\cite{Paulsen:2014}

Another strategy for encoding memory in soft glassy materials is based on {\it bistability}. There have been few reports to date of rheological bistability. One early example concerns lyotropic liquid crystalline phases of some aqueous surfactant solutions.\cite{Diat:1993b,Roux:1993,Bonn:1998a} There, stress first drives the solution from a low-viscosity lamellar phase to a high-viscosity phase made of an amorphous assembly of multilamellar spherulites, a.k.a., ``disordered onions.'' Depending on the surfactant concentration and temperature, these systems can further bifurcate to a low-viscosity phase with long-range order, either made of oriented lamellae or of a crystalline-like assembly of ``ordered onions.'' Such systems demonstrate ``weak'' bistability, in the sense that back-and-forth switching between the various states is not easy to control, as they involve complex, chaotic-like temporal dynamics and shear-banding flows.\cite{Wunenburger:2001,Salmon:2002,Manneville:2004b} Similar examples of bistability can be found in surfactant hexagonal phases,\cite{Ramos:2000,Ramos:2001a} and in the context of dense non-Brownian suspensions that exhibit discontinuous shear-thickening.\cite{Cates:2014} A recent demonstration of ``strong'' bistability, where the system is shown to be switchable at will between two different structural states, concerns dispersions of non-Brownian repulsive particles in a background gel of much smaller, attractive colloids.\cite{Jiang:2022} By tuning the applied stress and strain accumulation, the system can cycle between solid and fluid states [see Fig.~\ref{fig:Memory}(a)-(d)]. Such a behavior allows the ``writing'' of memory through particular shear protocols, and the ``reading'' of the written state through, e.g., oscillatory rheology. 

Soft amorphous systems with memory intrinsically show ``history dependence.'' Experimentally, this implies that specifying a well-defined protocol is vital. In this regard, one may usefully make contact with theoretical computer science,\cite{Kolmogorov:1963,Martiniani:2019} where the ``algorithmic complexity'' of a string is defined as the shortest algorithm that can be written to generate the string. Thus, `abababababab' is generated by the pseudocode ``[print `ab' $\times 6$],'' while  `aowgsiybmwsd' is generated by ``[print `a', print `o', \ldots, print `d'],'' and is more algorithmically complex. Along the same lines, one may perhaps define the concept of ``rheological complexity,'' which would encompass that of ``rheological memory,'' as the length of the shortest experimental protocol necessary for reproducing observations. Careful adherence to such protocol specifications is a prerequisite for enabling experiments to build on one another. It may also give some insight into the nature of the complexity in the systems under study. 

\subsubsection{Signatures of memory}

The extent to which a soft glassy material system recollects memory is largest for training amplitudes near the point where the dynamic moduli cross.\cite{Mukherji:2019,Donley:2020} This effect is a direct consequence of the spatial extent to which the system reorganizes during the encoding process, and correlates with a peak in the contribution of unrecoverable processes to the viscous modulus\cite{Donley:2020} These results raise the question of length scales associated with memory formation in jammed materials and gels. In practice, such a length scale could be determined experimentally via X-ray photon correlation spectroscopy (XPCS),\cite{Donley:2023} or multispeckle dynamics light scattering,\cite{Cipelletti:2016} which calls for a more systematic use of these techniques to investigate the yielding transition in various types of soft solids, from soft glasses to gels, in order to link their respective length scale to memory storage. 

Under continuous shear, yield stress fluids also display time-dependent features, often referred to as ``\textit{memory effects},'' which are closely related to thixotropy and/or viscoelasticity. \cite{Sharma:2023} So-called rheological hysteresis or hysteresis loops, which are obtained, e.g., by a decreasing ramp of shear rate, followed by the same increasing ramp, allow one to probe thixotropic time scales (for instance, by varying the sweep rate) that are characteristic of the material.\cite{Divoux:2013,Jamali:2019} Such time scales have been used to introduce a dimensionless number, the Mnemosyne number, defined as the product of the flow strength and the thixotropic time scale, and allows distinguishing thixotropic phenomena from other rheological responses such as viscoelasticity or plasticity.\cite{Jamali:2022} In that context, strong flows, i.e., large Mnemosyne numbers, are required to erase the memory from previous flows, e.g., inherited by loading the sample into the rheometer. Erasing memory has indeed long been studied in terms of defining good preshear protocols that allow for material behaviors to be studied irrespective of prior deformation.\cite{Choi:2020}

Last but not least, some memory can also be encoded via the direction of shear, as identified in a pioneering work on suspensions, \cite{Gadala:1980} and subsequently observed in cyclic shear experiments,\cite{Keim:2013,Paulsen:2014} and more recently quantified in shear start-up experiments performed along the same direction as a preshear, or the opposite [see Figs.~\ref{fig:Memory}(e)-(h)].\cite{Choi:2020,DiDio:2022} In practice, a moderate shear rejuvenates a yield stress fluid only partially and may imprint some level of anisotropy in the sample microstructure, e.g., through some degree of alignment along the flow direction. Such microstructural anisotropy is then frozen upon flow cessation through the rapid liquid-to-solid transition, as we will discuss below in Sect.~\ref{sec:Residual}. Direct evidence for ``{\it directional memory}'', also referred to as kinematic hardening or Bauschinger effect,\cite{Larson:2019} has been reported in a broad range of systems, e.g., soft glasses,\cite{DiDio:2022} colloidal gels,\cite{Grenard:2014} or waxy crude oil.\cite{Dimitriou:2014} In that context, the sample memory is directly linked to the structural anisotropy of the yield stress material. The latter memory can then be read by orthogonal superposition rheometry.\cite{Moghimi:2019} Finally, note that the sample memory has been also related to the recoverable strain that might be acquired prior to yielding, and which remains until deformation is applied in the opposite direction.

\subsubsection{Outlook: impact of noise on memory and connection with local yielding}

Memory effects in soft glassy materials are undoubtedly a flourishing area of research. 
Long-lasting memory of shear history plays a key role in defining appropriate conditioning protocols prior to rheological experiments, aimed at preparing the sample in a well-defined state. While simple, fixed-rate preshear protocols often provide robust and reproducible results, they sometimes encode memories, which may bias the outcome of the subsequent experiment (see section~\ref{sec:Flowcessation} below). A better control of memory encoding and erasing is needed to design smarter conditioning protocols.
Moving forward, future experiments need to delve deeper into establishing the connection between unrecoverable strain and memory storage. Moreover, efforts should be directed toward devising methods for the effective readout of stored memory, with a particular focus on employing orthogonal superposition rheometry.\cite{Schwen:2020} 
Among the most pressing challenges stands the influence of noise on memory effect. Although researchers have explored and discussed its impact within the context of dense non-Brownian suspensions and 2D binary Lennard-Jones systems,\cite{Keim:2013,Patinet:2020} there remains a significant gap in experimental studies, particularly in the case of colloidal gels and soft glasses.

Finally, a key area of focus should be the experimental exploration of the \textit{local} yielding properties. This entails examining the response at the microscopic or mesoscopic scales, following a loading phase at the macroscopic scale, and during the reverse experiment in which the shear is imposed in the opposite direction. Such investigations can be instrumental in identifying potential asymmetries in the local response, which may elucidate the mechanisms behind directional memory in soft glassy materials.\cite{Patinet:2020} Insights into strongly intermittent motion can be expected based on microrheology experiments performed in colloidal glass.\cite{Senbil:2019} A promising approach lies in leveraging exceptional experimental resolution to execute a systematic analysis akin to previous numerical studies.\cite{Mungan:2019,Richard:2020}

\subsection{\large Microstructure and mechanics inherited upon flow cessation}
\label{sec:Flowcessation}

While soft glassy materials flow under external shear, they proceed toward a complete stop upon flow cessation through a liquid-to-solid transition. Such a recovery of solid-like properties from a liquid-like behavior leads to the emergence of so-called residual stresses, which may display complex temporal dynamics linked to the memory effects discussed in the previous section.

\subsubsection{Residual stress stored upon flow cessation}
\label{sec:Residual}

\begin{figure}[!t]
    \centering
    \includegraphics[width=0.9\columnwidth]{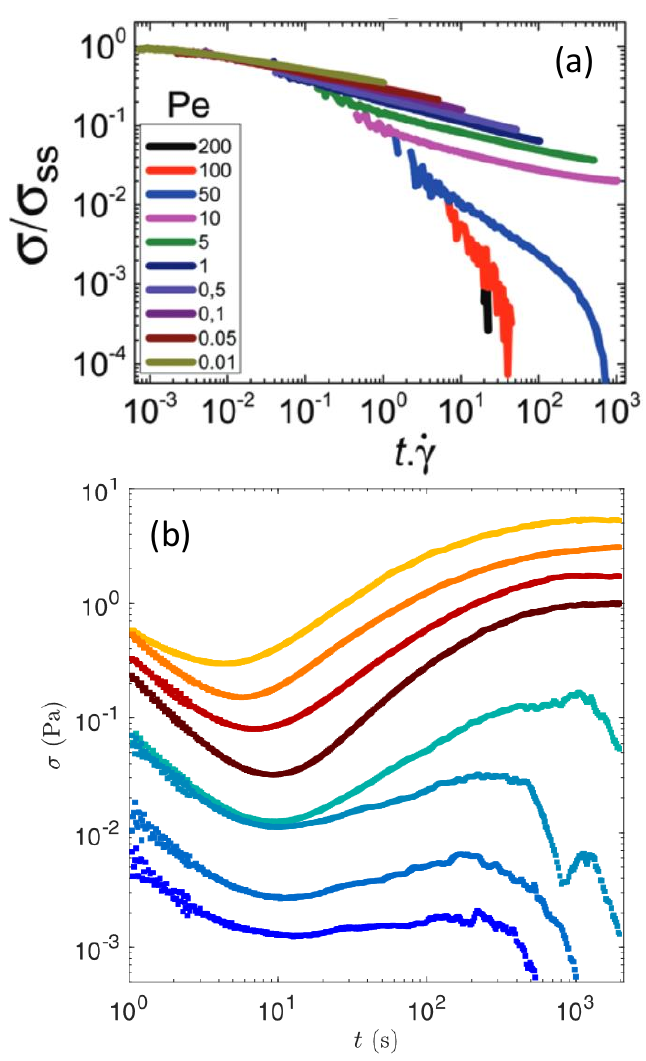}
    \caption{Stress relaxation following flow cessation. (a)~Monotonic stress relaxation vs.~strain following various shear rates, reported here as a particulate Peclet numbers defined as the ratio of the strength of the shear flow with the interparticle bond force. The sample stores a non-negligible residual stress for $Pe \leq 10$. Experiments performed on a gel of hard spheres of PMMA. Extracted from Ref.~\onlinecite{Moghimi:2017b}. (b)~Non-monotonic stress relaxation vs.~time in a boehmite gel following a period of shear at constant shear rates before flow cessation. Each color corresponds to a different shear rate: from 3~s$^{-1}$ (top) to 500~s$^{-1}$ (bottom). Extracted from Ref.~\onlinecite{Sudreau:2022b}.}
    \label{fig:StressRelax}
\end{figure}

In practice, flow cessation under a controlled shear rate, i.e., upon setting the shear rate to zero, leads to a decrease of the shear stress down to some residual stress, which depends in general upon the shear history, and decreases as a function of the shear rate applied before the flow cessation [see Fig.~\ref{fig:StressRelax}(a)]. The latter result has been reported experimentally and numerically in both hard-sphere and soft-sphere suspensions,\cite{Ballauff:2013,Mohan:2013,Mohan:2015,Lidon:2017,Vasisht:2021,chaudhuri2012inhomogeneous} and in colloidal gels.\cite{Osuji:2008,Negi:2009b,Moghimi:2017b} The stress decays following a compressed exponential or a similar function. \cite{Zausch:2009,Barik:2022} From a microscopic perspective, the material comes to a complete stop through a series of plastic events, which occur even long after the forcing is switched off, as evidenced by recent molecular dynamics simulations of densely packed non-Brownian particles.\cite{Vasisht:2021} Under a large shear rate (or under large Peclet number for Brownian systems), the microstructure is fully fluidized and reforms upon flow cessation without any external constraint, hence inheriting small, if not negligible, residual stresses. By contrast, low shear rates (or small Peclet numbers) only partially disrupt the microstructure, which bears some memory of the shear history. Some degree of structural anisotropy is observed in both gels and soft glasses, which correlates with reinforced mechanical properties. \cite{Moghimi:2017b,Sudreau:2022b,Sudreau:2023} In the case of gels, the anisotropy lies in anisotropic clusters involving several particles, whereas in soft glasses, the anisotropy is encoded at the particle scale through the asymmetry in the angular distribution of neighboring particles.\cite{Mohan:2013}

Future progress on residual stresses should come from \textit{local} flow cessation experiments in which a spherical intruder is dragged through the sample of interest before being brought to a complete stop.\cite{Gomez-Solano:2015} Numerical simulations of \textit{active} microrheology in a colloidal dispersion have already led to a micromechanical model of flow cessation, and have shown that the dispersion stress can be viewed as the duration of the microstructural memory.\cite{Zia:2013,Mohanty:2020} In general, additional experiments, particularly in gels, should be crucial in connecting the distorted microstructure inherited from flow cessation, the dynamics of local stress relaxation, and the macroscopic residual stress. 

\subsubsection{Non-monotonic stress relaxation}
    
Most soft glassy materials display a \textit{monotonic} stress relaxation upon flow cessation, irrespective of the shear rate applied prior to flow cessation. However, a few experimental systems were recently reported to display a surprising \textit{non-monotonic} stress relaxation [see, for instance, Fig.~\ref{fig:StressRelax}(b)].\cite{Murphy:2020,Hendricks:2019,Sudreau:2022b} Such an anomalous relaxation was attributed to two key ingredients, namely the alignment of the sample microstructure under flow, and the formation of bonds upon flow cessation due to interparticle attractive forces that locally drive bulk reorganization of the sample, hence inducing the growth of the macroscopic stress.\cite{Hendricks:2019,Sudreau:2023} Although counter-intuitive, such an anomalous relaxation was shown to be compatible with the second law of thermodynamics in the framework of a structural kinetic model.\cite{Joshi:2022} Yet, numerous issues remain open, including (i)~the minimal microscopic ingredients needed to observe a non-monotonic relaxation, (ii) the role of the shape of the colloids, their deformability,\cite{Bares:2022} and of the nature of interparticle (non-central) forces in such an anomalous relaxation, and (iii)~
the possibility for systems denser than gels and polymer solutions, e.g., soft repulsive glasses, to display similar stress relaxation. The interplay of these different ingredients could be naturally investigated in coarse-grained descriptions
such as elasto-plastic models.\cite{Nicolas:2018}

\begin{figure*}[t!]
\centering
\includegraphics[width=1\textwidth]{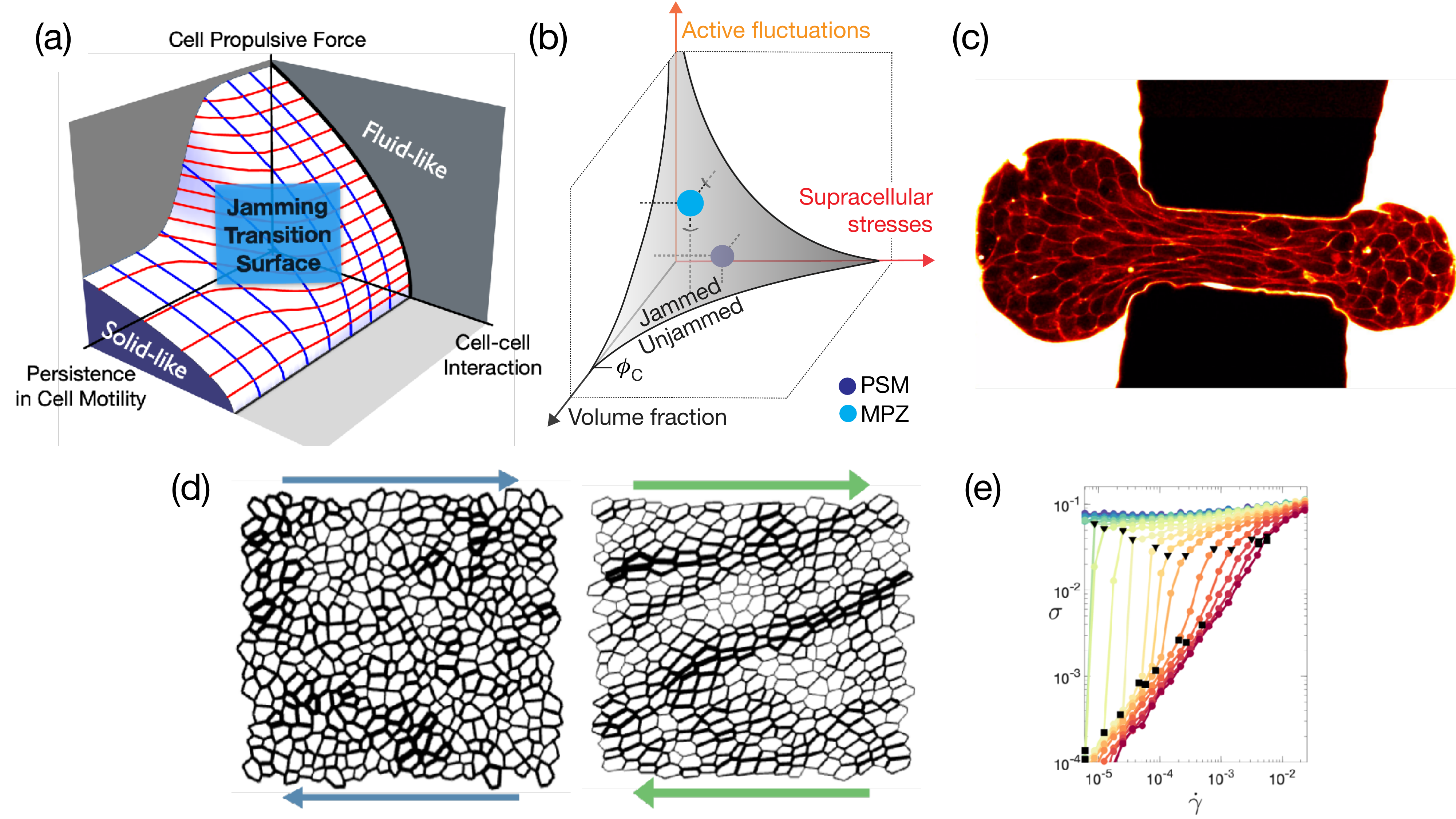}
\caption{(a) Phase diagram illustrating the jamming-glass transition within the framework of the self-propelled Voronoi model.\cite{bi2016motility,sussman:2017cellgpu,sussman:2018anomalous} This diagram explicates the conditions under which a confluent tissue transitions into a jammed state, maintaining a constant packing density. (b) explores the scenario where the packing density within a tissue is variable. Under these circumstances, the jamming-glass behavior exhibits sensitivity to fluctuations in packing density. A phase diagram that encapsulates this behavior, drawing parallels to the jamming transitions observed in particulate systems, has been recently introduced.\cite{mongera2018fluid} (a) and (b) are extracted from Ref.~\onlinecite{bi2016motility} and Ref.~\onlinecite{mongera2018fluid}, respectively. (c) Aspiration of an embryonic cellular aggregate through a constriction of size 50~$\mu$m used to generate large deformations to probe tissue elasto-capillary properties; picture extracted from Ref.\cite{tlili2022}
(d) The shear response of a two-dimensional tissue using the vertex model shows the emergence of system-spanning tension chains. These structures bear a striking resemblance to the force chains observed in granular materials, suggesting a commonality in the stress response mechanisms between these disparate systems.
(e) The flow curve resulting from a model tissue under shear, where the intricate balance between global external forces and local internal active drivers dictates the overall mechanical properties. In proximity to the jamming/unjamming transition, the tissue exhibits a spectrum of intriguing rheological behaviors. These include yielding, shear thinning, and both continuous (CST) and discontinuous shear thickening (DST). (d) and (e) are extracted from Ref.\cite{hertaeg:2022}
}
\label{fig:bio}
\end{figure*}

\subsubsection{Outlook: identifying the local scenario underpinning the non-local stress relaxation}

The renewed interest in flow cessation experiments illustrates the relevance of this rheological test for understanding the dynamics of soft amorphous materials and for shedding new light on the yielding transition. Indeed, flow cessation can be perceived as a \textit{method} to probe the sample properties at a given point in time, following a well-controlled shear history. In this context, flow cessation experiments in soft glassy materials partially unveil the \textit{memory} embedded in the sample after an extended period of shear for sufficiently low shear rates. Lastly, a relevant question arises regarding the potential correlation between flow cessation and thixotropy.\cite{Larson:2019}

To conclude, one of the key open issues pertains to the microscopic scenario linked with the cessation of shear. While numerical results are available, there is, to date, no direct experimental evidence for structural anisotropy driving the dynamics of residual stresses. Future experiments, e.g., involving rheo-confocal or rheo-scattering experiments, are anticipated to provide an essential validation of the existing numerical scenarios and to prompt theoretical developments.

\section{Yielding, ductility, and brittleness in biological soft materials}
\label{sec:Bio}

\subsection{\large Biological materials as soft glasses}

Over the past few years, there has been overwhelming evidence supporting the importance of jammed (solid-like) to unjammed (liquid-like) transition and glassy behavior across various kingdoms of living organisms and across scales.\cite{Janssen:2019} 
These systems are \textit{de facto} amorphous, hence it is relevant to ask whether the usual glassy physics known for passive systems may be relevant for biological functions.
Intracellular examples include bacterial cytoplasm fluidization due to cellular metabolism,\cite{parry2014bacterial} self-organization of disk-like chloroplasts into an active glassy state in plant cells under dim light,\cite{schramma2023chloroplasts} and glassy behavior of intracellular motion in eukaryotic cells.\cite{aaberg2021glass,corci2023extending} On a larger scale, the role of rheology, and particularly of the jamming-unjamming transition [see Figs.~\ref{fig:bio}(a) and (b)], is an important aspect of collective cell mechanics and morphogenesis in multicellular organisms. A bridge to the physics of glassy systems has especially been made for in-vitro epithelial cell-sheets in experiment\cite{Garcia:2015,angelini2011glass,malinverno:2017,atia2018geometric,nelson2022mechanical,park2015unjamming,mongera2018fluid,mitchel:2020,atia2021cell,lawson2021jamming} and theoretically,~\cite{bi:2015, bi2016motility,Das:2021} but also in embryo development,\cite{petridou2019tissue, hannezo2022rigidity} or in cancer cell invasion.\cite{oswald2017jamming,blauth2021jamming,gottheil2023state} Lessons from soft condensed matter physics have become essential to analyzing the mechanical behavior of these active biological systems.\cite{tlili2015colloquium} However, the additional complexity of biological systems, which include internal active forces that interact with external forces, including feedback, introduces new challenges (see Fig.~\ref{fig:bio}). This then requires incorporating these new microscopic mechanisms, such as local protein concentration and regulation through active transport inside the cell or mechanotransduction pathways\cite{sergides2021} into our approaches, which impose different time scales and length scales.

\subsection{\large Emergent and collective behavior in biological glassy systems} 

The mechanical properties of biological systems emerge from the molecular level to the cell level, and eventually to the tissue level. Despite great progress in recent years,\cite{Harris:2012,doubrovinski:2017,petridou2019tissue,hannezo2019mechanochemical,pinheiro2022morphogen} many questions remain open. For instance, how exactly do biomolecular interactions control the glassiness of the cytoplasm? What is the role of the cell cytoskeleton in the tissue's emergent collective behavior? How do the mechanical properties of organelles, such as nuclei, affect the mechanical properties of tissues on a much larger scale? How does the rheology of a tissue determine its role in morphogenesis? What changes in the computational models of biological systems should be adopted to incorporate these effects, and what experimental protocol must be developed to investigate this systematically?

Furthermore, biological samples often actively respond to various stimuli, including mechanical forces.\cite{hoffman2009cell,janmey2007cell,Sadeghipour:2018, ladoux_mechanobiology_2017} This poses a great challenge as the mechanical response of the system is also a function of the system's mechanosensing, i.e., there is feedback.\cite{Roca-Cusachs:2013,trepat:2011plithotaxis} This raises many open questions, among them: How do cell mechanosensing and active cell decision-making affect the emergent collective cell behavior and the mechanical properties on a tissue level? What is the detailed role of signaling pathways in such potential contributions? Accordingly, how can we computationally and experimentally study such an effect?
\\

\subsection{\large Challenges in comparing models and experiments}

Many models have been crafted to elucidate the behavior of epithelial tissues under stress, such as the Potts, Voronoi, and vertex models. See Ref.~\cite{beatrici2023} for an in-depth comparative analysis. The discussions at the Lorentz Centre were particularly centered on the intricacies of vertex-based models that were originally devised for foams.\cite{Alt:2017}

The vertex model is a sophisticated computational architecture designed to simulate and decode the mechanical attributes of tissue structures. It particularly focuses on the dynamics of cell packing, proliferation, and division.\cite{Nagai:2001,Farhadifar:2007,Staple:2010} This model conceptualizes tissues as a two-dimensional mosaic of polygonal cells, with vertices representing cell junctions and edges denoting cell boundaries. By imbuing edges with tensional forces and cells with elastic properties, the model adeptly mimics the mechanical conduct of cells within a tissue matrix. 

Vertex-based models have recently shed light on the phenomenon of glassiness in densely packed tissues.\cite{bi:2015,Das:2021} The self-propelled Voronoi model,\cite{bi2016motility,sussman:2017cellgpu,sussman:2018anomalous} in particular, has been noteworthy for its comprehensive integration of polarized cellular motion and the intricate web of cell-cell interactions within confluent tissues, where cells are tightly packed without gaps. This model, readily available as an open-source code,\cite{Barton:2017} has been a cornerstone in making quantitative predictions about the glass transition phenomena in tissue biology [Fig.~\ref{fig:bio}(a)].
The vertex model methodology, augmented by detail on biological feedback mechanisms, is becoming pivotal in dissecting the intricacies of morphogenesis --the developmental stages where tissues acquire their shape and structure-- and in shedding light on the underlying physical principles governing these biological events.\cite{fletcher2014vertex, sknepnek2023generating,boocock_interplay_2023,claussen_geometric_2023}

On the side of fundamentally understanding cell rheology, recent experiments on the aspiration of embryonic cell aggregates into a microfluidic constriction [see Fig.~\ref{fig:bio}(c)] have revealed the existence of characteristic time scales and critical shear rates, fixing the domains in between which rearrangements can be observed.\cite{tlili2022} These experiments also showed that a viscosity could be difficult to define and measure in this context. All these different ingredients still remain to be incorporated into vertex models.

Moreover, these models are, for now, mainly used for describing cell monolayers that are considered two-dimensional systems. Nevertheless, it is not clear whether such a 2D assumption is always appropriate. Indeed, in monolayers, cells interact with a substrate and present an apico-basal polarity perpendicular to the substrate.\cite{lo2012,ladoux_mechanobiology_2017} More generally, including properly the effects of boundary conditions, such as the differential role of substrate friction and pair dissipation between cells, is crucial to describe the dynamics of deforming tissues. In experiments, the interactions with the substrate and the occurrence of slip can be adjusted through adhesive proteins, allowing cells to form biological focal adhesions.\cite{zamir1999}  Furthermore, the role of boundary conditions is even more significant for developmental biology systems,\cite{karzbrun2021human} as their entire behavior and response may depend on the rigidity of the substrate,\cite{Lo2000,Plotnikov2012} the surrounding matrix in organoids or tumor spheroids, or on molecular properties,\cite{elosegui2014} exhibiting mechanical or chemical sensitivity.  This sensitivity mirrors the one of active systems, where the rigidity of the boundaries influences the dynamic response of active particles.\cite{solon2015}  This discovery raises intriguing questions about the interplay between boundary conditions and the dynamics of biological systems.

\subsection{\large Non-linear rheology of biological tissues}

The response of biological tissues to external or internal stresses can be relaxed through different cellular processes, such as cell rearrangements, divisions, and apoptosis, but also through active cell response that reorganizes the sub-cellular structure. Due to different modes of stress dissipation, as well as an inherent active mechanical noise, biological tissues are typically ductile, and the response becomes brittle-like when cellular adhesion fails to sustain imposed stresses and a tear in the tissue appears.\cite{Prakash:2021} Such tearing failure can occur in a ductile tissue when a small defect is introduced at which the stress is concentrated upon loading.\cite{Harris:2012,Chen:2022}

More generally, there are many open questions regarding the non-linear rheology of biological materials. For instance, do tissue models and biopolymer networks sit in the same stiffening universality class, and how does shear-stiffening influence ductility of tissues?~\cite{huang:2022,hertaeg:2022} What are the dominant/relevant length and time scales controlling the ductile-brittle transition in biological tissues? How does elastoplastic tissue flow differ from glassy elastoplastic flow, and how much information about such flow can be inferred from tissue cellular structure?\cite{Popovic:2021} How do cell division and apoptosis modify the mechanical noise compared to standard plasticity in passive systems, leading to non-linear rheological properties?\cite{Matoz:2017}

\section{Material design}
\label{sec:MatDe}    
Numerous hard materials from major industries, i.e., foodstuff, personal care, and building materials, are obtained from soft viscoelastic precursors. Representative examples include cement paste, a colloidal gel of calcium silicate hydrate, the “glue” that, once hardened, provides strength to concrete,\cite{Ioannidou:2016} and acid-induced gels of microcrystalline boehmite, which serve as soft precursors for alumina-derived materials of controlled porosity such as catalysts support.\cite{Trimm:1986} In practice, controlling the properties of these soft precursors is critical to controlling the terminal properties of the corresponding hardened material. As discussed in Sect.~\ref{sec:Residual}, the microstructure and mechanical properties of these soft precursors can be reinforced along the direction of shear through moderate shear intensity followed by flow cessation. At fixed sample composition, it is, in principle, possible to build complex shear protocols to control precisely the texture of soft precursors and potentially inherit that texture in the hardened state of the material. In the case of boehmite gels, for instance, it was recently shown that shear history can be used to confer some anisotropy to the gel microstructure, which then results in a strong increase of the gel elasticity upon flow cessation.\cite{Sudreau:2022} This effect could be potentially used to control the porosity of catalyst support obtained from these boehmite gels. Such a strategy was recently applied to the fabrication of strong and tough cellulose nanocrystals (CNC)-epoxy composites. The microstructure of a gel precursor comprising CNCs and epoxide oligomers dispersed in a solvent is printed by direct ink writing, before being crosslinked to form dry, solid nanocomposites whose properties are inherited from the soft precursor.\cite{Rao:2019,Rao:2022} Similar strategies can be employed to tune the optical properties of amorphous packings serving as soft precursors to make materials with \textit{structural colors}.\cite{Vynck:2023} 
 
The specific case of foodstuff is particularly relevant in terms of materials design, for edible materials are subject to intricate mechanical processes typically involving large deformations during production, mastication, and digestion. For example, pasta dough undergoes extrusion through a die to create long, thin strands, cheese curds are pressed to make cheese (with different textures, e.g., brittle, like the crumbly nature of feta cheese, or ductile, like the plastic flow of cream cheese), and bread dough is kneaded and folded to develop gluten structure and make a bread with a desirable texture.  In the context of dairy products, the processing incorporates various principles covered in this workshop, i.e., shear-induced yielding, either brittle or ductile, as well as residual stresses and shear-induced texture frozen upon flow cessation. For instance, the preparation of stirred yogurt initially entails a sol-gel transition based on lactose fermentation that yields a brittle soft solid, which is then broken down into a dispersion of gel particles that dictates the texture and the sensory quality of the yogurt.\cite{Guenard:2020} Another striking example is related to the fabrication of butter. In practice, the cream is churned to separate the butterfat from the buttermilk. As a result of large deformations, the butterfat clumps together and forms butter, whose final texture is controlled by the shear history.\cite{Mazzanti:2003} The same conclusions hold for crystallization under shear in general, as evidenced in cocoa butter,\cite{Sonwai:2006} and oil-in-water emulsions.\cite{Yang:2011} These examples illustrate the relevance of studying dairy products as model systems\cite{Bauland:2023} to elucidate some of the open questions highlighted in the framework of this workshop.  

\section{Conclusion and outlooks}
   
During the workshop that took place at the Lorentz Center, we identified several challenges and open questions that emerged during the discussion sessions. In the following, we highlight what we believe are the six main open challenges that should serve as a guideline for the community in the years to come. 

\begin{enumerate}
\item {{\bf Role of shear bands at the yielding transition --} Experiments, numerical and theoretical results highlight the relevance of shear bands in understanding the yielding transition. For the brittle transition, one has to understand the cause-effect relation between shear band formation and the observed brittleness, and whether a possible universal picture is emerging.  Somehow similar is the case for the ductile transition, where long-lasting shear bands are one possible mechanism to achieve complete fluidization.}

\item{{\bf Quantitative measure of brittleness --} Although some proposals have emerged, there is a clear need to quantify brittleness in experiments and/or numerical simulations. This is a critical first step if one wants to compare results using different theoretical approaches or different material properties and/or preparations in a quantitative way. }

\item{{\bf Theoretical Frameworks --} So far, there exist different theoretical frameworks to discuss and/or explain different material behavior at the yielding transition (e.g., DBT, creep flow, fluidization time, and rate dependence of the stress overshoot).  Most of these frameworks or models show qualitative agreement with one or more features observed at the yielding transition. However, it is imperative to progress by offering quantitative predictions that can be compared against empirical data. Moreover, it is important to understand how different frameworks are linked with different material properties, identify possible limitations and generalizations, and, last but not least, suggest new experiments and/or data analysis to validate underlying theories.
}

\item{{\bf Microscopic view --} Several different topics need deeper investigations at the microscopic level, both experimentally and numerically. Here, we provide a non-exhaustive list:
 \begin{itemize}
 \item {Identify microscopic signatures for the DBT, determine a microscopic scenario underpinning the existence of long-lasting shear bands, and investigate gels at the yield point.}
 \item{Investigate the role played by surface topology and microscopic interactions, if possible independently, on wall slip via numerical simulations and experiments.}
 \item{Identify particle-scale and/or mesoscopic dynamics relevant to memory effects and to the non-monotonic stress relaxation.}
 \end{itemize}
 Moreover, a recent study \cite{Martinelli:2023} indicates the intriguing possibility of reaching a state close to the yield point of an oxide glass by homogeneous irradiation rather than mechanical deformation. Comparing the macroscopic and microscopic properties of systems prepared in such different manners could lead to new insights concerning the yielding transition.
}

\item{{\bf Rheology of biological materials --} There are many open questions about the interaction between single-cell behaviour and the large-scale rheology of living organisms. Most of the above points need to be investigated in detail for the specific case of biological materials. Another critical concern involves exploring potential connections between memory effects observed at the rheological level and the biochemical dynamics occurring at the single-cell level, if any such interaction exists.}

\item{{\bf Materials design --} Use shear-induced memory effects to modify the properties of soft viscoelastic precursors employed to make hard materials, such as foodstuffs or cementitious materials.}

\end{enumerate}

\section*{Acknowledgements}
The authors thank Hugo L. Fran\c{c}a, Edan Lerner, and Nico Schramma for fruitful discussions and express their gratitude for the technical assistance provided by the Lorentz Center, with special acknowledgment to Marieke Brock and Daniëlle van Rijk for their efforts in coordinating the organization of the workshop. The organizers, namely C.~Barentin, R.~Benzi, T.~Divoux, S.~Manneville, M.~Sbragaglia, and F.~Toschi, extend their appreciation to Lorentz Center, Centro Ricerche Enrico Fermi, CNRS, Saint-Gobain Recherche, Groupe Français de Rhéologie, Université Lyon 1, Eindhoven University of Technology, and the J.M. Burgerscentrum for their financial support.

\section*{Conflicts of interest}
There are no conflicts to declare.



\begin{mcitethebibliography}{272}
\providecommand*{\natexlab}[1]{#1}
\providecommand*{\mciteSetBstSublistMode}[1]{}
\providecommand*{\mciteSetBstMaxWidthForm}[2]{}
\providecommand*{\mciteBstWouldAddEndPuncttrue}
  {\def\EndOfBibitem{\unskip.}}
\providecommand*{\mciteBstWouldAddEndPunctfalse}
  {\let\EndOfBibitem\relax}
\providecommand*{\mciteSetBstMidEndSepPunct}[3]{}
\providecommand*{\mciteSetBstSublistLabelBeginEnd}[3]{}
\providecommand*{\EndOfBibitem}{}
\mciteSetBstSublistMode{f}
\mciteSetBstMaxWidthForm{subitem}
{(\emph{\alph{mcitesubitemcount}})}
\mciteSetBstSublistLabelBeginEnd{\mcitemaxwidthsubitemform\space}
{\relax}{\relax}

\bibitem[C\'elari\'e \emph{et~al.}(2003)C\'elari\'e, Prades, Bonamy, Ferrero,
  Bouchaud, Guillot, and Marli\`ere]{Celarie2003Glass}
F.~C\'elari\'e, S.~Prades, D.~Bonamy, L.~Ferrero, E.~Bouchaud, C.~Guillot and
  C.~Marli\`ere, \emph{Phys. Rev. Lett.}, 2003, \textbf{90}, 075504\relax
\mciteBstWouldAddEndPuncttrue
\mciteSetBstMidEndSepPunct{\mcitedefaultmidpunct}
{\mcitedefaultendpunct}{\mcitedefaultseppunct}\relax
\EndOfBibitem
\bibitem[Bonn \emph{et~al.}(2017)Bonn, Denn, Berthier, Divoux, and
  Manneville]{Bonn:2017}
D.~Bonn, M.~M. Denn, L.~Berthier, T.~Divoux and S.~Manneville, \emph{Rev. Mod.
  Phys.}, 2017, \textbf{89}, 035005\relax
\mciteBstWouldAddEndPuncttrue
\mciteSetBstMidEndSepPunct{\mcitedefaultmidpunct}
{\mcitedefaultendpunct}{\mcitedefaultseppunct}\relax
\EndOfBibitem
\bibitem[Maloney and Lema{\^\i}tre(2004)]{MaloneyLemaitre2004a}
C.~Maloney and A.~Lema{\^\i}tre, \emph{Phys. Rev. Lett.}, 2004, \textbf{93},
  016001\relax
\mciteBstWouldAddEndPuncttrue
\mciteSetBstMidEndSepPunct{\mcitedefaultmidpunct}
{\mcitedefaultendpunct}{\mcitedefaultseppunct}\relax
\EndOfBibitem
\bibitem[Maloney and Lema{\^\i}tre(2006)]{MaloneyLemaitre2006}
C.~E. Maloney and A.~Lema{\^\i}tre, \emph{Phys. Rev. E}, 2006, \textbf{74},
  016118\relax
\mciteBstWouldAddEndPuncttrue
\mciteSetBstMidEndSepPunct{\mcitedefaultmidpunct}
{\mcitedefaultendpunct}{\mcitedefaultseppunct}\relax
\EndOfBibitem
\bibitem[Schall \emph{et~al.}(2007)Schall, Weitz, and Spaepen]{Schall:2007}
P.~Schall, D.~A. Weitz and F.~Spaepen, \emph{Science}, 2007, \textbf{318},
  1895--1899\relax
\mciteBstWouldAddEndPuncttrue
\mciteSetBstMidEndSepPunct{\mcitedefaultmidpunct}
{\mcitedefaultendpunct}{\mcitedefaultseppunct}\relax
\EndOfBibitem
\bibitem[Argon and Kuo(1979)]{ArgonKuo1979}
A.~S. Argon and H.~Y. Kuo, \emph{Materials Science and Engineering}, 1979,
  \textbf{39}, 101--109\relax
\mciteBstWouldAddEndPuncttrue
\mciteSetBstMidEndSepPunct{\mcitedefaultmidpunct}
{\mcitedefaultendpunct}{\mcitedefaultseppunct}\relax
\EndOfBibitem
\bibitem[Chattoraj \emph{et~al.}(2010)Chattoraj, Caroli, and
  Lema{\^\i}tre]{ChattorajCaroliLemaitre2010}
J.~Chattoraj, C.~Caroli and A.~Lema{\^\i}tre, \emph{Phys. Rev. Lett.}, 2010,
  \textbf{105}, 266001\relax
\mciteBstWouldAddEndPuncttrue
\mciteSetBstMidEndSepPunct{\mcitedefaultmidpunct}
{\mcitedefaultendpunct}{\mcitedefaultseppunct}\relax
\EndOfBibitem
\bibitem[Nicolas \emph{et~al.}(2018)Nicolas, Ferrero, Martens, and
  Barrat]{Nicolas:2018}
A.~Nicolas, E.~E. Ferrero, K.~Martens and J.-L. Barrat, \emph{Rev. Mod. Phys.},
  2018, \textbf{90}, 045006\relax
\mciteBstWouldAddEndPuncttrue
\mciteSetBstMidEndSepPunct{\mcitedefaultmidpunct}
{\mcitedefaultendpunct}{\mcitedefaultseppunct}\relax
\EndOfBibitem
\bibitem[Fielding(2014)]{Fielding:2014}
S.~Fielding, \emph{Rep. Prog. Phys.}, 2014, \textbf{77}, 102601\relax
\mciteBstWouldAddEndPuncttrue
\mciteSetBstMidEndSepPunct{\mcitedefaultmidpunct}
{\mcitedefaultendpunct}{\mcitedefaultseppunct}\relax
\EndOfBibitem
\bibitem[Divoux \emph{et~al.}(2016)Divoux, Fardin, Manneville, and
  Lerouge]{Divoux:2016}
T.~Divoux, M.-A. Fardin, S.~Manneville and S.~Lerouge, \emph{Annu. Rev. Fluid
  Mech.}, 2016, \textbf{48}, 81--103\relax
\mciteBstWouldAddEndPuncttrue
\mciteSetBstMidEndSepPunct{\mcitedefaultmidpunct}
{\mcitedefaultendpunct}{\mcitedefaultseppunct}\relax
\EndOfBibitem
\bibitem[Lu \emph{et~al.}(2003)Lu, Ravichandran, and Johnson]{Lu:2003}
J.~Lu, G.~Ravichandran and W.~L. Johnson, \emph{Acta Mater.}, 2003,
  \textbf{51}, 3429--3443\relax
\mciteBstWouldAddEndPuncttrue
\mciteSetBstMidEndSepPunct{\mcitedefaultmidpunct}
{\mcitedefaultendpunct}{\mcitedefaultseppunct}\relax
\EndOfBibitem
\bibitem[Dimitriou and McKinley(2014)]{Dimitriou:2014}
C.~J. Dimitriou and G.~H. McKinley, \emph{Soft Matter}, 2014, \textbf{10},
  6619--6644\relax
\mciteBstWouldAddEndPuncttrue
\mciteSetBstMidEndSepPunct{\mcitedefaultmidpunct}
{\mcitedefaultendpunct}{\mcitedefaultseppunct}\relax
\EndOfBibitem
\bibitem[Keshavarz \emph{et~al.}(2017)Keshavarz, Divoux, Manneville, and
  McKinley]{Keshavarz:2017}
B.~Keshavarz, T.~Divoux, S.~Manneville and G.~H. McKinley, \emph{ACS Macro
  Letters}, 2017, \textbf{6}, 663--667\relax
\mciteBstWouldAddEndPuncttrue
\mciteSetBstMidEndSepPunct{\mcitedefaultmidpunct}
{\mcitedefaultendpunct}{\mcitedefaultseppunct}\relax
\EndOfBibitem
\bibitem[Gibaud \emph{et~al.}(2010)Gibaud, Frelat, and Manneville]{Gibaud:2010}
T.~Gibaud, D.~Frelat and S.~Manneville, \emph{Soft Matter}, 2010, \textbf{6},
  3482--3488\relax
\mciteBstWouldAddEndPuncttrue
\mciteSetBstMidEndSepPunct{\mcitedefaultmidpunct}
{\mcitedefaultendpunct}{\mcitedefaultseppunct}\relax
\EndOfBibitem
\bibitem[Lindstr\"om \emph{et~al.}(2012)Lindstr\"om, Kodger, Sprakel, and
  Weitz]{Lindstrom:2012}
S.~Lindstr\"om, T.~Kodger, J.~Sprakel and D.~Weitz, \emph{Soft Matter}, 2012,
  \textbf{8}, 3657--3664\relax
\mciteBstWouldAddEndPuncttrue
\mciteSetBstMidEndSepPunct{\mcitedefaultmidpunct}
{\mcitedefaultendpunct}{\mcitedefaultseppunct}\relax
\EndOfBibitem
\bibitem[Colombo and {Del Gado}(2014)]{Colombo:2014}
J.~Colombo and E.~{Del Gado}, \emph{J. Rheol.}, 2014, \textbf{58},
  1089--1116\relax
\mciteBstWouldAddEndPuncttrue
\mciteSetBstMidEndSepPunct{\mcitedefaultmidpunct}
{\mcitedefaultendpunct}{\mcitedefaultseppunct}\relax
\EndOfBibitem
\bibitem[Perge \emph{et~al.}(2014)Perge, Taberlet, Gibaud, and
  Manneville]{Perge:2014b}
C.~Perge, N.~Taberlet, T.~Gibaud and S.~Manneville, \emph{J. Rheol.}, 2014,
  \textbf{58}, 1331--1357\relax
\mciteBstWouldAddEndPuncttrue
\mciteSetBstMidEndSepPunct{\mcitedefaultmidpunct}
{\mcitedefaultendpunct}{\mcitedefaultseppunct}\relax
\EndOfBibitem
\bibitem[Schuh \emph{et~al.}(2007)Schuh, Hufnagel, and Ramamurty]{Schuh:2007}
C.~A. Schuh, T.~C. Hufnagel and U.~Ramamurty, \emph{Acta Mater.}, 2007,
  \textbf{55}, 4067--4109\relax
\mciteBstWouldAddEndPuncttrue
\mciteSetBstMidEndSepPunct{\mcitedefaultmidpunct}
{\mcitedefaultendpunct}{\mcitedefaultseppunct}\relax
\EndOfBibitem
\bibitem[Gu \emph{et~al.}(2009)Gu, Poon, Shiflet, and Lewandowski]{Gu:2009}
X.~Gu, S.~Poon, G.~Shiflet and J.~Lewandowski, \emph{Scr. Mater.}, 2009,
  \textbf{60}, 1027--1030\relax
\mciteBstWouldAddEndPuncttrue
\mciteSetBstMidEndSepPunct{\mcitedefaultmidpunct}
{\mcitedefaultendpunct}{\mcitedefaultseppunct}\relax
\EndOfBibitem
\bibitem[Luo \emph{et~al.}(2016)Luo, Wang, Bitzek, Huang, Zheng, Tong, Yang,
  Li, and Mao]{Luo:2016}
J.~Luo, J.~Wang, E.~Bitzek, J.~Y. Huang, H.~Zheng, L.~Tong, Q.~Yang, J.~Li and
  S.~X. Mao, \emph{Nano Lett.}, 2016, \textbf{16}, 105--113\relax
\mciteBstWouldAddEndPuncttrue
\mciteSetBstMidEndSepPunct{\mcitedefaultmidpunct}
{\mcitedefaultendpunct}{\mcitedefaultseppunct}\relax
\EndOfBibitem
\bibitem[Macias-Rodriguez \emph{et~al.}(2018)Macias-Rodriguez, Ewoldt, and
  Marangoni]{Macias:2018}
B.~A. Macias-Rodriguez, R.~H. Ewoldt and A.~G. Marangoni, \emph{Rheol. Acta},
  2018, \textbf{57}, 251--266\relax
\mciteBstWouldAddEndPuncttrue
\mciteSetBstMidEndSepPunct{\mcitedefaultmidpunct}
{\mcitedefaultendpunct}{\mcitedefaultseppunct}\relax
\EndOfBibitem
\bibitem[Macias-Rodriguez and Marangoni(2018)]{Macias:2018b}
B.~A. Macias-Rodriguez and A.~G. Marangoni, \emph{Crit. Rev. Food. Sci. Nutr.},
  2018, \textbf{58}, 2398--2415\relax
\mciteBstWouldAddEndPuncttrue
\mciteSetBstMidEndSepPunct{\mcitedefaultmidpunct}
{\mcitedefaultendpunct}{\mcitedefaultseppunct}\relax
\EndOfBibitem
\bibitem[Meyers \emph{et~al.}(2006)Meyers, Lin, Seki, Chen, Kad, and
  Bodde]{Meyers:2006}
M.~A. Meyers, A.~Y. Lin, Y.~Seki, P.-Y. Chen, B.~K. Kad and S.~Bodde,
  \emph{JOM}, 2006, \textbf{58}, 35--41\relax
\mciteBstWouldAddEndPuncttrue
\mciteSetBstMidEndSepPunct{\mcitedefaultmidpunct}
{\mcitedefaultendpunct}{\mcitedefaultseppunct}\relax
\EndOfBibitem
\bibitem[Peterlik \emph{et~al.}(2006)Peterlik, Roschger, Klaushofer, and
  Fratzl]{Peterlik:2006}
H.~Peterlik, P.~Roschger, K.~Klaushofer and P.~Fratzl, \emph{Nat. Mater.},
  2006, \textbf{5}, 52--55\relax
\mciteBstWouldAddEndPuncttrue
\mciteSetBstMidEndSepPunct{\mcitedefaultmidpunct}
{\mcitedefaultendpunct}{\mcitedefaultseppunct}\relax
\EndOfBibitem
\bibitem[Barthelat \emph{et~al.}(2016)Barthelat, Yin, and
  Buehler]{Barthelat:2016}
F.~Barthelat, Z.~Yin and M.~J. Buehler, \emph{Nat. Rev. Mater.}, 2016,
  \textbf{1}, 1--16\relax
\mciteBstWouldAddEndPuncttrue
\mciteSetBstMidEndSepPunct{\mcitedefaultmidpunct}
{\mcitedefaultendpunct}{\mcitedefaultseppunct}\relax
\EndOfBibitem
\bibitem[Ozawa \emph{et~al.}(2022)Ozawa, Berthier, Biroli, and
  Tarjus]{Ozawa:2022}
M.~Ozawa, L.~Berthier, G.~Biroli and G.~Tarjus, \emph{Phys. Rev. Res.}, 2022,
  \textbf{4}, 023227\relax
\mciteBstWouldAddEndPuncttrue
\mciteSetBstMidEndSepPunct{\mcitedefaultmidpunct}
{\mcitedefaultendpunct}{\mcitedefaultseppunct}\relax
\EndOfBibitem
\bibitem[Vasisht and Del~Gado(2020)]{Vasisht:2020}
V.~V. Vasisht and E.~Del~Gado, \emph{Phys. Rev. E}, 2020, \textbf{102},
  012603\relax
\mciteBstWouldAddEndPuncttrue
\mciteSetBstMidEndSepPunct{\mcitedefaultmidpunct}
{\mcitedefaultendpunct}{\mcitedefaultseppunct}\relax
\EndOfBibitem
\bibitem[Leocmach \emph{et~al.}(2014)Leocmach, Perge, Divoux, and
  Manneville]{Leocmach:2014}
M.~Leocmach, C.~Perge, T.~Divoux and S.~Manneville, \emph{Phys. Rev. Lett.},
  2014, \textbf{113}, 038303\relax
\mciteBstWouldAddEndPuncttrue
\mciteSetBstMidEndSepPunct{\mcitedefaultmidpunct}
{\mcitedefaultendpunct}{\mcitedefaultseppunct}\relax
\EndOfBibitem
\bibitem[Aime \emph{et~al.}(2018)Aime, Ramos, and Cipelletti]{Aime:2018}
S.~Aime, L.~Ramos and L.~Cipelletti, \emph{Proc. Natl. Acad. Sci. U.S.A.},
  2018, \textbf{115}, 3587--3592\relax
\mciteBstWouldAddEndPuncttrue
\mciteSetBstMidEndSepPunct{\mcitedefaultmidpunct}
{\mcitedefaultendpunct}{\mcitedefaultseppunct}\relax
\EndOfBibitem
\bibitem[Barlow \emph{et~al.}(2020)Barlow, Cochran, and Fielding]{Barlow:2020}
H.~J. Barlow, J.~O. Cochran and S.~M. Fielding, \emph{Phys. Rev. Lett.}, 2020,
  \textbf{125}, 168003\relax
\mciteBstWouldAddEndPuncttrue
\mciteSetBstMidEndSepPunct{\mcitedefaultmidpunct}
{\mcitedefaultendpunct}{\mcitedefaultseppunct}\relax
\EndOfBibitem
\bibitem[Ozawa \emph{et~al.}(2018)Ozawa, Berthier, Biroli, Rosso, and
  Tarjus]{Ozawa:2018}
M.~Ozawa, L.~Berthier, G.~Biroli, A.~Rosso and G.~Tarjus, \emph{Proc. Natl.
  Acad. Sci. USA}, 2018, \textbf{115}, 6656--6661\relax
\mciteBstWouldAddEndPuncttrue
\mciteSetBstMidEndSepPunct{\mcitedefaultmidpunct}
{\mcitedefaultendpunct}{\mcitedefaultseppunct}\relax
\EndOfBibitem
\bibitem[Rossi \emph{et~al.}(2022)Rossi, Biroli, Ozawa, Tarjus, and
  Zamponi]{Rossi:2022}
S.~Rossi, G.~Biroli, M.~Ozawa, G.~Tarjus and F.~Zamponi, \emph{Phys. Rev.
  Lett.}, 2022, \textbf{129}, 228002\relax
\mciteBstWouldAddEndPuncttrue
\mciteSetBstMidEndSepPunct{\mcitedefaultmidpunct}
{\mcitedefaultendpunct}{\mcitedefaultseppunct}\relax
\EndOfBibitem
\bibitem[Kamani \emph{et~al.}(2021)Kamani, Donley, and Rogers]{Kamani:2021}
K.~Kamani, G.~J. Donley and S.~A. Rogers, \emph{Phys. Rev. Lett.}, 2021,
  \textbf{126}, 218002\relax
\mciteBstWouldAddEndPuncttrue
\mciteSetBstMidEndSepPunct{\mcitedefaultmidpunct}
{\mcitedefaultendpunct}{\mcitedefaultseppunct}\relax
\EndOfBibitem
\bibitem[Singh \emph{et~al.}(2020)Singh, Ozawa, and Berthier]{Singh:2020}
M.~Singh, M.~Ozawa and L.~Berthier, \emph{Phys. Rev. Mater.}, 2020, \textbf{4},
  025603\relax
\mciteBstWouldAddEndPuncttrue
\mciteSetBstMidEndSepPunct{\mcitedefaultmidpunct}
{\mcitedefaultendpunct}{\mcitedefaultseppunct}\relax
\EndOfBibitem
\bibitem[Benzi \emph{et~al.}(2021)Benzi, Divoux, Barentin, Manneville,
  Sbragaglia, and Toschi]{Benzi:2021PRE}
R.~Benzi, T.~Divoux, C.~Barentin, S.~Manneville, M.~Sbragaglia and F.~Toschi,
  \emph{Phys. Rev. E}, 2021, \textbf{104}, 034612\relax
\mciteBstWouldAddEndPuncttrue
\mciteSetBstMidEndSepPunct{\mcitedefaultmidpunct}
{\mcitedefaultendpunct}{\mcitedefaultseppunct}\relax
\EndOfBibitem
\bibitem[Barbot \emph{et~al.}(2020)Barbot, Lerbinger, Lema{\^\i}tre,
  Vandembroucq, and Patinet]{BarbotLerbingerLemaitreVandembroucqPatinet2020}
A.~Barbot, M.~Lerbinger, A.~Lema{\^\i}tre, D.~Vandembroucq and S.~Patinet,
  \emph{Phys. Rev. E}, 2020, \textbf{101}, 033001\relax
\mciteBstWouldAddEndPuncttrue
\mciteSetBstMidEndSepPunct{\mcitedefaultmidpunct}
{\mcitedefaultendpunct}{\mcitedefaultseppunct}\relax
\EndOfBibitem
\bibitem[Martens \emph{et~al.}(2012)Martens, Bocquet, and Barrat]{Martens:2012}
K.~Martens, L.~Bocquet and J.-L. Barrat, \emph{Soft Matter}, 2012, \textbf{8},
  4197--4205\relax
\mciteBstWouldAddEndPuncttrue
\mciteSetBstMidEndSepPunct{\mcitedefaultmidpunct}
{\mcitedefaultendpunct}{\mcitedefaultseppunct}\relax
\EndOfBibitem
\bibitem[Liu \emph{et~al.}(2018)Liu, Martens, and Barrat]{Liu:2018}
C.~Liu, K.~Martens and J.-L. Barrat, \emph{Phys. Rev. Lett.}, 2018,
  \textbf{120}, 028004\relax
\mciteBstWouldAddEndPuncttrue
\mciteSetBstMidEndSepPunct{\mcitedefaultmidpunct}
{\mcitedefaultendpunct}{\mcitedefaultseppunct}\relax
\EndOfBibitem
\bibitem[Benzi \emph{et~al.}(2023)Benzi, Divoux, Barentin, Manneville,
  Sbragaglia, and Toschi]{Benzi:2023}
R.~Benzi, T.~Divoux, C.~Barentin, S.~Manneville, M.~Sbragaglia and F.~Toschi,
  \emph{Europhys. Lett.}, 2023, \textbf{141}, 56001\relax
\mciteBstWouldAddEndPuncttrue
\mciteSetBstMidEndSepPunct{\mcitedefaultmidpunct}
{\mcitedefaultendpunct}{\mcitedefaultseppunct}\relax
\EndOfBibitem
\bibitem[Ozawa \emph{et~al.}(2020)Ozawa, Berthier, Biroli, and
  Tarjus]{Ozawa:2020}
M.~Ozawa, L.~Berthier, G.~Biroli and G.~Tarjus, \emph{Phys. Rev. Res.}, 2020,
  \textbf{2}, 023203\relax
\mciteBstWouldAddEndPuncttrue
\mciteSetBstMidEndSepPunct{\mcitedefaultmidpunct}
{\mcitedefaultendpunct}{\mcitedefaultseppunct}\relax
\EndOfBibitem
\bibitem[Falk and Langer(1998)]{FalkLanger1998}
M.~L. Falk and J.~S. Langer, \emph{Phys. Rev. E}, 1998, \textbf{57}, 7192\relax
\mciteBstWouldAddEndPuncttrue
\mciteSetBstMidEndSepPunct{\mcitedefaultmidpunct}
{\mcitedefaultendpunct}{\mcitedefaultseppunct}\relax
\EndOfBibitem
\bibitem[Schuh and Lund(2003)]{SchuhLund2003}
C.~A. Schuh and A.~C. Lund, \emph{Nature Materials}, 2003, \textbf{2},
  449--452\relax
\mciteBstWouldAddEndPuncttrue
\mciteSetBstMidEndSepPunct{\mcitedefaultmidpunct}
{\mcitedefaultendpunct}{\mcitedefaultseppunct}\relax
\EndOfBibitem
\bibitem[Eshelby(1957)]{eshelby1957determination}
J.~D. Eshelby, \emph{Proc. R. Soc. A: Math. Phys. Eng. Sci.}, 1957,
  \textbf{241}, 376--396\relax
\mciteBstWouldAddEndPuncttrue
\mciteSetBstMidEndSepPunct{\mcitedefaultmidpunct}
{\mcitedefaultendpunct}{\mcitedefaultseppunct}\relax
\EndOfBibitem
\bibitem[Picard \emph{et~al.}(2002)Picard, Ajdari, Bocquet, and
  Lequeux]{Picard:2002}
G.~Picard, A.~Ajdari, L.~Bocquet and F.~Lequeux, \emph{Phys. Rev. E}, 2002,
  \textbf{66}, 051501\relax
\mciteBstWouldAddEndPuncttrue
\mciteSetBstMidEndSepPunct{\mcitedefaultmidpunct}
{\mcitedefaultendpunct}{\mcitedefaultseppunct}\relax
\EndOfBibitem
\bibitem[Maloney and Lema{\^\i}tre(2004)]{MaloneyLemaitre2004b}
C.~Maloney and A.~Lema{\^\i}tre, \emph{Phys. Rev. Lett.}, 2004, \textbf{93},
  195501\relax
\mciteBstWouldAddEndPuncttrue
\mciteSetBstMidEndSepPunct{\mcitedefaultmidpunct}
{\mcitedefaultendpunct}{\mcitedefaultseppunct}\relax
\EndOfBibitem
\bibitem[Tanguy \emph{et~al.}(2006)Tanguy, Leonforte, and
  Barrat]{TanguyLeonforteBarrat2006}
A.~Tanguy, F.~Leonforte and J.~L. Barrat, \emph{Eur. Phys. J. E}, 2006,
  \textbf{20}, 355--364\relax
\mciteBstWouldAddEndPuncttrue
\mciteSetBstMidEndSepPunct{\mcitedefaultmidpunct}
{\mcitedefaultendpunct}{\mcitedefaultseppunct}\relax
\EndOfBibitem
\bibitem[Lema{\^\i}tre and Caroli(2009)]{LemaitreCaroli2009}
A.~Lema{\^\i}tre and C.~Caroli, \emph{Phys. Rev. Lett.}, 2009, \textbf{103},
  065501\relax
\mciteBstWouldAddEndPuncttrue
\mciteSetBstMidEndSepPunct{\mcitedefaultmidpunct}
{\mcitedefaultendpunct}{\mcitedefaultseppunct}\relax
\EndOfBibitem
\bibitem[Le~Bouil \emph{et~al.}(2014)Le~Bouil, Amon, McNamara, and
  Crassous]{le2014emergence}
A.~Le~Bouil, A.~Amon, S.~McNamara and J.~Crassous, \emph{Phys. Rev. Lett.},
  2014, \textbf{112}, 246001\relax
\mciteBstWouldAddEndPuncttrue
\mciteSetBstMidEndSepPunct{\mcitedefaultmidpunct}
{\mcitedefaultendpunct}{\mcitedefaultseppunct}\relax
\EndOfBibitem
\bibitem[Karmakar \emph{et~al.}(2010)Karmakar, Lerner, and Procaccia]{10KLPb}
S.~Karmakar, E.~Lerner and I.~Procaccia, \emph{Phys. Rev. E}, 2010,
  \textbf{82}, 055103\relax
\mciteBstWouldAddEndPuncttrue
\mciteSetBstMidEndSepPunct{\mcitedefaultmidpunct}
{\mcitedefaultendpunct}{\mcitedefaultseppunct}\relax
\EndOfBibitem
\bibitem[Bailey \emph{et~al.}(2007)Bailey, Schiotz, Lema{\^\i}tre, and
  Jacobsen]{BaileySchiotzLemaitreJacobsen2007}
N.~P. Bailey, J.~Schiotz, A.~Lema{\^\i}tre and K.~W. Jacobsen, \emph{Phys. Rev.
  Lett.}, 2007, \textbf{98}, 095501\relax
\mciteBstWouldAddEndPuncttrue
\mciteSetBstMidEndSepPunct{\mcitedefaultmidpunct}
{\mcitedefaultendpunct}{\mcitedefaultseppunct}\relax
\EndOfBibitem
\bibitem[Maloney and Robbins(2008)]{MaloneyRobbins2008}
C.~E. Maloney and M.~O. Robbins, \emph{J. Phys. Chem.}, 2008, \textbf{20},
  244128\relax
\mciteBstWouldAddEndPuncttrue
\mciteSetBstMidEndSepPunct{\mcitedefaultmidpunct}
{\mcitedefaultendpunct}{\mcitedefaultseppunct}\relax
\EndOfBibitem
\bibitem[Lerner and Procaccia(2009)]{LernerProcaccia2009}
E.~Lerner and I.~Procaccia, \emph{Physical Review E}, 2009, \textbf{79},
  066109\relax
\mciteBstWouldAddEndPuncttrue
\mciteSetBstMidEndSepPunct{\mcitedefaultmidpunct}
{\mcitedefaultendpunct}{\mcitedefaultseppunct}\relax
\EndOfBibitem
\bibitem[Hentschel \emph{et~al.}(2011)Hentschel, Karmakar, Lerner, and
  Procaccia]{11HKLP}
H.~G.~E. Hentschel, S.~Karmakar, E.~Lerner and I.~Procaccia, \emph{Phys. Rev.
  E}, 2011, \textbf{83}, 061101\relax
\mciteBstWouldAddEndPuncttrue
\mciteSetBstMidEndSepPunct{\mcitedefaultmidpunct}
{\mcitedefaultendpunct}{\mcitedefaultseppunct}\relax
\EndOfBibitem
\bibitem[Salerno and Robbins(2013)]{SalernoRobbins2013}
K.~M. Salerno and M.~O. Robbins, \emph{Physical Review E}, 2013, \textbf{88},
  062206\relax
\mciteBstWouldAddEndPuncttrue
\mciteSetBstMidEndSepPunct{\mcitedefaultmidpunct}
{\mcitedefaultendpunct}{\mcitedefaultseppunct}\relax
\EndOfBibitem
\bibitem[Chattoraj and Lema{\^\i}tre(2013)]{ChattorajLemaitre2013}
J.~Chattoraj and A.~Lema{\^\i}tre, \emph{Phys. Rev. Lett.}, 2013, \textbf{111},
  066001\relax
\mciteBstWouldAddEndPuncttrue
\mciteSetBstMidEndSepPunct{\mcitedefaultmidpunct}
{\mcitedefaultendpunct}{\mcitedefaultseppunct}\relax
\EndOfBibitem
\bibitem[Houdoux \emph{et~al.}(2018)Houdoux, Nguyen, Amon, and
  Crassous]{HoudouxNguyenAmonCrassous2018}
D.~Houdoux, T.~B. Nguyen, A.~Amon and J.~Crassous, \emph{Phys. Rev. E}, 2018,
  \textbf{98}, 022905\relax
\mciteBstWouldAddEndPuncttrue
\mciteSetBstMidEndSepPunct{\mcitedefaultmidpunct}
{\mcitedefaultendpunct}{\mcitedefaultseppunct}\relax
\EndOfBibitem
\bibitem[Richard \emph{et~al.}(2020)Richard, Ozawa, Patinet, Stanifer, Shang,
  Ridout, Xu, Zhang, Morse, Barrat, Berthier, Falk, Guan, Liu, Martens, Sastry,
  Vandembroucq, Lerner, and Manning]{Richard:2020}
D.~Richard, M.~Ozawa, S.~Patinet, E.~Stanifer, B.~Shang, S.~A. Ridout, B.~Xu,
  G.~Zhang, P.~K. Morse, J.-L. Barrat, L.~Berthier, M.~L. Falk, P.~Guan, A.~J.
  Liu, K.~Martens, S.~Sastry, D.~Vandembroucq, E.~Lerner and M.~L. Manning,
  \emph{Phys. Rev. Mater.}, 2020, \textbf{4}, 113609\relax
\mciteBstWouldAddEndPuncttrue
\mciteSetBstMidEndSepPunct{\mcitedefaultmidpunct}
{\mcitedefaultendpunct}{\mcitedefaultseppunct}\relax
\EndOfBibitem
\bibitem[Puosi \emph{et~al.}(2015)Puosi, Olivier, and
  Martens]{PuosiOlivierMartens2015}
F.~Puosi, J.~Olivier and K.~Martens, \emph{Soft Matter}, 2015, \textbf{11},
  7639--7647\relax
\mciteBstWouldAddEndPuncttrue
\mciteSetBstMidEndSepPunct{\mcitedefaultmidpunct}
{\mcitedefaultendpunct}{\mcitedefaultseppunct}\relax
\EndOfBibitem
\bibitem[Patinet \emph{et~al.}(2016)Patinet, Vandembroucq, and
  Falk]{Patinet:2016}
S.~Patinet, D.~Vandembroucq and M.~L. Falk, \emph{Phys. Rev. Lett.}, 2016,
  \textbf{117}, 045501\relax
\mciteBstWouldAddEndPuncttrue
\mciteSetBstMidEndSepPunct{\mcitedefaultmidpunct}
{\mcitedefaultendpunct}{\mcitedefaultseppunct}\relax
\EndOfBibitem
\bibitem[Dasgupta \emph{et~al.}(2012)Dasgupta, Hentschel, and Procaccia]{12DHP}
R.~Dasgupta, H.~G.~E. Hentschel and I.~Procaccia, \emph{Phys. Rev. Lett.},
  2012, \textbf{109}, 255502\relax
\mciteBstWouldAddEndPuncttrue
\mciteSetBstMidEndSepPunct{\mcitedefaultmidpunct}
{\mcitedefaultendpunct}{\mcitedefaultseppunct}\relax
\EndOfBibitem
\bibitem[Dasgupta \emph{et~al.}(2013)Dasgupta, Hentschel, and Procaccia]{13DHP}
R.~Dasgupta, H.~G.~E. Hentschel and I.~Procaccia, \emph{Phys. Rev. E}, 2013,
  \textbf{87}, 022810\relax
\mciteBstWouldAddEndPuncttrue
\mciteSetBstMidEndSepPunct{\mcitedefaultmidpunct}
{\mcitedefaultendpunct}{\mcitedefaultseppunct}\relax
\EndOfBibitem
\bibitem[Lema\^{\i}tre \emph{et~al.}(2021)Lema\^{\i}tre, Mondal, Moshe,
  Procaccia, Roy, and Screiber-Re'em]{21LMMPRS}
A.~Lema\^{\i}tre, C.~Mondal, M.~Moshe, I.~Procaccia, S.~Roy and
  K.~Screiber-Re'em, \emph{Phys. Rev. E}, 2021, \textbf{104}, 024904\relax
\mciteBstWouldAddEndPuncttrue
\mciteSetBstMidEndSepPunct{\mcitedefaultmidpunct}
{\mcitedefaultendpunct}{\mcitedefaultseppunct}\relax
\EndOfBibitem
\bibitem[Charan \emph{et~al.}(2023)Charan, Moshe, and Procaccia]{23CMP}
H.~Charan, M.~Moshe and I.~Procaccia, \emph{Phys. Rev. E}, 2023, \textbf{107},
  055005\relax
\mciteBstWouldAddEndPuncttrue
\mciteSetBstMidEndSepPunct{\mcitedefaultmidpunct}
{\mcitedefaultendpunct}{\mcitedefaultseppunct}\relax
\EndOfBibitem
\bibitem[Mondal \emph{et~al.}(2022)Mondal, Moshe, Procaccia, Roy, Shang, and
  Zhang]{22MMPRSZ}
C.~Mondal, M.~Moshe, I.~Procaccia, S.~Roy, J.~Shang and J.~Zhang, \emph{Chaos
  Solitons Fractals}, 2022, \textbf{164}, 112609\relax
\mciteBstWouldAddEndPuncttrue
\mciteSetBstMidEndSepPunct{\mcitedefaultmidpunct}
{\mcitedefaultendpunct}{\mcitedefaultseppunct}\relax
\EndOfBibitem
\bibitem[Bhowmik \emph{et~al.}(2022)Bhowmik, Moshe, and Procaccia]{22BMP}
B.~P. Bhowmik, M.~Moshe and I.~Procaccia, \emph{Phys. Rev. E}, 2022,
  \textbf{105}, L043001\relax
\mciteBstWouldAddEndPuncttrue
\mciteSetBstMidEndSepPunct{\mcitedefaultmidpunct}
{\mcitedefaultendpunct}{\mcitedefaultseppunct}\relax
\EndOfBibitem
\bibitem[Mondal \emph{et~al.}(2023)Mondal, Moshe, Procaccia, and Roy]{23CMPR}
C.~Mondal, M.~Moshe, I.~Procaccia and S.~Roy, \emph{Dipole Screening in Pure
  Shear Strain Protocols of Amorphous Solids}, 2023\relax
\mciteBstWouldAddEndPuncttrue
\mciteSetBstMidEndSepPunct{\mcitedefaultmidpunct}
{\mcitedefaultendpunct}{\mcitedefaultseppunct}\relax
\EndOfBibitem
\bibitem[Jin \emph{et~al.}(2023)Jin, Procaccia, and Samanta]{23JPS}
Y.~Jin, I.~Procaccia and T.~Samanta, \emph{An intermediate phase between jammed
  and un-jammed amorphous solids}, 2023\relax
\mciteBstWouldAddEndPuncttrue
\mciteSetBstMidEndSepPunct{\mcitedefaultmidpunct}
{\mcitedefaultendpunct}{\mcitedefaultseppunct}\relax
\EndOfBibitem
\bibitem[Bar-Sinai \emph{et~al.}(2020)Bar-Sinai, Librandi, Bertoldi, and
  Moshe]{20BLBM}
Y.~Bar-Sinai, G.~Librandi, K.~Bertoldi and M.~Moshe, \emph{Proc. Natl. Acad.
  Sci. U.S.A.}, 2020, \textbf{117}, 10195--10202\relax
\mciteBstWouldAddEndPuncttrue
\mciteSetBstMidEndSepPunct{\mcitedefaultmidpunct}
{\mcitedefaultendpunct}{\mcitedefaultseppunct}\relax
\EndOfBibitem
\bibitem[Saint-Michel \emph{et~al.}(2017)Saint-Michel, Gibaud, and
  Manneville]{Saintmichel:2017}
B.~Saint-Michel, T.~Gibaud and S.~Manneville, \emph{Soft Matter}, 2017,
  \textbf{13}, 2643--2653\relax
\mciteBstWouldAddEndPuncttrue
\mciteSetBstMidEndSepPunct{\mcitedefaultmidpunct}
{\mcitedefaultendpunct}{\mcitedefaultseppunct}\relax
\EndOfBibitem
\bibitem[Cho and Bischofberger(2022)]{Cho:2022}
J.~H. Cho and I.~Bischofberger, \emph{Soft Matter}, 2022, \textbf{18},
  7612--7620\relax
\mciteBstWouldAddEndPuncttrue
\mciteSetBstMidEndSepPunct{\mcitedefaultmidpunct}
{\mcitedefaultendpunct}{\mcitedefaultseppunct}\relax
\EndOfBibitem
\bibitem[Pommella \emph{et~al.}(2020)Pommella, Cipelletti, and
  Ramos]{Pommella:2020}
A.~Pommella, L.~Cipelletti and L.~Ramos, \emph{Phys. Rev. Lett.}, 2020,
  \textbf{125}, 268006\relax
\mciteBstWouldAddEndPuncttrue
\mciteSetBstMidEndSepPunct{\mcitedefaultmidpunct}
{\mcitedefaultendpunct}{\mcitedefaultseppunct}\relax
\EndOfBibitem
\bibitem[Koivisto \emph{et~al.}(2016)Koivisto, Ovaska, Miksic, Laurson, and
  Alava]{Koivisto:2016}
J.~Koivisto, M.~Ovaska, A.~Miksic, L.~Laurson and M.~J. Alava, \emph{Phys. Rev.
  E}, 2016, \textbf{94}, 023002\relax
\mciteBstWouldAddEndPuncttrue
\mciteSetBstMidEndSepPunct{\mcitedefaultmidpunct}
{\mcitedefaultendpunct}{\mcitedefaultseppunct}\relax
\EndOfBibitem
\bibitem[Pommella \emph{et~al.}(2019)Pommella, Philippe, Phou, Ramos, and
  Cipelletti]{Pommella:2019}
A.~Pommella, A.-M. Philippe, T.~Phou, L.~Ramos and L.~Cipelletti, \emph{Phys.
  Rev. Applied}, 2019, \textbf{11}, 034073\relax
\mciteBstWouldAddEndPuncttrue
\mciteSetBstMidEndSepPunct{\mcitedefaultmidpunct}
{\mcitedefaultendpunct}{\mcitedefaultseppunct}\relax
\EndOfBibitem
\bibitem[Helal \emph{et~al.}(2016)Helal, Divoux, and McKinley]{Helal:2016}
A.~Helal, T.~Divoux and G.~H. McKinley, \emph{Phys. Rev. Applied}, 2016,
  \textbf{6}, 064004\relax
\mciteBstWouldAddEndPuncttrue
\mciteSetBstMidEndSepPunct{\mcitedefaultmidpunct}
{\mcitedefaultendpunct}{\mcitedefaultseppunct}\relax
\EndOfBibitem
\bibitem[Garcimartin \emph{et~al.}(1997)Garcimartin, Guarino, Bellon, and
  Ciliberto]{Garcimartin:1997}
A.~Garcimartin, A.~Guarino, L.~Bellon and S.~Ciliberto, \emph{Phys. Rev.
  Lett.}, 1997, \textbf{79}, 3202\relax
\mciteBstWouldAddEndPuncttrue
\mciteSetBstMidEndSepPunct{\mcitedefaultmidpunct}
{\mcitedefaultendpunct}{\mcitedefaultseppunct}\relax
\EndOfBibitem
\bibitem[Scuderi \emph{et~al.}(2016)Scuderi, Marone, Tinti, Di~Stefano, and
  Collettini]{Scuderi:2016}
M.~Scuderi, C.~Marone, E.~Tinti, G.~Di~Stefano and C.~Collettini, \emph{Nat.
  Geosci.}, 2016, \textbf{9}, 695--700\relax
\mciteBstWouldAddEndPuncttrue
\mciteSetBstMidEndSepPunct{\mcitedefaultmidpunct}
{\mcitedefaultendpunct}{\mcitedefaultseppunct}\relax
\EndOfBibitem
\bibitem[Slootman \emph{et~al.}(2020)Slootman, Waltz, Yeh, Baumann, G\"ostl,
  Comtet, and Creton]{Slootman:2020}
J.~Slootman, V.~Waltz, C.~J. Yeh, C.~Baumann, R.~G\"ostl, J.~Comtet and
  C.~Creton, \emph{Phys. Rev. X}, 2020, \textbf{10}, 041045\relax
\mciteBstWouldAddEndPuncttrue
\mciteSetBstMidEndSepPunct{\mcitedefaultmidpunct}
{\mcitedefaultendpunct}{\mcitedefaultseppunct}\relax
\EndOfBibitem
\bibitem[Bakun \emph{et~al.}(2005)Bakun, Aagaard, Dost, Ellsworth, Hardebeck,
  Harris, Ji, Johnston, Langbein, Lienkaemper,\emph{et~al.}]{Bakun:2005}
W.~Bakun, B.~Aagaard, B.~Dost, W.~L. Ellsworth, J.~L. Hardebeck, R.~A. Harris,
  C.~Ji, M.~J. Johnston, J.~Langbein, J.~J. Lienkaemper \emph{et~al.},
  \emph{Nature}, 2005, \textbf{437}, 969--974\relax
\mciteBstWouldAddEndPuncttrue
\mciteSetBstMidEndSepPunct{\mcitedefaultmidpunct}
{\mcitedefaultendpunct}{\mcitedefaultseppunct}\relax
\EndOfBibitem
\bibitem[Zhang \emph{et~al.}(2021)Zhang, Ridout, and Liu]{Zhang:2021}
G.~Zhang, S.~A. Ridout and A.~J. Liu, \emph{Phys. Rev. X}, 2021, \textbf{11},
  041019\relax
\mciteBstWouldAddEndPuncttrue
\mciteSetBstMidEndSepPunct{\mcitedefaultmidpunct}
{\mcitedefaultendpunct}{\mcitedefaultseppunct}\relax
\EndOfBibitem
\bibitem[Bouzid and Del~Gado(2017)]{Bouzid2017network}
M.~Bouzid and E.~Del~Gado, \emph{Langmuir}, 2017, \textbf{34}, 773--781\relax
\mciteBstWouldAddEndPuncttrue
\mciteSetBstMidEndSepPunct{\mcitedefaultmidpunct}
{\mcitedefaultendpunct}{\mcitedefaultseppunct}\relax
\EndOfBibitem
\bibitem[Tauber \emph{et~al.}(2020)Tauber, Dussi, and Van
  Der~Gucht]{tauber2020microscopic}
J.~Tauber, S.~Dussi and J.~Van Der~Gucht, \emph{Phys. Rev. Mater.}, 2020,
  \textbf{4}, 063603\relax
\mciteBstWouldAddEndPuncttrue
\mciteSetBstMidEndSepPunct{\mcitedefaultmidpunct}
{\mcitedefaultendpunct}{\mcitedefaultseppunct}\relax
\EndOfBibitem
\bibitem[Berthier \emph{et~al.}(2019)Berthier, Kollmer, Henkes, Liu, Schwarz,
  and Daniels]{berthier2019rigidity}
E.~Berthier, J.~E. Kollmer, S.~E. Henkes, K.~Liu, J.~M. Schwarz and K.~E.
  Daniels, \emph{Phys. Rev. Mater.}, 2019, \textbf{3}, 075602\relax
\mciteBstWouldAddEndPuncttrue
\mciteSetBstMidEndSepPunct{\mcitedefaultmidpunct}
{\mcitedefaultendpunct}{\mcitedefaultseppunct}\relax
\EndOfBibitem
\bibitem[Javerzat and Bouzid(2023)]{javerzat2023evidences}
N.~Javerzat and M.~Bouzid, \emph{Phys. Rev. Lett.}, 2023, \textbf{130},
  268201\relax
\mciteBstWouldAddEndPuncttrue
\mciteSetBstMidEndSepPunct{\mcitedefaultmidpunct}
{\mcitedefaultendpunct}{\mcitedefaultseppunct}\relax
\EndOfBibitem
\bibitem[Lewandowski \emph{et~al.}(2005)Lewandowski, Wang, and
  Greer]{Lewandowski:2005}
J.~Lewandowski, W.~Wang and A.~Greer, \emph{Philos. Mag. Lett.}, 2005,
  \textbf{85}, 77--87\relax
\mciteBstWouldAddEndPuncttrue
\mciteSetBstMidEndSepPunct{\mcitedefaultmidpunct}
{\mcitedefaultendpunct}{\mcitedefaultseppunct}\relax
\EndOfBibitem
\bibitem[Aime \emph{et~al.}(2023)Aime, Truzzolillo, Pine, Ramos, and
  Cipelletti]{aime2023}
S.~Aime, D.~Truzzolillo, D.~J. Pine, L.~Ramos and L.~Cipelletti, \emph{Nat.
  Phys.}, 2023,  1--7\relax
\mciteBstWouldAddEndPuncttrue
\mciteSetBstMidEndSepPunct{\mcitedefaultmidpunct}
{\mcitedefaultendpunct}{\mcitedefaultseppunct}\relax
\EndOfBibitem
\bibitem[Divoux \emph{et~al.}(2010)Divoux, Tamarii, Barentin, and
  Manneville]{Divoux:2010}
T.~Divoux, D.~Tamarii, C.~Barentin and S.~Manneville, \emph{Phys. Rev. Lett.},
  2010, \textbf{104}, 208301\relax
\mciteBstWouldAddEndPuncttrue
\mciteSetBstMidEndSepPunct{\mcitedefaultmidpunct}
{\mcitedefaultendpunct}{\mcitedefaultseppunct}\relax
\EndOfBibitem
\bibitem[Divoux \emph{et~al.}(2012)Divoux, Tamarii, Barentin, Teitel, and
  Manneville]{Divoux:2012}
T.~Divoux, D.~Tamarii, C.~Barentin, S.~Teitel and S.~Manneville, \emph{Soft
  Matter}, 2012, \textbf{8}, 4151--4164\relax
\mciteBstWouldAddEndPuncttrue
\mciteSetBstMidEndSepPunct{\mcitedefaultmidpunct}
{\mcitedefaultendpunct}{\mcitedefaultseppunct}\relax
\EndOfBibitem
\bibitem[Martin and Hu(2012)]{Martin:2012}
J.~Martin and Y.~Hu, \emph{Soft Matter}, 2012, \textbf{8}, 6940--6949\relax
\mciteBstWouldAddEndPuncttrue
\mciteSetBstMidEndSepPunct{\mcitedefaultmidpunct}
{\mcitedefaultendpunct}{\mcitedefaultseppunct}\relax
\EndOfBibitem
\bibitem[Divoux \emph{et~al.}(2011)Divoux, Barentin, and
  Manneville]{Divoux:2011b}
T.~Divoux, C.~Barentin and S.~Manneville, \emph{Soft Matter}, 2011, \textbf{7},
  8409--8418\relax
\mciteBstWouldAddEndPuncttrue
\mciteSetBstMidEndSepPunct{\mcitedefaultmidpunct}
{\mcitedefaultendpunct}{\mcitedefaultseppunct}\relax
\EndOfBibitem
\bibitem[Grenard \emph{et~al.}(2014)Grenard, Divoux, Taberlet, and
  Manneville]{Grenard:2014}
V.~Grenard, T.~Divoux, N.~Taberlet and S.~Manneville, \emph{Soft Matter}, 2014,
  \textbf{10}, 1555--1571\relax
\mciteBstWouldAddEndPuncttrue
\mciteSetBstMidEndSepPunct{\mcitedefaultmidpunct}
{\mcitedefaultendpunct}{\mcitedefaultseppunct}\relax
\EndOfBibitem
\bibitem[Caton and Baravian(2008)]{Caton:2008}
F.~Caton and C.~Baravian, \emph{Rheol. Acta}, 2008, \textbf{47}, 601--607\relax
\mciteBstWouldAddEndPuncttrue
\mciteSetBstMidEndSepPunct{\mcitedefaultmidpunct}
{\mcitedefaultendpunct}{\mcitedefaultseppunct}\relax
\EndOfBibitem
\bibitem[Siebenb\"urger \emph{et~al.}(2012)Siebenb\"urger, Ballauf, and
  Voigtmann]{Siebenburger:2012a}
M.~Siebenb\"urger, M.~Ballauf and T.~Voigtmann, \emph{Phys. Rev. Lett.}, 2012,
  \textbf{108}, 255701\relax
\mciteBstWouldAddEndPuncttrue
\mciteSetBstMidEndSepPunct{\mcitedefaultmidpunct}
{\mcitedefaultendpunct}{\mcitedefaultseppunct}\relax
\EndOfBibitem
\bibitem[Moorcroft and Fielding(2013)]{Moorcroft:2013}
R.~Moorcroft and S.~Fielding, \emph{Phys. Rev. Lett.}, 2013, \textbf{110},
  086001\relax
\mciteBstWouldAddEndPuncttrue
\mciteSetBstMidEndSepPunct{\mcitedefaultmidpunct}
{\mcitedefaultendpunct}{\mcitedefaultseppunct}\relax
\EndOfBibitem
\bibitem[Fielding(2016)]{Fielding:2016}
S.~M. Fielding, \emph{J. Rheol.}, 2016, \textbf{60}, 821--834\relax
\mciteBstWouldAddEndPuncttrue
\mciteSetBstMidEndSepPunct{\mcitedefaultmidpunct}
{\mcitedefaultendpunct}{\mcitedefaultseppunct}\relax
\EndOfBibitem
\bibitem[Moorcroft and Fielding(2014)]{Moorcroft:2014}
R.~L. Moorcroft and S.~M. Fielding, \emph{J. Rheol.}, 2014, \textbf{58},
  103--147\relax
\mciteBstWouldAddEndPuncttrue
\mciteSetBstMidEndSepPunct{\mcitedefaultmidpunct}
{\mcitedefaultendpunct}{\mcitedefaultseppunct}\relax
\EndOfBibitem
\bibitem[Sharma \emph{et~al.}(2021)Sharma, Shankar, and Joshi]{Sharma:2021}
S.~Sharma, V.~Shankar and Y.~M. Joshi, \emph{J. Rheol.}, 2021, \textbf{65},
  1391--1412\relax
\mciteBstWouldAddEndPuncttrue
\mciteSetBstMidEndSepPunct{\mcitedefaultmidpunct}
{\mcitedefaultendpunct}{\mcitedefaultseppunct}\relax
\EndOfBibitem
\bibitem[Sharma \emph{et~al.}(2023)Sharma, Joshi, and
  Shankar]{Sharma:2023preprint}
S.~Sharma, Y.~M. Joshi and V.~Shankar, \emph{arXiv preprint arXiv:2302.06129},
  2023\relax
\mciteBstWouldAddEndPuncttrue
\mciteSetBstMidEndSepPunct{\mcitedefaultmidpunct}
{\mcitedefaultendpunct}{\mcitedefaultseppunct}\relax
\EndOfBibitem
\bibitem[Briole \emph{et~al.}(2021)Briole, Casanellas, Fardin, Py, Cardoso,
  Browaeys, and Lerouge]{Briole:2021}
A.~Briole, L.~Casanellas, M.-A. Fardin, C.~Py, O.~Cardoso, J.~Browaeys and
  S.~Lerouge, \emph{J. Rheol.}, 2021, \textbf{65}, 1201--1217\relax
\mciteBstWouldAddEndPuncttrue
\mciteSetBstMidEndSepPunct{\mcitedefaultmidpunct}
{\mcitedefaultendpunct}{\mcitedefaultseppunct}\relax
\EndOfBibitem
\bibitem[B{\'e}cu \emph{et~al.}(2006)B{\'e}cu, Manneville, and
  Colin]{Becu:2006}
L.~B{\'e}cu, S.~Manneville and A.~Colin, \emph{Phys. Rev. Lett.}, 2006,
  \textbf{96}, 138302\relax
\mciteBstWouldAddEndPuncttrue
\mciteSetBstMidEndSepPunct{\mcitedefaultmidpunct}
{\mcitedefaultendpunct}{\mcitedefaultseppunct}\relax
\EndOfBibitem
\bibitem[Schall and van Hecke(2010)]{Schall:2010}
P.~Schall and M.~van Hecke, \emph{Annu. Rev. Fluid Mech.}, 2010, \textbf{42},
  67--88\relax
\mciteBstWouldAddEndPuncttrue
\mciteSetBstMidEndSepPunct{\mcitedefaultmidpunct}
{\mcitedefaultendpunct}{\mcitedefaultseppunct}\relax
\EndOfBibitem
\bibitem[Besseling \emph{et~al.}(2010)Besseling, Ballesta, Petekidis, Cates,
  and Poon]{Besseling:2010}
R.~Besseling, L.~I.~P. Ballesta, G.~Petekidis, M.~Cates and W.~Poon,
  \emph{Phys. Rev. Lett.}, 2010, \textbf{105}, 268301\relax
\mciteBstWouldAddEndPuncttrue
\mciteSetBstMidEndSepPunct{\mcitedefaultmidpunct}
{\mcitedefaultendpunct}{\mcitedefaultseppunct}\relax
\EndOfBibitem
\bibitem[Fall \emph{et~al.}(2010)Fall, Paredes, and Bonn]{Fall:2010b}
A.~Fall, J.~Paredes and D.~Bonn, \emph{Phys. Rev. Lett.}, 2010, \textbf{105},
  225502\relax
\mciteBstWouldAddEndPuncttrue
\mciteSetBstMidEndSepPunct{\mcitedefaultmidpunct}
{\mcitedefaultendpunct}{\mcitedefaultseppunct}\relax
\EndOfBibitem
\bibitem[Coussot and Ovarlez(2010)]{Coussot:2010}
P.~Coussot and G.~Ovarlez, \emph{Eur. Phys. J. E}, 2010, \textbf{33},
  183--188\relax
\mciteBstWouldAddEndPuncttrue
\mciteSetBstMidEndSepPunct{\mcitedefaultmidpunct}
{\mcitedefaultendpunct}{\mcitedefaultseppunct}\relax
\EndOfBibitem
\bibitem[Chikkadi \emph{et~al.}(2014)Chikkadi, Miedema, Dang, Nienhuis, and
  Schall]{Chikkadi:2014}
V.~Chikkadi, D.~Miedema, M.~Dang, B.~Nienhuis and P.~Schall, \emph{Phys. Rev.
  Lett.}, 2014, \textbf{113}, 208301\relax
\mciteBstWouldAddEndPuncttrue
\mciteSetBstMidEndSepPunct{\mcitedefaultmidpunct}
{\mcitedefaultendpunct}{\mcitedefaultseppunct}\relax
\EndOfBibitem
\bibitem[Shi and Falk(2005)]{Shi:2005}
Y.~Shi and M.~L. Falk, \emph{Phys. Rev. Lett.}, 2005, \textbf{95}, 095502\relax
\mciteBstWouldAddEndPuncttrue
\mciteSetBstMidEndSepPunct{\mcitedefaultmidpunct}
{\mcitedefaultendpunct}{\mcitedefaultseppunct}\relax
\EndOfBibitem
\bibitem[Greer \emph{et~al.}(2013)Greer, Cheng, and Ma]{Greer:2013}
A.~Greer, Y.~Cheng and E.~Ma, \emph{Mater. Sci. Eng. R Rep.}, 2013,
  \textbf{74}, 71--132\relax
\mciteBstWouldAddEndPuncttrue
\mciteSetBstMidEndSepPunct{\mcitedefaultmidpunct}
{\mcitedefaultendpunct}{\mcitedefaultseppunct}\relax
\EndOfBibitem
\bibitem[Vanel \emph{et~al.}(2009)Vanel, Ciliberto, Cortet, and
  Santucci]{Vanel:2009}
L.~Vanel, S.~Ciliberto, P.-P. Cortet and S.~Santucci, \emph{J. Phys. D: Appl.
  Phys.}, 2009, \textbf{42}, 214007\relax
\mciteBstWouldAddEndPuncttrue
\mciteSetBstMidEndSepPunct{\mcitedefaultmidpunct}
{\mcitedefaultendpunct}{\mcitedefaultseppunct}\relax
\EndOfBibitem
\bibitem[Popovi\ifmmode~\acute{c}\else \'{c}\fi{}
  \emph{et~al.}(2022)Popovi\ifmmode~\acute{c}\else \'{c}\fi{}, de~Geus, Ji,
  Rosso, and Wyart]{Popovic:2022}
M.~Popovi\ifmmode~\acute{c}\else \'{c}\fi{}, T.~W.~J. de~Geus, W.~Ji, A.~Rosso
  and M.~Wyart, \emph{Phys. Rev. Lett.}, 2022, \textbf{129}, 208001\relax
\mciteBstWouldAddEndPuncttrue
\mciteSetBstMidEndSepPunct{\mcitedefaultmidpunct}
{\mcitedefaultendpunct}{\mcitedefaultseppunct}\relax
\EndOfBibitem
\bibitem[Bray(2002)]{Bray:2002}
A.~J. Bray, \emph{Adv. Phys.}, 2002, \textbf{51}, 481--587\relax
\mciteBstWouldAddEndPuncttrue
\mciteSetBstMidEndSepPunct{\mcitedefaultmidpunct}
{\mcitedefaultendpunct}{\mcitedefaultseppunct}\relax
\EndOfBibitem
\bibitem[Moorcroft \emph{et~al.}(2011)Moorcroft, Cates, and
  Fielding]{Moorcroft:2011}
R.~Moorcroft, M.~Cates and S.~Fielding, \emph{Phys. Rev. Lett.}, 2011,
  \textbf{106}, 055502\relax
\mciteBstWouldAddEndPuncttrue
\mciteSetBstMidEndSepPunct{\mcitedefaultmidpunct}
{\mcitedefaultendpunct}{\mcitedefaultseppunct}\relax
\EndOfBibitem
\bibitem[Benzi \emph{et~al.}(2021)Benzi, Divoux, Barentin, Manneville,
  Sbragaglia, and Toschi]{Benzi:2021PRL}
R.~Benzi, T.~Divoux, C.~Barentin, S.~Manneville, M.~Sbragaglia and F.~Toschi,
  \emph{Phys. Rev. Lett.}, 2021, \textbf{127}, 148003\relax
\mciteBstWouldAddEndPuncttrue
\mciteSetBstMidEndSepPunct{\mcitedefaultmidpunct}
{\mcitedefaultendpunct}{\mcitedefaultseppunct}\relax
\EndOfBibitem
\bibitem[Nicolas \emph{et~al.}(2016)Nicolas, Barrat, and Rottler]{Nicolas:2016}
A.~Nicolas, J.-L. Barrat and J.~Rottler, \emph{Phys. Rev. Lett.}, 2016,
  \textbf{116}, 058303\relax
\mciteBstWouldAddEndPuncttrue
\mciteSetBstMidEndSepPunct{\mcitedefaultmidpunct}
{\mcitedefaultendpunct}{\mcitedefaultseppunct}\relax
\EndOfBibitem
\bibitem[Jain \emph{et~al.}(2018)Jain, Singh, Kushwaha, Shankar, and
  Joshi]{Jain:2018}
A.~Jain, R.~Singh, L.~Kushwaha, V.~Shankar and Y.~M. Joshi, \emph{J. Rheol.},
  2018, \textbf{62}, 1001--1016\relax
\mciteBstWouldAddEndPuncttrue
\mciteSetBstMidEndSepPunct{\mcitedefaultmidpunct}
{\mcitedefaultendpunct}{\mcitedefaultseppunct}\relax
\EndOfBibitem
\bibitem[Kushwaha \emph{et~al.}(2022)Kushwaha, Shankar, and
  Joshi]{Kushwaha:2022}
L.~Kushwaha, V.~Shankar and Y.~M. Joshi, \emph{Rheol. Acta}, 2022, \textbf{61},
  355 – 371\relax
\mciteBstWouldAddEndPuncttrue
\mciteSetBstMidEndSepPunct{\mcitedefaultmidpunct}
{\mcitedefaultendpunct}{\mcitedefaultseppunct}\relax
\EndOfBibitem
\bibitem[Colombo \emph{et~al.}(2019)Colombo, Massaro, Coleman, L{\"a}uger,
  Puyvelde, and Vermant]{Colombo:2019}
G.~Colombo, R.~Massaro, S.~Coleman, J.~L{\"a}uger, P.~V. Puyvelde and
  J.~Vermant, \emph{Korea Aust. Rheol. J.}, 2019, \textbf{31}, 229--240\relax
\mciteBstWouldAddEndPuncttrue
\mciteSetBstMidEndSepPunct{\mcitedefaultmidpunct}
{\mcitedefaultendpunct}{\mcitedefaultseppunct}\relax
\EndOfBibitem
\bibitem[P{\'e}m{\'e}ja \emph{et~al.}(2019)P{\'e}m{\'e}ja, G{\'e}raud,
  Barentin, and Le~Merrer]{Pemeja:2019}
J.~P{\'e}m{\'e}ja, B.~G{\'e}raud, C.~Barentin and M.~Le~Merrer, \emph{Phys.
  Rev. Fluids}, 2019, \textbf{4}, 033301\relax
\mciteBstWouldAddEndPuncttrue
\mciteSetBstMidEndSepPunct{\mcitedefaultmidpunct}
{\mcitedefaultendpunct}{\mcitedefaultseppunct}\relax
\EndOfBibitem
\bibitem[Benzi \emph{et~al.}(2019)Benzi, Divoux, Barentin, Manneville,
  Sbragaglia, and Toschi]{Benzi:2019}
R.~Benzi, T.~Divoux, C.~Barentin, S.~Manneville, M.~Sbragaglia and F.~Toschi,
  \emph{Phys. Rev. Lett.}, 2019, \textbf{123}, 248001\relax
\mciteBstWouldAddEndPuncttrue
\mciteSetBstMidEndSepPunct{\mcitedefaultmidpunct}
{\mcitedefaultendpunct}{\mcitedefaultseppunct}\relax
\EndOfBibitem
\bibitem[Divoux \emph{et~al.}(2011)Divoux, Barentin, and
  Manneville]{Divoux:2011}
T.~Divoux, C.~Barentin and S.~Manneville, \emph{Soft Matter}, 2011, \textbf{7},
  9335--9349\relax
\mciteBstWouldAddEndPuncttrue
\mciteSetBstMidEndSepPunct{\mcitedefaultmidpunct}
{\mcitedefaultendpunct}{\mcitedefaultseppunct}\relax
\EndOfBibitem
\bibitem[M{\"u}ller \emph{et~al.}(2023)M{\"u}ller, Isa, and
  Vermant]{Muller:2023}
F.~J. M{\"u}ller, L.~Isa and J.~Vermant, \emph{Nat. Commun.}, 2023,
  \textbf{14}, 5309\relax
\mciteBstWouldAddEndPuncttrue
\mciteSetBstMidEndSepPunct{\mcitedefaultmidpunct}
{\mcitedefaultendpunct}{\mcitedefaultseppunct}\relax
\EndOfBibitem
\bibitem[Chaudhuri \emph{et~al.}(2012)Chaudhuri, Berthier, and
  Bocquet]{chaudhuri2012inhomogeneous}
P.~Chaudhuri, L.~Berthier and L.~Bocquet, \emph{Phys. Rev. E}, 2012,
  \textbf{85}, 021503\relax
\mciteBstWouldAddEndPuncttrue
\mciteSetBstMidEndSepPunct{\mcitedefaultmidpunct}
{\mcitedefaultendpunct}{\mcitedefaultseppunct}\relax
\EndOfBibitem
\bibitem[Vasisht \emph{et~al.}(2020)Vasisht, Roberts, and
  Del~Gado]{Vasisht:2020b}
V.~V. Vasisht, G.~Roberts and E.~Del~Gado, \emph{Phys. Rev. E}, 2020,
  \textbf{102}, 010604\relax
\mciteBstWouldAddEndPuncttrue
\mciteSetBstMidEndSepPunct{\mcitedefaultmidpunct}
{\mcitedefaultendpunct}{\mcitedefaultseppunct}\relax
\EndOfBibitem
\bibitem[Rajaram and Mohraz(2010)]{Rajaram:2010}
B.~Rajaram and A.~Mohraz, \emph{Soft Matter}, 2010, \textbf{6},
  2246--2259\relax
\mciteBstWouldAddEndPuncttrue
\mciteSetBstMidEndSepPunct{\mcitedefaultmidpunct}
{\mcitedefaultendpunct}{\mcitedefaultseppunct}\relax
\EndOfBibitem
\bibitem[Rajaram and Mohraz(2011)]{Rajaram:2011}
B.~Rajaram and A.~Mohraz, \emph{Phys. Rev. E}, 2011, \textbf{84}, 011405\relax
\mciteBstWouldAddEndPuncttrue
\mciteSetBstMidEndSepPunct{\mcitedefaultmidpunct}
{\mcitedefaultendpunct}{\mcitedefaultseppunct}\relax
\EndOfBibitem
\bibitem[Chan and Mohraz(2013)]{Chan:2013}
H.~Chan and A.~Mohraz, \emph{Rheol. Acta}, 2013, \textbf{52}, 383--394\relax
\mciteBstWouldAddEndPuncttrue
\mciteSetBstMidEndSepPunct{\mcitedefaultmidpunct}
{\mcitedefaultendpunct}{\mcitedefaultseppunct}\relax
\EndOfBibitem
\bibitem[Hatzikiriakos(2012)]{hatzikiriakos:2012}
S.~G. Hatzikiriakos, \emph{Progress in Polymer Science}, 2012, \textbf{37},
  624--643\relax
\mciteBstWouldAddEndPuncttrue
\mciteSetBstMidEndSepPunct{\mcitedefaultmidpunct}
{\mcitedefaultendpunct}{\mcitedefaultseppunct}\relax
\EndOfBibitem
\bibitem[Cloitre and Bonnecaze(2017)]{Cloitre:2017}
M.~Cloitre and R.~T. Bonnecaze, \emph{Rheol. Acta}, 2017, \textbf{56},
  283--305\relax
\mciteBstWouldAddEndPuncttrue
\mciteSetBstMidEndSepPunct{\mcitedefaultmidpunct}
{\mcitedefaultendpunct}{\mcitedefaultseppunct}\relax
\EndOfBibitem
\bibitem[Malkin and Patlazhan(2018)]{Malkin:2018}
A.~Y. Malkin and S.~Patlazhan, \emph{Adv. Colloid Interface Sci.}, 2018,
  \textbf{257}, 42--57\relax
\mciteBstWouldAddEndPuncttrue
\mciteSetBstMidEndSepPunct{\mcitedefaultmidpunct}
{\mcitedefaultendpunct}{\mcitedefaultseppunct}\relax
\EndOfBibitem
\bibitem[Gibaud \emph{et~al.}(2008)Gibaud, Barentin, and
  Manneville]{Gibaud:2008}
T.~Gibaud, C.~Barentin and S.~Manneville, \emph{Phys. Rev. Lett.}, 2008,
  \textbf{101}, 258302\relax
\mciteBstWouldAddEndPuncttrue
\mciteSetBstMidEndSepPunct{\mcitedefaultmidpunct}
{\mcitedefaultendpunct}{\mcitedefaultseppunct}\relax
\EndOfBibitem
\bibitem[Kalyon(2005)]{Kalyon:2005}
D.~Kalyon, \emph{J. Rheol.}, 2005, \textbf{49}, 621--640\relax
\mciteBstWouldAddEndPuncttrue
\mciteSetBstMidEndSepPunct{\mcitedefaultmidpunct}
{\mcitedefaultendpunct}{\mcitedefaultseppunct}\relax
\EndOfBibitem
\bibitem[Ballesta \emph{et~al.}(2012)Ballesta, Petekidis, Isa, Poon, and
  Besseling]{Ballesta:2012}
P.~Ballesta, G.~Petekidis, L.~Isa, W.~Poon and R.~Besseling, \emph{J. Rheol.},
  2012, \textbf{56}, 1005--1037\relax
\mciteBstWouldAddEndPuncttrue
\mciteSetBstMidEndSepPunct{\mcitedefaultmidpunct}
{\mcitedefaultendpunct}{\mcitedefaultseppunct}\relax
\EndOfBibitem
\bibitem[Seth \emph{et~al.}(2008)Seth, Cloitre, and Bonnecaze]{Seth:2008}
J.~Seth, M.~Cloitre and R.~Bonnecaze, \emph{J. Rheol.}, 2008, \textbf{52},
  1241--1268\relax
\mciteBstWouldAddEndPuncttrue
\mciteSetBstMidEndSepPunct{\mcitedefaultmidpunct}
{\mcitedefaultendpunct}{\mcitedefaultseppunct}\relax
\EndOfBibitem
\bibitem[Meeker \emph{et~al.}(2004)Meeker, Bonnecaze, and
  Cloitre]{Meeker:2004a}
S.~P. Meeker, R.~T. Bonnecaze and M.~Cloitre, \emph{Phys. Rev. Lett.}, 2004,
  \textbf{92}, 198302\relax
\mciteBstWouldAddEndPuncttrue
\mciteSetBstMidEndSepPunct{\mcitedefaultmidpunct}
{\mcitedefaultendpunct}{\mcitedefaultseppunct}\relax
\EndOfBibitem
\bibitem[Divoux \emph{et~al.}(2015)Divoux, Lapeyre, Ravaine, and
  Manneville]{Divoux:2015}
T.~Divoux, V.~Lapeyre, V.~Ravaine and S.~Manneville, \emph{Phys. Rev. E}, 2015,
  \textbf{92}, 060301\relax
\mciteBstWouldAddEndPuncttrue
\mciteSetBstMidEndSepPunct{\mcitedefaultmidpunct}
{\mcitedefaultendpunct}{\mcitedefaultseppunct}\relax
\EndOfBibitem
\bibitem[Le~Merrer \emph{et~al.}(2015)Le~Merrer, Lespiat, H{\"o}hler, and
  Cohen-Addad]{Lemerrer:2015}
M.~Le~Merrer, R.~Lespiat, R.~H{\"o}hler and S.~Cohen-Addad, \emph{Soft matter},
  2015, \textbf{11}, 368--381\relax
\mciteBstWouldAddEndPuncttrue
\mciteSetBstMidEndSepPunct{\mcitedefaultmidpunct}
{\mcitedefaultendpunct}{\mcitedefaultseppunct}\relax
\EndOfBibitem
\bibitem[Zhang \emph{et~al.}(2017)Zhang, Lorenceau, Basset, Bourouina, Rouyer,
  Goyon, and Coussot]{Zhang:2017}
X.~Zhang, E.~Lorenceau, P.~Basset, T.~Bourouina, F.~Rouyer, J.~Goyon and
  P.~Coussot, \emph{Phys. Rev. Lett.}, 2017, \textbf{119}, 208004\relax
\mciteBstWouldAddEndPuncttrue
\mciteSetBstMidEndSepPunct{\mcitedefaultmidpunct}
{\mcitedefaultendpunct}{\mcitedefaultseppunct}\relax
\EndOfBibitem
\bibitem[Walls \emph{et~al.}(2003)Walls, Caines, Sanchez, and Khan]{Walls:2003}
H.~Walls, S.~Caines, A.~Sanchez and S.~Khan, \emph{J. Rheol.}, 2003,
  \textbf{47}, 847--868\relax
\mciteBstWouldAddEndPuncttrue
\mciteSetBstMidEndSepPunct{\mcitedefaultmidpunct}
{\mcitedefaultendpunct}{\mcitedefaultseppunct}\relax
\EndOfBibitem
\bibitem[Gibaud \emph{et~al.}(2009)Gibaud, Barentin, Taberlet, and
  Manneville]{Gibaud:2009}
T.~Gibaud, C.~Barentin, N.~Taberlet and S.~Manneville, \emph{Soft Matter},
  2009, \textbf{5}, 3026--3037\relax
\mciteBstWouldAddEndPuncttrue
\mciteSetBstMidEndSepPunct{\mcitedefaultmidpunct}
{\mcitedefaultendpunct}{\mcitedefaultseppunct}\relax
\EndOfBibitem
\bibitem[Seth \emph{et~al.}(2012)Seth, Locatelli-Champagne, Monti, Bonnecaze,
  and Cloitre]{Seth:2012}
J.~Seth, C.~Locatelli-Champagne, F.~Monti, R.~Bonnecaze and M.~Cloitre,
  \emph{Soft Matter}, 2012, \textbf{8}, 140--148\relax
\mciteBstWouldAddEndPuncttrue
\mciteSetBstMidEndSepPunct{\mcitedefaultmidpunct}
{\mcitedefaultendpunct}{\mcitedefaultseppunct}\relax
\EndOfBibitem
\bibitem[Mansard \emph{et~al.}(2014)Mansard, Bocquet, and Colin]{Mansard:2014}
V.~Mansard, L.~Bocquet and A.~Colin, \emph{Soft Matter}, 2014, \textbf{10},
  6984--6989\relax
\mciteBstWouldAddEndPuncttrue
\mciteSetBstMidEndSepPunct{\mcitedefaultmidpunct}
{\mcitedefaultendpunct}{\mcitedefaultseppunct}\relax
\EndOfBibitem
\bibitem[Jalaal \emph{et~al.}(2015)Jalaal, Balmforth, and Stoeber]{jalaal:2015}
M.~Jalaal, N.~J. Balmforth and B.~Stoeber, \emph{Langmuir}, 2015, \textbf{31},
  12071--12075\relax
\mciteBstWouldAddEndPuncttrue
\mciteSetBstMidEndSepPunct{\mcitedefaultmidpunct}
{\mcitedefaultendpunct}{\mcitedefaultseppunct}\relax
\EndOfBibitem
\bibitem[Mackay(2018)]{mackay:2018}
M.~E. Mackay, \emph{Journal of Rheology}, 2018, \textbf{62}, 1549--1561\relax
\mciteBstWouldAddEndPuncttrue
\mciteSetBstMidEndSepPunct{\mcitedefaultmidpunct}
{\mcitedefaultendpunct}{\mcitedefaultseppunct}\relax
\EndOfBibitem
\bibitem[Martouzet \emph{et~al.}(2021)Martouzet, J{\o}rgensen, Pelet, Biance,
  and Barentin]{martouzet:2021}
G.~Martouzet, L.~J{\o}rgensen, Y.~Pelet, A.-L. Biance and C.~Barentin,
  \emph{Physical Review Fluids}, 2021, \textbf{6}, 044006\relax
\mciteBstWouldAddEndPuncttrue
\mciteSetBstMidEndSepPunct{\mcitedefaultmidpunct}
{\mcitedefaultendpunct}{\mcitedefaultseppunct}\relax
\EndOfBibitem
\bibitem[van~der Kolk \emph{et~al.}(2023)van~der Kolk, Tieman, and
  Jalaal]{vander:2023}
J.~van~der Kolk, D.~Tieman and M.~Jalaal, \emph{Journal of Fluid Mechanics},
  2023, \textbf{958}, A34\relax
\mciteBstWouldAddEndPuncttrue
\mciteSetBstMidEndSepPunct{\mcitedefaultmidpunct}
{\mcitedefaultendpunct}{\mcitedefaultseppunct}\relax
\EndOfBibitem
\bibitem[Jung and Fielding(2021)]{Jung:2021}
G.~Jung and S.~M. Fielding, \emph{J. Rheol.}, 2021, \textbf{65}, 199--212\relax
\mciteBstWouldAddEndPuncttrue
\mciteSetBstMidEndSepPunct{\mcitedefaultmidpunct}
{\mcitedefaultendpunct}{\mcitedefaultseppunct}\relax
\EndOfBibitem
\bibitem[Nicolas and Barrat(2013)]{Nicolas:2013}
A.~Nicolas and J.-L. Barrat, \emph{Faraday Discussions}, 2013, \textbf{167},
  567\relax
\mciteBstWouldAddEndPuncttrue
\mciteSetBstMidEndSepPunct{\mcitedefaultmidpunct}
{\mcitedefaultendpunct}{\mcitedefaultseppunct}\relax
\EndOfBibitem
\bibitem[Jiang \emph{et~al.}(2022)Jiang, Makino, Royer, and Poon]{Jiang:2022}
Y.~Jiang, S.~Makino, J.~R. Royer and W.~C.~K. Poon, \emph{Phys. Rev. Lett.},
  2022, \textbf{128}, 248002\relax
\mciteBstWouldAddEndPuncttrue
\mciteSetBstMidEndSepPunct{\mcitedefaultmidpunct}
{\mcitedefaultendpunct}{\mcitedefaultseppunct}\relax
\EndOfBibitem
\bibitem[Di~Dio \emph{et~al.}(2022)Di~Dio, Khabaz, Bonnecaze, and
  Cloitre]{DiDio:2022}
B.~F. Di~Dio, F.~Khabaz, R.~T. Bonnecaze and M.~Cloitre, \emph{J. Rheol.},
  2022, \textbf{66}, 717--730\relax
\mciteBstWouldAddEndPuncttrue
\mciteSetBstMidEndSepPunct{\mcitedefaultmidpunct}
{\mcitedefaultendpunct}{\mcitedefaultseppunct}\relax
\EndOfBibitem
\bibitem[H{\"o}hler \emph{et~al.}(1999)H{\"o}hler, Cohen-Addad, and
  Asnacios]{Hohler:1999}
R.~H{\"o}hler, S.~Cohen-Addad and A.~Asnacios, \emph{Europhys. Lett.}, 1999,
  \textbf{48}, 93\relax
\mciteBstWouldAddEndPuncttrue
\mciteSetBstMidEndSepPunct{\mcitedefaultmidpunct}
{\mcitedefaultendpunct}{\mcitedefaultseppunct}\relax
\EndOfBibitem
\bibitem[Fuchs and Becker(2022)]{Fuchs:2022}
L.~Fuchs and T.~W. Becker, \emph{Geophys. Res. Lett.}, 2022, \textbf{49},
  e2022GL099574\relax
\mciteBstWouldAddEndPuncttrue
\mciteSetBstMidEndSepPunct{\mcitedefaultmidpunct}
{\mcitedefaultendpunct}{\mcitedefaultseppunct}\relax
\EndOfBibitem
\bibitem[Keim \emph{et~al.}(2019)Keim, Paulsen, Zeravcic, Sastry, and
  Nagel]{Keim:2019}
N.~C. Keim, J.~D. Paulsen, Z.~Zeravcic, S.~Sastry and S.~R. Nagel, \emph{Rev.
  Mod. Phys.}, 2019, \textbf{91}, 035002\relax
\mciteBstWouldAddEndPuncttrue
\mciteSetBstMidEndSepPunct{\mcitedefaultmidpunct}
{\mcitedefaultendpunct}{\mcitedefaultseppunct}\relax
\EndOfBibitem
\bibitem[Keim and Nagel(2011)]{Keim:2011}
N.~C. Keim and S.~R. Nagel, \emph{Phys. Rev. Lett.}, 2011, \textbf{107},
  010603\relax
\mciteBstWouldAddEndPuncttrue
\mciteSetBstMidEndSepPunct{\mcitedefaultmidpunct}
{\mcitedefaultendpunct}{\mcitedefaultseppunct}\relax
\EndOfBibitem
\bibitem[Paulsen \emph{et~al.}(2014)Paulsen, Keim, and Nagel]{Paulsen:2014}
J.~D. Paulsen, N.~C. Keim and S.~R. Nagel, \emph{Phys. Rev. Lett.}, 2014,
  \textbf{113}, 068301\relax
\mciteBstWouldAddEndPuncttrue
\mciteSetBstMidEndSepPunct{\mcitedefaultmidpunct}
{\mcitedefaultendpunct}{\mcitedefaultseppunct}\relax
\EndOfBibitem
\bibitem[Diat \emph{et~al.}(1993)Diat, Roux, and Nallet]{Diat:1993b}
O.~Diat, D.~Roux and F.~Nallet, \emph{J. Phys. II France}, 1993, \textbf{3},
  1427\relax
\mciteBstWouldAddEndPuncttrue
\mciteSetBstMidEndSepPunct{\mcitedefaultmidpunct}
{\mcitedefaultendpunct}{\mcitedefaultseppunct}\relax
\EndOfBibitem
\bibitem[Roux \emph{et~al.}(1993)Roux, Nallet, and Diat]{Roux:1993}
D.~Roux, F.~Nallet and O.~Diat, \emph{Europhys. Lett.}, 1993, \textbf{24},
  53--58\relax
\mciteBstWouldAddEndPuncttrue
\mciteSetBstMidEndSepPunct{\mcitedefaultmidpunct}
{\mcitedefaultendpunct}{\mcitedefaultseppunct}\relax
\EndOfBibitem
\bibitem[Bonn \emph{et~al.}(1998)Bonn, Meunier, Greffier, Al-Kahwaji, and
  Kellay]{Bonn:1998a}
D.~Bonn, J.~Meunier, O.~Greffier, A.~Al-Kahwaji and H.~Kellay, \emph{Phys. Rev.
  E}, 1998, \textbf{58}, 2215--2218\relax
\mciteBstWouldAddEndPuncttrue
\mciteSetBstMidEndSepPunct{\mcitedefaultmidpunct}
{\mcitedefaultendpunct}{\mcitedefaultseppunct}\relax
\EndOfBibitem
\bibitem[Wunenburger \emph{et~al.}(2001)Wunenburger, Colin, Leng, Arneodo, and
  Roux]{Wunenburger:2001}
A.-S. Wunenburger, A.~Colin, J.~Leng, A.~Arneodo and D.~Roux, \emph{Phys. Rev.
  Lett.}, 2001, \textbf{86}, 1374--1377\relax
\mciteBstWouldAddEndPuncttrue
\mciteSetBstMidEndSepPunct{\mcitedefaultmidpunct}
{\mcitedefaultendpunct}{\mcitedefaultseppunct}\relax
\EndOfBibitem
\bibitem[Salmon \emph{et~al.}(2002)Salmon, Colin, and Roux]{Salmon:2002}
J.-B. Salmon, A.~Colin and D.~Roux, \emph{Phys. Rev. E}, 2002, \textbf{66},
  031505\relax
\mciteBstWouldAddEndPuncttrue
\mciteSetBstMidEndSepPunct{\mcitedefaultmidpunct}
{\mcitedefaultendpunct}{\mcitedefaultseppunct}\relax
\EndOfBibitem
\bibitem[Manneville \emph{et~al.}(2004)Manneville, Salmon, and
  Colin]{Manneville:2004b}
S.~Manneville, J.-B. Salmon and A.~Colin, \emph{Eur. Phys. J. E}, 2004,
  \textbf{13}, 197--212\relax
\mciteBstWouldAddEndPuncttrue
\mciteSetBstMidEndSepPunct{\mcitedefaultmidpunct}
{\mcitedefaultendpunct}{\mcitedefaultseppunct}\relax
\EndOfBibitem
\bibitem[Ramos \emph{et~al.}(2000)Ramos, Molino, and Porte]{Ramos:2000}
L.~Ramos, F.~Molino and G.~Porte, \emph{Langmuir}, 2000, \textbf{16},
  5846--5848\relax
\mciteBstWouldAddEndPuncttrue
\mciteSetBstMidEndSepPunct{\mcitedefaultmidpunct}
{\mcitedefaultendpunct}{\mcitedefaultseppunct}\relax
\EndOfBibitem
\bibitem[Ramos(2001)]{Ramos:2001a}
L.~Ramos, \emph{Phys. Rev. E}, 2001, \textbf{64}, 061502\relax
\mciteBstWouldAddEndPuncttrue
\mciteSetBstMidEndSepPunct{\mcitedefaultmidpunct}
{\mcitedefaultendpunct}{\mcitedefaultseppunct}\relax
\EndOfBibitem
\bibitem[Cates and Wyart(2014)]{Cates:2014}
M.~E. Cates and M.~Wyart, \emph{Rheol. acta}, 2014, \textbf{53}, 755--764\relax
\mciteBstWouldAddEndPuncttrue
\mciteSetBstMidEndSepPunct{\mcitedefaultmidpunct}
{\mcitedefaultendpunct}{\mcitedefaultseppunct}\relax
\EndOfBibitem
\bibitem[Kolmogorov(1963)]{Kolmogorov:1963}
A.~N. Kolmogorov, \emph{Sankhy{\=a}: The Indian Journal of Statistics, Series
  A}, 1963,  369--376\relax
\mciteBstWouldAddEndPuncttrue
\mciteSetBstMidEndSepPunct{\mcitedefaultmidpunct}
{\mcitedefaultendpunct}{\mcitedefaultseppunct}\relax
\EndOfBibitem
\bibitem[Martiniani \emph{et~al.}(2019)Martiniani, Chaikin, and
  Levine]{Martiniani:2019}
S.~Martiniani, P.~M. Chaikin and D.~Levine, \emph{Phys. Rev. X}, 2019,
  \textbf{9}, 011031\relax
\mciteBstWouldAddEndPuncttrue
\mciteSetBstMidEndSepPunct{\mcitedefaultmidpunct}
{\mcitedefaultendpunct}{\mcitedefaultseppunct}\relax
\EndOfBibitem
\bibitem[Mukherji \emph{et~al.}(2019)Mukherji, Kandula, Sood, and
  Ganapathy]{Mukherji:2019}
S.~Mukherji, N.~Kandula, A.~K. Sood and R.~Ganapathy, \emph{Phys. Rev. Lett.},
  2019, \textbf{122}, 158001\relax
\mciteBstWouldAddEndPuncttrue
\mciteSetBstMidEndSepPunct{\mcitedefaultmidpunct}
{\mcitedefaultendpunct}{\mcitedefaultseppunct}\relax
\EndOfBibitem
\bibitem[Donley \emph{et~al.}(2020)Donley, Singh, Shetty, and
  Rogers]{Donley:2020}
G.~J. Donley, P.~K. Singh, A.~Shetty and S.~A. Rogers, \emph{Proc. Natl. Acad.
  Sci. U.S.A.}, 2020, \textbf{117}, 21945--21952\relax
\mciteBstWouldAddEndPuncttrue
\mciteSetBstMidEndSepPunct{\mcitedefaultmidpunct}
{\mcitedefaultendpunct}{\mcitedefaultseppunct}\relax
\EndOfBibitem
\bibitem[Donley \emph{et~al.}(2023)Donley, Narayanan, Wade, Park, Leheny,
  Harden, and Rogers]{Donley:2023}
G.~J. Donley, S.~Narayanan, M.~A. Wade, J.~D. Park, R.~L. Leheny, J.~L. Harden
  and S.~A. Rogers, \emph{Proc. Natl. Acad. Sci. U.S.A.}, 2023, \textbf{120},
  e2215517120\relax
\mciteBstWouldAddEndPuncttrue
\mciteSetBstMidEndSepPunct{\mcitedefaultmidpunct}
{\mcitedefaultendpunct}{\mcitedefaultseppunct}\relax
\EndOfBibitem
\bibitem[Cipelletti \emph{et~al.}(2016)Cipelletti, Trappe, and
  Pine]{Cipelletti:2016}
L.~Cipelletti, V.~Trappe and D.~J. Pine, \emph{Fluids, Colloids and Soft
  Materials: An Introduction to Soft Matter Physics}, 2016,  131--148\relax
\mciteBstWouldAddEndPuncttrue
\mciteSetBstMidEndSepPunct{\mcitedefaultmidpunct}
{\mcitedefaultendpunct}{\mcitedefaultseppunct}\relax
\EndOfBibitem
\bibitem[Sharma \emph{et~al.}(2023)Sharma, Shankar, and Joshi]{Sharma:2023}
S.~Sharma, V.~Shankar and Y.~M. Joshi, \emph{J. Rheol.}, 2023, \textbf{67},
  139--155\relax
\mciteBstWouldAddEndPuncttrue
\mciteSetBstMidEndSepPunct{\mcitedefaultmidpunct}
{\mcitedefaultendpunct}{\mcitedefaultseppunct}\relax
\EndOfBibitem
\bibitem[Divoux \emph{et~al.}(2013)Divoux, Grenard, and
  Manneville]{Divoux:2013}
T.~Divoux, V.~Grenard and S.~Manneville, \emph{Phys. Rev. Lett.}, 2013,
  \textbf{110}, 018304\relax
\mciteBstWouldAddEndPuncttrue
\mciteSetBstMidEndSepPunct{\mcitedefaultmidpunct}
{\mcitedefaultendpunct}{\mcitedefaultseppunct}\relax
\EndOfBibitem
\bibitem[Jamali \emph{et~al.}(2019)Jamali, Armstrong, and
  McKinley]{Jamali:2019}
S.~Jamali, R.~C. Armstrong and G.~H. McKinley, \emph{Phys. Rev. Lett.}, 2019,
  \textbf{123}, 248003\relax
\mciteBstWouldAddEndPuncttrue
\mciteSetBstMidEndSepPunct{\mcitedefaultmidpunct}
{\mcitedefaultendpunct}{\mcitedefaultseppunct}\relax
\EndOfBibitem
\bibitem[Jamali and McKinley(2022)]{Jamali:2022}
S.~Jamali and G.~H. McKinley, \emph{J. Rheol.}, 2022, \textbf{66},
  1027--1039\relax
\mciteBstWouldAddEndPuncttrue
\mciteSetBstMidEndSepPunct{\mcitedefaultmidpunct}
{\mcitedefaultendpunct}{\mcitedefaultseppunct}\relax
\EndOfBibitem
\bibitem[Choi and Rogers(2020)]{Choi:2020}
J.~Choi and S.~A. Rogers, \emph{Rheol. Acta}, 2020, \textbf{59}, 921--934\relax
\mciteBstWouldAddEndPuncttrue
\mciteSetBstMidEndSepPunct{\mcitedefaultmidpunct}
{\mcitedefaultendpunct}{\mcitedefaultseppunct}\relax
\EndOfBibitem
\bibitem[Gadala-Maria and Acrivos(1980)]{Gadala:1980}
F.~Gadala-Maria and A.~Acrivos, \emph{J. Rheol.}, 1980, \textbf{24},
  799--814\relax
\mciteBstWouldAddEndPuncttrue
\mciteSetBstMidEndSepPunct{\mcitedefaultmidpunct}
{\mcitedefaultendpunct}{\mcitedefaultseppunct}\relax
\EndOfBibitem
\bibitem[Keim \emph{et~al.}(2013)Keim, Paulsen, and Nagel]{Keim:2013}
N.~C. Keim, J.~D. Paulsen and S.~R. Nagel, \emph{Phys. Rev. E}, 2013,
  \textbf{88}, 032306\relax
\mciteBstWouldAddEndPuncttrue
\mciteSetBstMidEndSepPunct{\mcitedefaultmidpunct}
{\mcitedefaultendpunct}{\mcitedefaultseppunct}\relax
\EndOfBibitem
\bibitem[Larson and Wei(2019)]{Larson:2019}
R.~G. Larson and Y.~Wei, \emph{J. Rheol.}, 2019, \textbf{63}, 477--501\relax
\mciteBstWouldAddEndPuncttrue
\mciteSetBstMidEndSepPunct{\mcitedefaultmidpunct}
{\mcitedefaultendpunct}{\mcitedefaultseppunct}\relax
\EndOfBibitem
\bibitem[Moghimi \emph{et~al.}(2019)Moghimi, Vermant, and
  Petekidis]{Moghimi:2019}
E.~Moghimi, J.~Vermant and G.~Petekidis, \emph{J. Rheol.}, 2019, \textbf{63},
  533--546\relax
\mciteBstWouldAddEndPuncttrue
\mciteSetBstMidEndSepPunct{\mcitedefaultmidpunct}
{\mcitedefaultendpunct}{\mcitedefaultseppunct}\relax
\EndOfBibitem
\bibitem[Schwen \emph{et~al.}(2020)Schwen, Ramaswamy, Cheng, Jan, and
  Cohen]{Schwen:2020}
E.~M. Schwen, M.~Ramaswamy, C.-M. Cheng, L.~Jan and I.~Cohen, \emph{Soft
  Matter}, 2020, \textbf{16}, 3746--3752\relax
\mciteBstWouldAddEndPuncttrue
\mciteSetBstMidEndSepPunct{\mcitedefaultmidpunct}
{\mcitedefaultendpunct}{\mcitedefaultseppunct}\relax
\EndOfBibitem
\bibitem[Patinet \emph{et~al.}(2020)Patinet, Barbot, Lerbinger, Vandembroucq,
  and Lema\^{\i}tre]{Patinet:2020}
S.~Patinet, A.~Barbot, M.~Lerbinger, D.~Vandembroucq and A.~Lema\^{\i}tre,
  \emph{Phys. Rev. Lett.}, 2020, \textbf{124}, 205503\relax
\mciteBstWouldAddEndPuncttrue
\mciteSetBstMidEndSepPunct{\mcitedefaultmidpunct}
{\mcitedefaultendpunct}{\mcitedefaultseppunct}\relax
\EndOfBibitem
\bibitem[{\c{S}}enbil \emph{et~al.}(2019){\c{S}}enbil, Gruber, Zhang, Fuchs,
  and Scheffold]{Senbil:2019}
N.~{\c{S}}enbil, M.~Gruber, C.~Zhang, M.~Fuchs and F.~Scheffold, \emph{Phys.
  Rev. Lett.}, 2019, \textbf{122}, 108002\relax
\mciteBstWouldAddEndPuncttrue
\mciteSetBstMidEndSepPunct{\mcitedefaultmidpunct}
{\mcitedefaultendpunct}{\mcitedefaultseppunct}\relax
\EndOfBibitem
\bibitem[Mungan \emph{et~al.}(2019)Mungan, Sastry, Dahmen, and
  Regev]{Mungan:2019}
M.~Mungan, S.~Sastry, K.~Dahmen and I.~Regev, \emph{Phys. Rev. Lett.}, 2019,
  \textbf{123}, 178002\relax
\mciteBstWouldAddEndPuncttrue
\mciteSetBstMidEndSepPunct{\mcitedefaultmidpunct}
{\mcitedefaultendpunct}{\mcitedefaultseppunct}\relax
\EndOfBibitem
\bibitem[Ballauff \emph{et~al.}(2013)Ballauff, Brader, Egelhaaf, Fuchs,
  Horbach, Koumakis, Kr{\"u}ger, Laurati, Mutch, Petekidis, Siebenburger,
  Voigtmann, and Zausch]{Ballauff:2013}
M.~Ballauff, J.~Brader, S.~Egelhaaf, M.~Fuchs, J.~Horbach, N.~Koumakis,
  M.~Kr{\"u}ger, M.~Laurati, K.~Mutch, G.~Petekidis, M.~Siebenburger,
  T.~Voigtmann and J.~Zausch, \emph{Phys. Rev. Lett.}, 2013, \textbf{110},
  215701\relax
\mciteBstWouldAddEndPuncttrue
\mciteSetBstMidEndSepPunct{\mcitedefaultmidpunct}
{\mcitedefaultendpunct}{\mcitedefaultseppunct}\relax
\EndOfBibitem
\bibitem[Mohan \emph{et~al.}(2013)Mohan, Bonnecaze, and Cloitre]{Mohan:2013}
L.~Mohan, R.~Bonnecaze and M.~Cloitre, \emph{Phys. Rev. Lett.}, 2013,
  \textbf{111}, 268301\relax
\mciteBstWouldAddEndPuncttrue
\mciteSetBstMidEndSepPunct{\mcitedefaultmidpunct}
{\mcitedefaultendpunct}{\mcitedefaultseppunct}\relax
\EndOfBibitem
\bibitem[Mohan \emph{et~al.}(2015)Mohan, Cloitre, and Bonnecaze]{Mohan:2015}
L.~Mohan, M.~Cloitre and R.~T. Bonnecaze, \emph{J. Rheol.}, 2015, \textbf{59},
  63--84\relax
\mciteBstWouldAddEndPuncttrue
\mciteSetBstMidEndSepPunct{\mcitedefaultmidpunct}
{\mcitedefaultendpunct}{\mcitedefaultseppunct}\relax
\EndOfBibitem
\bibitem[Lidon \emph{et~al.}(2017)Lidon, Villa, and Manneville]{Lidon:2017}
P.~Lidon, L.~Villa and S.~Manneville, \emph{Rheol. Acta}, 2017, \textbf{56},
  307--323\relax
\mciteBstWouldAddEndPuncttrue
\mciteSetBstMidEndSepPunct{\mcitedefaultmidpunct}
{\mcitedefaultendpunct}{\mcitedefaultseppunct}\relax
\EndOfBibitem
\bibitem[Vasisht \emph{et~al.}(2021)Vasisht, Chaudhuri, and
  Martens]{Vasisht:2021}
V.~V. Vasisht, P.~Chaudhuri and K.~Martens, \emph{arXiv:2108.12782}, 2021\relax
\mciteBstWouldAddEndPuncttrue
\mciteSetBstMidEndSepPunct{\mcitedefaultmidpunct}
{\mcitedefaultendpunct}{\mcitedefaultseppunct}\relax
\EndOfBibitem
\bibitem[Osuji \emph{et~al.}(2008)Osuji, Kim, and Weitz]{Osuji:2008}
C.~O. Osuji, C.~Kim and D.~A. Weitz, \emph{Phys. Rev. E}, 2008, \textbf{77},
  060402(R)\relax
\mciteBstWouldAddEndPuncttrue
\mciteSetBstMidEndSepPunct{\mcitedefaultmidpunct}
{\mcitedefaultendpunct}{\mcitedefaultseppunct}\relax
\EndOfBibitem
\bibitem[Negi and Osuji(2009)]{Negi:2009b}
A.~Negi and C.~Osuji, \emph{Phys. Rev. E}, 2009, \textbf{80}, 010404\relax
\mciteBstWouldAddEndPuncttrue
\mciteSetBstMidEndSepPunct{\mcitedefaultmidpunct}
{\mcitedefaultendpunct}{\mcitedefaultseppunct}\relax
\EndOfBibitem
\bibitem[Moghimi \emph{et~al.}(2017)Moghimi, Jacob, and
  Petekidis]{Moghimi:2017b}
E.~Moghimi, A.~R. Jacob and G.~Petekidis, \emph{Soft Matter}, 2017,
  \textbf{13}, 7824--7833\relax
\mciteBstWouldAddEndPuncttrue
\mciteSetBstMidEndSepPunct{\mcitedefaultmidpunct}
{\mcitedefaultendpunct}{\mcitedefaultseppunct}\relax
\EndOfBibitem
\bibitem[Zausch and Horbach(2009)]{Zausch:2009}
J.~Zausch and J.~Horbach, \emph{Europhys. Lett.}, 2009, \textbf{88},
  60001\relax
\mciteBstWouldAddEndPuncttrue
\mciteSetBstMidEndSepPunct{\mcitedefaultmidpunct}
{\mcitedefaultendpunct}{\mcitedefaultseppunct}\relax
\EndOfBibitem
\bibitem[Barik and Majumdar(2022)]{Barik:2022}
S.~Barik and S.~Majumdar, \emph{Phys. Rev. Lett.}, 2022, \textbf{128},
  258002\relax
\mciteBstWouldAddEndPuncttrue
\mciteSetBstMidEndSepPunct{\mcitedefaultmidpunct}
{\mcitedefaultendpunct}{\mcitedefaultseppunct}\relax
\EndOfBibitem
\bibitem[Sudreau \emph{et~al.}(2022)Sudreau, Auxois, Servel, L\'ecolier,
  Manneville, and Divoux]{Sudreau:2022b}
I.~Sudreau, M.~Auxois, M.~Servel, E.~L\'ecolier, S.~Manneville and T.~Divoux,
  \emph{Phys. Rev. Materials}, 2022, \textbf{6}, L042601\relax
\mciteBstWouldAddEndPuncttrue
\mciteSetBstMidEndSepPunct{\mcitedefaultmidpunct}
{\mcitedefaultendpunct}{\mcitedefaultseppunct}\relax
\EndOfBibitem
\bibitem[Sudreau \emph{et~al.}(2023)Sudreau, Servel, Freyssingeas, Li\'enard,
  Karpati, Parola, Jaurand, Dugas, Matthews, Gibaud, Divoux, and
  Manneville]{Sudreau:2023}
I.~Sudreau, M.~Servel, E.~Freyssingeas, F.~m.~c. Li\'enard, S.~Karpati,
  S.~Parola, X.~Jaurand, P.-Y. Dugas, L.~Matthews, T.~Gibaud, T.~Divoux and
  S.~Manneville, \emph{Phys. Rev. Mater.}, 2023, \textbf{7}, 115603\relax
\mciteBstWouldAddEndPuncttrue
\mciteSetBstMidEndSepPunct{\mcitedefaultmidpunct}
{\mcitedefaultendpunct}{\mcitedefaultseppunct}\relax
\EndOfBibitem
\bibitem[Gomez-Solano and Bechinger(2015)]{Gomez-Solano:2015}
J.~R. Gomez-Solano and C.~Bechinger, \emph{New J. Phys.}, 2015, \textbf{17},
  103032\relax
\mciteBstWouldAddEndPuncttrue
\mciteSetBstMidEndSepPunct{\mcitedefaultmidpunct}
{\mcitedefaultendpunct}{\mcitedefaultseppunct}\relax
\EndOfBibitem
\bibitem[Zia and Brady(2013)]{Zia:2013}
R.~N. Zia and J.~F. Brady, \emph{J. Rheol.}, 2013, \textbf{57}, 457--492\relax
\mciteBstWouldAddEndPuncttrue
\mciteSetBstMidEndSepPunct{\mcitedefaultmidpunct}
{\mcitedefaultendpunct}{\mcitedefaultseppunct}\relax
\EndOfBibitem
\bibitem[Mohanty and Zia(2020)]{Mohanty:2020}
R.~P. Mohanty and R.~N. Zia, \emph{Journal of Fluid Mechanics}, 2020,
  \textbf{884}, A14\relax
\mciteBstWouldAddEndPuncttrue
\mciteSetBstMidEndSepPunct{\mcitedefaultmidpunct}
{\mcitedefaultendpunct}{\mcitedefaultseppunct}\relax
\EndOfBibitem
\bibitem[Murphy \emph{et~al.}(2020)Murphy, Kruppe, and Jaeger]{Murphy:2020}
K.~A. Murphy, J.~W. Kruppe and H.~M. Jaeger, \emph{Phys. Rev. Lett.}, 2020,
  \textbf{124}, 168002\relax
\mciteBstWouldAddEndPuncttrue
\mciteSetBstMidEndSepPunct{\mcitedefaultmidpunct}
{\mcitedefaultendpunct}{\mcitedefaultseppunct}\relax
\EndOfBibitem
\bibitem[Hendricks \emph{et~al.}(2019)Hendricks, Louhichi, Metri, Fournier,
  Reddy, Bouteiller, Cloitre, Clasen, Vlassopoulos, and Briels]{Hendricks:2019}
J.~Hendricks, A.~Louhichi, V.~Metri, R.~Fournier, N.~Reddy, L.~Bouteiller,
  M.~Cloitre, C.~Clasen, D.~Vlassopoulos and W.~J. Briels, \emph{Phys. Rev.
  Lett.}, 2019, \textbf{123}, 218003\relax
\mciteBstWouldAddEndPuncttrue
\mciteSetBstMidEndSepPunct{\mcitedefaultmidpunct}
{\mcitedefaultendpunct}{\mcitedefaultseppunct}\relax
\EndOfBibitem
\bibitem[Joshi(2022)]{Joshi:2022}
Y.~M. Joshi, \emph{J. Rheol.}, 2022, \textbf{66}, 111--123\relax
\mciteBstWouldAddEndPuncttrue
\mciteSetBstMidEndSepPunct{\mcitedefaultmidpunct}
{\mcitedefaultendpunct}{\mcitedefaultseppunct}\relax
\EndOfBibitem
\bibitem[Bar{\'e}s \emph{et~al.}(2022)Bar{\'e}s, C{\'a}rdenas-Barrantes,
  Cantor, Renouf, and Az{\'e}ma]{Bares:2022}
J.~Bar{\'e}s, M.~C{\'a}rdenas-Barrantes, D.~Cantor, M.~Renouf and
  {\'E}.~Az{\'e}ma, \emph{Pap. Phys.}, 2022, \textbf{14}, 140009\relax
\mciteBstWouldAddEndPuncttrue
\mciteSetBstMidEndSepPunct{\mcitedefaultmidpunct}
{\mcitedefaultendpunct}{\mcitedefaultseppunct}\relax
\EndOfBibitem
\bibitem[Bi \emph{et~al.}(2016)Bi, Yang, Marchetti, and
  Manning]{bi2016motility}
D.~Bi, X.~Yang, M.~C. Marchetti and M.~L. Manning, \emph{Phys. Rev. X}, 2016,
  \textbf{6}, 021011\relax
\mciteBstWouldAddEndPuncttrue
\mciteSetBstMidEndSepPunct{\mcitedefaultmidpunct}
{\mcitedefaultendpunct}{\mcitedefaultseppunct}\relax
\EndOfBibitem
\bibitem[Sussman(2017)]{sussman:2017cellgpu}
D.~M. Sussman, \emph{Comput. Phys. Commun.}, 2017, \textbf{219}, 400--406\relax
\mciteBstWouldAddEndPuncttrue
\mciteSetBstMidEndSepPunct{\mcitedefaultmidpunct}
{\mcitedefaultendpunct}{\mcitedefaultseppunct}\relax
\EndOfBibitem
\bibitem[Sussman \emph{et~al.}(2018)Sussman, Paoluzzi, Marchetti, and
  Manning]{sussman:2018anomalous}
D.~M. Sussman, M.~Paoluzzi, M.~C. Marchetti and M.~L. Manning, \emph{Europhys.
  Lett.}, 2018, \textbf{121}, 36001\relax
\mciteBstWouldAddEndPuncttrue
\mciteSetBstMidEndSepPunct{\mcitedefaultmidpunct}
{\mcitedefaultendpunct}{\mcitedefaultseppunct}\relax
\EndOfBibitem
\bibitem[Mongera \emph{et~al.}(2018)Mongera, Rowghanian, Gustafson, Shelton,
  Kealhofer, Carn, Serwane, Lucio, Giammona, and Camp{\`a}s]{mongera2018fluid}
A.~Mongera, P.~Rowghanian, H.~J. Gustafson, E.~Shelton, D.~A. Kealhofer, E.~K.
  Carn, F.~Serwane, A.~A. Lucio, J.~Giammona and O.~Camp{\`a}s, \emph{Nature},
  2018, \textbf{561}, 401--405\relax
\mciteBstWouldAddEndPuncttrue
\mciteSetBstMidEndSepPunct{\mcitedefaultmidpunct}
{\mcitedefaultendpunct}{\mcitedefaultseppunct}\relax
\EndOfBibitem
\bibitem[Tlili \emph{et~al.}(2022)Tlili, Graner, and
  Delano{\"e}-Ayari]{tlili2022}
S.~L. Tlili, F.~Graner and H.~Delano{\"e}-Ayari, \emph{Development}, 2022,
  \textbf{149}, dev200774\relax
\mciteBstWouldAddEndPuncttrue
\mciteSetBstMidEndSepPunct{\mcitedefaultmidpunct}
{\mcitedefaultendpunct}{\mcitedefaultseppunct}\relax
\EndOfBibitem
\bibitem[Hertaeg \emph{et~al.}(2023)Hertaeg, Fielding, and Bi]{hertaeg:2022}
M.~J. Hertaeg, S.~M. Fielding and D.~Bi, \emph{arXiv preprint
  arXiv:2211.15015}, 2023\relax
\mciteBstWouldAddEndPuncttrue
\mciteSetBstMidEndSepPunct{\mcitedefaultmidpunct}
{\mcitedefaultendpunct}{\mcitedefaultseppunct}\relax
\EndOfBibitem
\bibitem[Janssen(2019)]{Janssen:2019}
L.~M. Janssen, \emph{J. Condens. Matter Phys.}, 2019, \textbf{31}, 503002\relax
\mciteBstWouldAddEndPuncttrue
\mciteSetBstMidEndSepPunct{\mcitedefaultmidpunct}
{\mcitedefaultendpunct}{\mcitedefaultseppunct}\relax
\EndOfBibitem
\bibitem[Parry \emph{et~al.}(2014)Parry, Surovtsev, Cabeen, O’Hern, Dufresne,
  and Jacobs-Wagner]{parry2014bacterial}
B.~R. Parry, I.~V. Surovtsev, M.~T. Cabeen, C.~S. O’Hern, E.~R. Dufresne and
  C.~Jacobs-Wagner, \emph{Cell}, 2014, \textbf{156}, 183--194\relax
\mciteBstWouldAddEndPuncttrue
\mciteSetBstMidEndSepPunct{\mcitedefaultmidpunct}
{\mcitedefaultendpunct}{\mcitedefaultseppunct}\relax
\EndOfBibitem
\bibitem[Schramma \emph{et~al.}(2023)Schramma, Perugachi~Isra{\"e}ls, and
  Jalaal]{schramma2023chloroplasts}
N.~Schramma, C.~Perugachi~Isra{\"e}ls and M.~Jalaal, \emph{Proc. Natl. Acad.
  Sci. U.S.A.}, 2023, \textbf{120}, e2216497120\relax
\mciteBstWouldAddEndPuncttrue
\mciteSetBstMidEndSepPunct{\mcitedefaultmidpunct}
{\mcitedefaultendpunct}{\mcitedefaultseppunct}\relax
\EndOfBibitem
\bibitem[{\AA}berg and Poolman(2021)]{aaberg2021glass}
C.~{\AA}berg and B.~Poolman, \emph{Biophys. J.}, 2021, \textbf{120},
  2355--2366\relax
\mciteBstWouldAddEndPuncttrue
\mciteSetBstMidEndSepPunct{\mcitedefaultmidpunct}
{\mcitedefaultendpunct}{\mcitedefaultseppunct}\relax
\EndOfBibitem
\bibitem[Corci \emph{et~al.}(2023)Corci, Hooiveld, Dolga, and
  {\AA}berg]{corci2023extending}
B.~Corci, O.~Hooiveld, A.~M. Dolga and C.~{\AA}berg, \emph{Soft Matter}, 2023,
  \textbf{19}, 2529--2538\relax
\mciteBstWouldAddEndPuncttrue
\mciteSetBstMidEndSepPunct{\mcitedefaultmidpunct}
{\mcitedefaultendpunct}{\mcitedefaultseppunct}\relax
\EndOfBibitem
\bibitem[Garcia \emph{et~al.}(2015)Garcia, Hannezo, Elgeti, Joanny, Silberzan,
  and Gov]{Garcia:2015}
S.~Garcia, E.~Hannezo, J.~Elgeti, J.-F. Joanny, P.~Silberzan and N.~S. Gov,
  \emph{Proc. Natl. Acad. Sci. U.S.A.}, 2015, \textbf{112}, 15314--15319\relax
\mciteBstWouldAddEndPuncttrue
\mciteSetBstMidEndSepPunct{\mcitedefaultmidpunct}
{\mcitedefaultendpunct}{\mcitedefaultseppunct}\relax
\EndOfBibitem
\bibitem[Angelini \emph{et~al.}(2011)Angelini, Hannezo, Trepat, Marquez,
  Fredberg, and Weitz]{angelini2011glass}
T.~E. Angelini, E.~Hannezo, X.~Trepat, M.~Marquez, J.~J. Fredberg and D.~A.
  Weitz, \emph{Proc. Natl. Acad. Sci. U.S.A.}, 2011, \textbf{108},
  4714--4719\relax
\mciteBstWouldAddEndPuncttrue
\mciteSetBstMidEndSepPunct{\mcitedefaultmidpunct}
{\mcitedefaultendpunct}{\mcitedefaultseppunct}\relax
\EndOfBibitem
\bibitem[Malinverno \emph{et~al.}(2017)Malinverno, Corallino, Giavazzi,
  Bergert, Li, Leoni, Disanza, Frittoli, Oldani,
  Martini,\emph{et~al.}]{malinverno:2017}
C.~Malinverno, S.~Corallino, F.~Giavazzi, M.~Bergert, Q.~Li, M.~Leoni,
  A.~Disanza, E.~Frittoli, A.~Oldani, E.~Martini \emph{et~al.}, \emph{Nat.
  Mater}, 2017, \textbf{16}, 587--596\relax
\mciteBstWouldAddEndPuncttrue
\mciteSetBstMidEndSepPunct{\mcitedefaultmidpunct}
{\mcitedefaultendpunct}{\mcitedefaultseppunct}\relax
\EndOfBibitem
\bibitem[Atia \emph{et~al.}(2018)Atia, Bi, Sharma, Mitchel, Gweon, A.~Koehler,
  DeCamp, Lan, Kim, Hirsch,\emph{et~al.}]{atia2018geometric}
L.~Atia, D.~Bi, Y.~Sharma, J.~A. Mitchel, B.~Gweon, S.~A.~Koehler, S.~J.
  DeCamp, B.~Lan, J.~H. Kim, R.~Hirsch \emph{et~al.}, \emph{Nat. Phys.}, 2018,
  \textbf{14}, 613--620\relax
\mciteBstWouldAddEndPuncttrue
\mciteSetBstMidEndSepPunct{\mcitedefaultmidpunct}
{\mcitedefaultendpunct}{\mcitedefaultseppunct}\relax
\EndOfBibitem
\bibitem[Nelson(2022)]{nelson2022mechanical}
C.~M. Nelson, \emph{Annual Review of Biomedical Engineering}, 2022,
  \textbf{24}, 307--322\relax
\mciteBstWouldAddEndPuncttrue
\mciteSetBstMidEndSepPunct{\mcitedefaultmidpunct}
{\mcitedefaultendpunct}{\mcitedefaultseppunct}\relax
\EndOfBibitem
\bibitem[Park \emph{et~al.}(2015)Park, Kim, Bi, Mitchel, Qazvini, Tantisira,
  Park, McGill, Kim, Gweon,\emph{et~al.}]{park2015unjamming}
J.-A. Park, J.~H. Kim, D.~Bi, J.~A. Mitchel, N.~T. Qazvini, K.~Tantisira, C.~Y.
  Park, M.~McGill, S.-H. Kim, B.~Gweon \emph{et~al.}, \emph{Nat. Mater}, 2015,
  \textbf{14}, 1040--1048\relax
\mciteBstWouldAddEndPuncttrue
\mciteSetBstMidEndSepPunct{\mcitedefaultmidpunct}
{\mcitedefaultendpunct}{\mcitedefaultseppunct}\relax
\EndOfBibitem
\bibitem[Mitchel \emph{et~al.}(2020)Mitchel, Das, O’Sullivan, Stancil,
  DeCamp, Koehler, Oca{\~n}a, Butler, Fredberg,
  Nieto,\emph{et~al.}]{mitchel:2020}
J.~A. Mitchel, A.~Das, M.~J. O’Sullivan, I.~T. Stancil, S.~J. DeCamp,
  S.~Koehler, O.~H. Oca{\~n}a, J.~P. Butler, J.~J. Fredberg, M.~A. Nieto
  \emph{et~al.}, \emph{Nat. Commun.}, 2020, \textbf{11}, 5053\relax
\mciteBstWouldAddEndPuncttrue
\mciteSetBstMidEndSepPunct{\mcitedefaultmidpunct}
{\mcitedefaultendpunct}{\mcitedefaultseppunct}\relax
\EndOfBibitem
\bibitem[Atia \emph{et~al.}(2021)Atia, Fredberg, Gov, and
  Pegoraro]{atia2021cell}
L.~Atia, J.~J. Fredberg, N.~S. Gov and A.~F. Pegoraro, \emph{Cells \&
  development}, 2021, \textbf{168}, 203727\relax
\mciteBstWouldAddEndPuncttrue
\mciteSetBstMidEndSepPunct{\mcitedefaultmidpunct}
{\mcitedefaultendpunct}{\mcitedefaultseppunct}\relax
\EndOfBibitem
\bibitem[Lawson-Keister and Manning(2021)]{lawson2021jamming}
E.~Lawson-Keister and M.~L. Manning, \emph{Curr. Opin. Cell Biol.}, 2021,
  \textbf{72}, 146--155\relax
\mciteBstWouldAddEndPuncttrue
\mciteSetBstMidEndSepPunct{\mcitedefaultmidpunct}
{\mcitedefaultendpunct}{\mcitedefaultseppunct}\relax
\EndOfBibitem
\bibitem[Bi \emph{et~al.}(2015)Bi, Lopez, Schwarz, and Manning]{bi:2015}
D.~Bi, J.~Lopez, J.~M. Schwarz and M.~L. Manning, \emph{Nat. Phys.}, 2015,
  \textbf{11}, 1074--1079\relax
\mciteBstWouldAddEndPuncttrue
\mciteSetBstMidEndSepPunct{\mcitedefaultmidpunct}
{\mcitedefaultendpunct}{\mcitedefaultseppunct}\relax
\EndOfBibitem
\bibitem[Das \emph{et~al.}(2021)Das, Sastry, and Bi]{Das:2021}
A.~Das, S.~Sastry and D.~Bi, \emph{Phys. Rev. X}, 2021, \textbf{11},
  041037\relax
\mciteBstWouldAddEndPuncttrue
\mciteSetBstMidEndSepPunct{\mcitedefaultmidpunct}
{\mcitedefaultendpunct}{\mcitedefaultseppunct}\relax
\EndOfBibitem
\bibitem[Petridou and Heisenberg(2019)]{petridou2019tissue}
N.~I. Petridou and C.-P. Heisenberg, \emph{The EMBO journal}, 2019,
  \textbf{38}, e102497\relax
\mciteBstWouldAddEndPuncttrue
\mciteSetBstMidEndSepPunct{\mcitedefaultmidpunct}
{\mcitedefaultendpunct}{\mcitedefaultseppunct}\relax
\EndOfBibitem
\bibitem[Hannezo and Heisenberg(2022)]{hannezo2022rigidity}
E.~Hannezo and C.-P. Heisenberg, \emph{Trends in Cell Biology}, 2022\relax
\mciteBstWouldAddEndPuncttrue
\mciteSetBstMidEndSepPunct{\mcitedefaultmidpunct}
{\mcitedefaultendpunct}{\mcitedefaultseppunct}\relax
\EndOfBibitem
\bibitem[Oswald \emph{et~al.}(2017)Oswald, Grosser, Smith, and
  K{\"a}s]{oswald2017jamming}
L.~Oswald, S.~Grosser, D.~M. Smith and J.~A. K{\"a}s, \emph{J. Phys. D: Appl.
  Phys.}, 2017, \textbf{50}, 483001\relax
\mciteBstWouldAddEndPuncttrue
\mciteSetBstMidEndSepPunct{\mcitedefaultmidpunct}
{\mcitedefaultendpunct}{\mcitedefaultseppunct}\relax
\EndOfBibitem
\bibitem[Blauth \emph{et~al.}(2021)Blauth, Kubitschke, Gottheil, Grosser, and
  K{\"a}s]{blauth2021jamming}
E.~Blauth, H.~Kubitschke, P.~Gottheil, S.~Grosser and J.~A. K{\"a}s,
  \emph{Front. Phys.}, 2021, \textbf{9}, 666709\relax
\mciteBstWouldAddEndPuncttrue
\mciteSetBstMidEndSepPunct{\mcitedefaultmidpunct}
{\mcitedefaultendpunct}{\mcitedefaultseppunct}\relax
\EndOfBibitem
\bibitem[Gottheil \emph{et~al.}(2023)Gottheil, Lippoldt, Grosser, Renner,
  Saibah, Tschodu, Po{\ss}{\"o}gel, Wegscheider, Ulm,
  Friedrichs,\emph{et~al.}]{gottheil2023state}
P.~Gottheil, J.~Lippoldt, S.~Grosser, F.~Renner, M.~Saibah, D.~Tschodu, A.-K.
  Po{\ss}{\"o}gel, A.-S. Wegscheider, B.~Ulm, K.~Friedrichs \emph{et~al.},
  \emph{Phys. Rev. X}, 2023, \textbf{13}, 031003\relax
\mciteBstWouldAddEndPuncttrue
\mciteSetBstMidEndSepPunct{\mcitedefaultmidpunct}
{\mcitedefaultendpunct}{\mcitedefaultseppunct}\relax
\EndOfBibitem
\bibitem[Tlili \emph{et~al.}(2015)Tlili, Gay, Graner, Marcq, Molino, and
  Saramito]{tlili2015colloquium}
S.~Tlili, C.~Gay, F.~Graner, P.~Marcq, F.~Molino and P.~Saramito, \emph{The
  European Physical Journal E}, 2015, \textbf{38}, 1--31\relax
\mciteBstWouldAddEndPuncttrue
\mciteSetBstMidEndSepPunct{\mcitedefaultmidpunct}
{\mcitedefaultendpunct}{\mcitedefaultseppunct}\relax
\EndOfBibitem
\bibitem[Sergides \emph{et~al.}(2021)Sergides, Perego, Galgani, Arbore, Pavone,
  and Capitanio]{sergides2021}
M.~Sergides, L.~Perego, T.~Galgani, C.~Arbore, F.~Pavone and M.~Capitanio,
  \emph{Eur. Phys. J. Plus}, 2021, \textbf{136}, 316\relax
\mciteBstWouldAddEndPuncttrue
\mciteSetBstMidEndSepPunct{\mcitedefaultmidpunct}
{\mcitedefaultendpunct}{\mcitedefaultseppunct}\relax
\EndOfBibitem
\bibitem[Harris \emph{et~al.}(2012)Harris, Peter, Bellis, Baum, Kabla, and
  Charras]{Harris:2012}
A.~R. Harris, L.~Peter, J.~Bellis, B.~Baum, A.~J. Kabla and G.~T. Charras,
  \emph{Proc. Natl. Acad. Sci. U.S.A.}, 2012, \textbf{109}, 16449–16454\relax
\mciteBstWouldAddEndPuncttrue
\mciteSetBstMidEndSepPunct{\mcitedefaultmidpunct}
{\mcitedefaultendpunct}{\mcitedefaultseppunct}\relax
\EndOfBibitem
\bibitem[Doubrovinski \emph{et~al.}(2017)Doubrovinski, Swan, Polyakov, and
  Wieschaus]{doubrovinski:2017}
K.~Doubrovinski, M.~Swan, O.~Polyakov and E.~F. Wieschaus, \emph{Proc. Natl.
  Acad. Sci. U.S.A.}, 2017, \textbf{114}, 1051--1056\relax
\mciteBstWouldAddEndPuncttrue
\mciteSetBstMidEndSepPunct{\mcitedefaultmidpunct}
{\mcitedefaultendpunct}{\mcitedefaultseppunct}\relax
\EndOfBibitem
\bibitem[Hannezo and Heisenberg(2019)]{hannezo2019mechanochemical}
E.~Hannezo and C.-P. Heisenberg, \emph{Cell}, 2019, \textbf{178}, 12--25\relax
\mciteBstWouldAddEndPuncttrue
\mciteSetBstMidEndSepPunct{\mcitedefaultmidpunct}
{\mcitedefaultendpunct}{\mcitedefaultseppunct}\relax
\EndOfBibitem
\bibitem[Pinheiro \emph{et~al.}(2022)Pinheiro, Kardos, Hannezo, and
  Heisenberg]{pinheiro2022morphogen}
D.~Pinheiro, R.~Kardos, {\'E}.~Hannezo and C.-P. Heisenberg, \emph{Nature
  Physics}, 2022, \textbf{18}, 1482--1493\relax
\mciteBstWouldAddEndPuncttrue
\mciteSetBstMidEndSepPunct{\mcitedefaultmidpunct}
{\mcitedefaultendpunct}{\mcitedefaultseppunct}\relax
\EndOfBibitem
\bibitem[Hoffman and Crocker(2009)]{hoffman2009cell}
B.~D. Hoffman and J.~C. Crocker, \emph{Annu. Rev. Biomed. Eng.}, 2009,
  \textbf{11}, 259--288\relax
\mciteBstWouldAddEndPuncttrue
\mciteSetBstMidEndSepPunct{\mcitedefaultmidpunct}
{\mcitedefaultendpunct}{\mcitedefaultseppunct}\relax
\EndOfBibitem
\bibitem[Janmey and McCulloch(2007)]{janmey2007cell}
P.~A. Janmey and C.~A. McCulloch, \emph{Annu. Rev. Biomed. Eng.}, 2007,
  \textbf{9}, 1--34\relax
\mciteBstWouldAddEndPuncttrue
\mciteSetBstMidEndSepPunct{\mcitedefaultmidpunct}
{\mcitedefaultendpunct}{\mcitedefaultseppunct}\relax
\EndOfBibitem
\bibitem[Sadeghipour \emph{et~al.}(2018)Sadeghipour, Garcia, Nelson, and
  Pruitt]{Sadeghipour:2018}
E.~Sadeghipour, M.~A. Garcia, W.~J. Nelson and B.~L. Pruitt, \emph{eLife},
  2018, \textbf{7}, e39640\relax
\mciteBstWouldAddEndPuncttrue
\mciteSetBstMidEndSepPunct{\mcitedefaultmidpunct}
{\mcitedefaultendpunct}{\mcitedefaultseppunct}\relax
\EndOfBibitem
\bibitem[Ladoux and Mège(2017)]{ladoux_mechanobiology_2017}
B.~Ladoux and R.-M. Mège, \emph{Nature Reviews Molecular Cell Biology}, 2017,
  \textbf{18}, 743--757\relax
\mciteBstWouldAddEndPuncttrue
\mciteSetBstMidEndSepPunct{\mcitedefaultmidpunct}
{\mcitedefaultendpunct}{\mcitedefaultseppunct}\relax
\EndOfBibitem
\bibitem[Roca-Cusachs \emph{et~al.}(2013)Roca-Cusachs, Sunyer, and
  Trepat]{Roca-Cusachs:2013}
P.~Roca-Cusachs, R.~Sunyer and X.~Trepat, \emph{Curr. Opin. Cell Biol.}, 2013,
  \textbf{25}, 543--549\relax
\mciteBstWouldAddEndPuncttrue
\mciteSetBstMidEndSepPunct{\mcitedefaultmidpunct}
{\mcitedefaultendpunct}{\mcitedefaultseppunct}\relax
\EndOfBibitem
\bibitem[Trepat and Fredberg(2011)]{trepat:2011plithotaxis}
X.~Trepat and J.~J. Fredberg, \emph{Trends Cell Biol.}, 2011, \textbf{21},
  638--646\relax
\mciteBstWouldAddEndPuncttrue
\mciteSetBstMidEndSepPunct{\mcitedefaultmidpunct}
{\mcitedefaultendpunct}{\mcitedefaultseppunct}\relax
\EndOfBibitem
\bibitem[Beatrici \emph{et~al.}(2023)Beatrici, Kirch, Henkes, Graner, and
  Brunnet]{beatrici2023}
C.~Beatrici, C.~Kirch, S.~Henkes, F.~Graner and L.~Brunnet, \emph{Soft Matter},
  2023, \textbf{19}, 5583--5601\relax
\mciteBstWouldAddEndPuncttrue
\mciteSetBstMidEndSepPunct{\mcitedefaultmidpunct}
{\mcitedefaultendpunct}{\mcitedefaultseppunct}\relax
\EndOfBibitem
\bibitem[Alt \emph{et~al.}(2017)Alt, Ganguly, and Salbreux]{Alt:2017}
S.~Alt, P.~Ganguly and G.~Salbreux, \emph{Philos. Trans. R. Soc. B: Biol.},
  2017, \textbf{372}, 20150520\relax
\mciteBstWouldAddEndPuncttrue
\mciteSetBstMidEndSepPunct{\mcitedefaultmidpunct}
{\mcitedefaultendpunct}{\mcitedefaultseppunct}\relax
\EndOfBibitem
\bibitem[Nagai and Honda(2001)]{Nagai:2001}
T.~Nagai and H.~Honda, \emph{Philosophical Magazine B}, 2001, \textbf{81},
  699--719\relax
\mciteBstWouldAddEndPuncttrue
\mciteSetBstMidEndSepPunct{\mcitedefaultmidpunct}
{\mcitedefaultendpunct}{\mcitedefaultseppunct}\relax
\EndOfBibitem
\bibitem[Farhadifar \emph{et~al.}(2007)Farhadifar, Röper, Aigouy, Eaton, and
  Jülicher]{Farhadifar:2007}
R.~Farhadifar, J.-C. Röper, B.~Aigouy, S.~Eaton and F.~Jülicher,
  \emph{Current Biology}, 2007, \textbf{17}, 2095 -- 2104\relax
\mciteBstWouldAddEndPuncttrue
\mciteSetBstMidEndSepPunct{\mcitedefaultmidpunct}
{\mcitedefaultendpunct}{\mcitedefaultseppunct}\relax
\EndOfBibitem
\bibitem[Staple \emph{et~al.}(2010)Staple, Farhadifar, R{\"o}per, Aigouy,
  Eaton, and J{\"u}licher]{Staple:2010}
D.~B. Staple, R.~Farhadifar, J.~C. R{\"o}per, B.~Aigouy, S.~Eaton and
  F.~J{\"u}licher, \emph{Eur. Phys. J. E.}, 2010, \textbf{33}, 117--127\relax
\mciteBstWouldAddEndPuncttrue
\mciteSetBstMidEndSepPunct{\mcitedefaultmidpunct}
{\mcitedefaultendpunct}{\mcitedefaultseppunct}\relax
\EndOfBibitem
\bibitem[Barton \emph{et~al.}(2017)Barton, Henkes, Weijer, and
  Sknepnek]{Barton:2017}
D.~L. Barton, S.~Henkes, C.~J. Weijer and R.~Sknepnek, \emph{PLoS Comput.
  Biol.}, 2017, \textbf{13}, e1005569\relax
\mciteBstWouldAddEndPuncttrue
\mciteSetBstMidEndSepPunct{\mcitedefaultmidpunct}
{\mcitedefaultendpunct}{\mcitedefaultseppunct}\relax
\EndOfBibitem
\bibitem[Fletcher \emph{et~al.}(2014)Fletcher, Osterfield, Baker, and
  Shvartsman]{fletcher2014vertex}
A.~G. Fletcher, M.~Osterfield, R.~E. Baker and S.~Y. Shvartsman,
  \emph{Biophysical journal}, 2014, \textbf{106}, 2291--2304\relax
\mciteBstWouldAddEndPuncttrue
\mciteSetBstMidEndSepPunct{\mcitedefaultmidpunct}
{\mcitedefaultendpunct}{\mcitedefaultseppunct}\relax
\EndOfBibitem
\bibitem[Sknepnek \emph{et~al.}(2023)Sknepnek, Djafer-Cherif, Chuai, Weijer,
  and Henkes]{sknepnek2023generating}
R.~Sknepnek, I.~Djafer-Cherif, M.~Chuai, C.~Weijer and S.~Henkes, \emph{Elife},
  2023, \textbf{12}, e79862\relax
\mciteBstWouldAddEndPuncttrue
\mciteSetBstMidEndSepPunct{\mcitedefaultmidpunct}
{\mcitedefaultendpunct}{\mcitedefaultseppunct}\relax
\EndOfBibitem
\bibitem[Boocock \emph{et~al.}(2023)Boocock, Hirashima, and
  Hannezo]{boocock_interplay_2023}
D.~Boocock, T.~Hirashima and E.~Hannezo, \emph{bioRxiv}, 2023,  2023--03\relax
\mciteBstWouldAddEndPuncttrue
\mciteSetBstMidEndSepPunct{\mcitedefaultmidpunct}
{\mcitedefaultendpunct}{\mcitedefaultseppunct}\relax
\EndOfBibitem
\bibitem[Claussen \emph{et~al.}(2023)Claussen, Brauns, and
  Shraiman]{claussen_geometric_2023}
N.~H. Claussen, F.~Brauns and B.~I. Shraiman, \emph{A {Geometric} {Tension}
  {Dynamics} {Model} of {Epithelial} {Convergent} {Extension}}, 2023,
  \url{http://arxiv.org/abs/2311.16384}, arXiv:2311.16384 [cond-mat,
  physics:physics, q-bio]\relax
\mciteBstWouldAddEndPuncttrue
\mciteSetBstMidEndSepPunct{\mcitedefaultmidpunct}
{\mcitedefaultendpunct}{\mcitedefaultseppunct}\relax
\EndOfBibitem
\bibitem[Lo \emph{et~al.}(2012)Lo, Hawrot, and Georgiou]{lo2012}
P.~Lo, H.~Hawrot and M.~Georgiou, \emph{Biomolecular Concepts}, 2012,
  \textbf{3}, 505--521\relax
\mciteBstWouldAddEndPuncttrue
\mciteSetBstMidEndSepPunct{\mcitedefaultmidpunct}
{\mcitedefaultendpunct}{\mcitedefaultseppunct}\relax
\EndOfBibitem
\bibitem[Zamir \emph{et~al.}(1999)Zamir, Katz, Aota, Yamada, Geiger, and
  Kam]{zamir1999}
E.~Zamir, B.-Z. Katz, S.-i. Aota, K.~M. Yamada, B.~Geiger and Z.~Kam, \emph{J.
  Cell Sci.}, 1999, \textbf{112}, 1655--1669\relax
\mciteBstWouldAddEndPuncttrue
\mciteSetBstMidEndSepPunct{\mcitedefaultmidpunct}
{\mcitedefaultendpunct}{\mcitedefaultseppunct}\relax
\EndOfBibitem
\bibitem[Karzbrun \emph{et~al.}(2021)Karzbrun, Khankhel, Megale, Glasauer,
  Wyle, Britton, Warmflash, Kosik, Siggia,
  Shraiman,\emph{et~al.}]{karzbrun2021human}
E.~Karzbrun, A.~H. Khankhel, H.~C. Megale, S.~M. Glasauer, Y.~Wyle, G.~Britton,
  A.~Warmflash, K.~S. Kosik, E.~D. Siggia, B.~I. Shraiman \emph{et~al.},
  \emph{Nature}, 2021, \textbf{599}, 268--272\relax
\mciteBstWouldAddEndPuncttrue
\mciteSetBstMidEndSepPunct{\mcitedefaultmidpunct}
{\mcitedefaultendpunct}{\mcitedefaultseppunct}\relax
\EndOfBibitem
\bibitem[Lo \emph{et~al.}(2000)Lo, Wang, Dembo, and Wang]{Lo2000}
C.~M. Lo, H.~B. Wang, M.~Dembo and Y.~L. Wang, \emph{Biophys J}, 2000,
  \textbf{79}, 144--152\relax
\mciteBstWouldAddEndPuncttrue
\mciteSetBstMidEndSepPunct{\mcitedefaultmidpunct}
{\mcitedefaultendpunct}{\mcitedefaultseppunct}\relax
\EndOfBibitem
\bibitem[Plotnikov \emph{et~al.}(2012)Plotnikov, Pasapera, Sabass, and
  Waterman]{Plotnikov2012}
S.~V. Plotnikov, A.~M. Pasapera, B.~Sabass and C.~M. Waterman, \emph{Cell},
  2012, \textbf{151}, 1513--1527\relax
\mciteBstWouldAddEndPuncttrue
\mciteSetBstMidEndSepPunct{\mcitedefaultmidpunct}
{\mcitedefaultendpunct}{\mcitedefaultseppunct}\relax
\EndOfBibitem
\bibitem[Elosegui-Artola \emph{et~al.}(2014)Elosegui-Artola, Bazelli{\`e}res,
  Allen, Andreu, Oria, Sunyer, Gomm, Marshall, Jones,
  Trepat,\emph{et~al.}]{elosegui2014}
A.~Elosegui-Artola, E.~Bazelli{\`e}res, M.~D. Allen, I.~Andreu, R.~Oria,
  R.~Sunyer, J.~J. Gomm, J.~F. Marshall, J.~L. Jones, X.~Trepat \emph{et~al.},
  \emph{Nat. Mater.}, 2014, \textbf{13}, 631--637\relax
\mciteBstWouldAddEndPuncttrue
\mciteSetBstMidEndSepPunct{\mcitedefaultmidpunct}
{\mcitedefaultendpunct}{\mcitedefaultseppunct}\relax
\EndOfBibitem
\bibitem[Solon \emph{et~al.}(2015)Solon, Fily, Baskaran, Cates, Kafri, Kardar,
  and Tailleur]{solon2015}
A.~P. Solon, Y.~Fily, A.~Baskaran, M.~E. Cates, Y.~Kafri, M.~Kardar and
  J.~Tailleur, \emph{Nat. Phys.}, 2015, \textbf{11}, 673--678\relax
\mciteBstWouldAddEndPuncttrue
\mciteSetBstMidEndSepPunct{\mcitedefaultmidpunct}
{\mcitedefaultendpunct}{\mcitedefaultseppunct}\relax
\EndOfBibitem
\bibitem[Prakash \emph{et~al.}(2021)Prakash, Bull, and Prakash]{Prakash:2021}
V.~N. Prakash, M.~S. Bull and M.~Prakash, \emph{Nat. Phys.}, 2021, \textbf{17},
  504–511\relax
\mciteBstWouldAddEndPuncttrue
\mciteSetBstMidEndSepPunct{\mcitedefaultmidpunct}
{\mcitedefaultendpunct}{\mcitedefaultseppunct}\relax
\EndOfBibitem
\bibitem[Chen \emph{et~al.}(2022)Chen, Gao, Li, Mao, Tang, and
  Jiang]{Chen:2022}
Y.~Chen, Q.~Gao, J.~Li, F.~Mao, R.~Tang and H.~Jiang, \emph{Phys. Rev. Lett.},
  2022\relax
\mciteBstWouldAddEndPuncttrue
\mciteSetBstMidEndSepPunct{\mcitedefaultmidpunct}
{\mcitedefaultendpunct}{\mcitedefaultseppunct}\relax
\EndOfBibitem
\bibitem[Huang \emph{et~al.}(2022)Huang, Cochran, Fielding, Marchetti, and
  Bi]{huang:2022}
J.~Huang, J.~O. Cochran, S.~M. Fielding, M.~C. Marchetti and D.~Bi, \emph{Phys.
  Rev. Lett.}, 2022, \textbf{128}, 178001\relax
\mciteBstWouldAddEndPuncttrue
\mciteSetBstMidEndSepPunct{\mcitedefaultmidpunct}
{\mcitedefaultendpunct}{\mcitedefaultseppunct}\relax
\EndOfBibitem
\bibitem[Popovi\'c \emph{et~al.}(2021)Popovi\'c, Druelle, Dye, J\"ulicher, and
  Wyart]{Popovic:2021}
M.~Popovi\'c, V.~Druelle, N.~A. Dye, F.~J\"ulicher and M.~Wyart, \emph{New J.
  Phys.}, 2021, \textbf{23}, 033004\relax
\mciteBstWouldAddEndPuncttrue
\mciteSetBstMidEndSepPunct{\mcitedefaultmidpunct}
{\mcitedefaultendpunct}{\mcitedefaultseppunct}\relax
\EndOfBibitem
\bibitem[Matoz-Fernandez \emph{et~al.}(2017)Matoz-Fernandez, Agoritsas, Barrat,
  Bertin, and Martens]{Matoz:2017}
D.~A. Matoz-Fernandez, E.~Agoritsas, J.-L. Barrat, E.~Bertin and K.~Martens,
  \emph{Phys. Rev. Lett.}, 2017, \textbf{118}, 158105\relax
\mciteBstWouldAddEndPuncttrue
\mciteSetBstMidEndSepPunct{\mcitedefaultmidpunct}
{\mcitedefaultendpunct}{\mcitedefaultseppunct}\relax
\EndOfBibitem
\bibitem[Ioannidou \emph{et~al.}(2016)Ioannidou, Kanduc, Li, Frenkel, Dobnikar,
  and Gado]{Ioannidou:2016}
K.~Ioannidou, M.~Kanduc, L.~Li, D.~Frenkel, J.~Dobnikar and E.~D. Gado,
  \emph{Nat. Commun.}, 2016, \textbf{7}, 12106\relax
\mciteBstWouldAddEndPuncttrue
\mciteSetBstMidEndSepPunct{\mcitedefaultmidpunct}
{\mcitedefaultendpunct}{\mcitedefaultseppunct}\relax
\EndOfBibitem
\bibitem[Trimm and Stanislaus(1986)]{Trimm:1986}
D.~Trimm and A.~Stanislaus, \emph{Applied Catalysis}, 1986, \textbf{21},
  215--238\relax
\mciteBstWouldAddEndPuncttrue
\mciteSetBstMidEndSepPunct{\mcitedefaultmidpunct}
{\mcitedefaultendpunct}{\mcitedefaultseppunct}\relax
\EndOfBibitem
\bibitem[Sudreau \emph{et~al.}(2022)Sudreau, Manneville, Servel, and
  Divoux]{Sudreau:2022}
I.~Sudreau, S.~Manneville, M.~Servel and T.~Divoux, \emph{J. Rheol.}, 2022,
  \textbf{66}, 91--104\relax
\mciteBstWouldAddEndPuncttrue
\mciteSetBstMidEndSepPunct{\mcitedefaultmidpunct}
{\mcitedefaultendpunct}{\mcitedefaultseppunct}\relax
\EndOfBibitem
\bibitem[Rao \emph{et~al.}(2019)Rao, Divoux, McKinley, and Hart]{Rao:2019}
A.~Rao, T.~Divoux, G.~H. McKinley and A.~J. Hart, \emph{Soft Matter}, 2019,
  \textbf{15}, 4401--4412\relax
\mciteBstWouldAddEndPuncttrue
\mciteSetBstMidEndSepPunct{\mcitedefaultmidpunct}
{\mcitedefaultendpunct}{\mcitedefaultseppunct}\relax
\EndOfBibitem
\bibitem[Rao \emph{et~al.}(2022)Rao, Divoux, Owens, and Hart]{Rao:2022}
A.~Rao, T.~Divoux, C.~E. Owens and A.~J. Hart, \emph{Cellulose}, 2022,
  \textbf{29}, 2387--2398\relax
\mciteBstWouldAddEndPuncttrue
\mciteSetBstMidEndSepPunct{\mcitedefaultmidpunct}
{\mcitedefaultendpunct}{\mcitedefaultseppunct}\relax
\EndOfBibitem
\bibitem[Vynck \emph{et~al.}(2023)Vynck, Pierrat, Carminati, Froufe-P\'erez,
  Scheffold, Sapienza, Vignolini, and S\'aenz]{Vynck:2023}
K.~Vynck, R.~Pierrat, R.~Carminati, L.~S. Froufe-P\'erez, F.~Scheffold,
  R.~Sapienza, S.~Vignolini and J.~J. S\'aenz, \emph{Rev. Mod. Phys.}, 2023,
  \textbf{95}, 045003\relax
\mciteBstWouldAddEndPuncttrue
\mciteSetBstMidEndSepPunct{\mcitedefaultmidpunct}
{\mcitedefaultendpunct}{\mcitedefaultseppunct}\relax
\EndOfBibitem
\bibitem[Gu{\'e}nard-Lampron \emph{et~al.}(2020)Gu{\'e}nard-Lampron,
  Villeneuve, St-Gelais, and Turgeon]{Guenard:2020}
V.~Gu{\'e}nard-Lampron, S.~Villeneuve, D.~St-Gelais and S.~L. Turgeon,
  \emph{International Dairy Journal}, 2020, \textbf{109}, 104742\relax
\mciteBstWouldAddEndPuncttrue
\mciteSetBstMidEndSepPunct{\mcitedefaultmidpunct}
{\mcitedefaultendpunct}{\mcitedefaultseppunct}\relax
\EndOfBibitem
\bibitem[Mazzanti \emph{et~al.}(2003)Mazzanti, Guthrie, Sirota, Marangoni, and
  Idziak]{Mazzanti:2003}
G.~Mazzanti, S.~E. Guthrie, E.~B. Sirota, A.~G. Marangoni and S.~H. Idziak,
  \emph{Cryst. Growth Des.}, 2003, \textbf{3}, 721--725\relax
\mciteBstWouldAddEndPuncttrue
\mciteSetBstMidEndSepPunct{\mcitedefaultmidpunct}
{\mcitedefaultendpunct}{\mcitedefaultseppunct}\relax
\EndOfBibitem
\bibitem[Sonwai and Mackley(2006)]{Sonwai:2006}
S.~Sonwai and M.~Mackley, \emph{J. Am. Oil Chem. Soc.}, 2006, \textbf{83},
  583--596\relax
\mciteBstWouldAddEndPuncttrue
\mciteSetBstMidEndSepPunct{\mcitedefaultmidpunct}
{\mcitedefaultendpunct}{\mcitedefaultseppunct}\relax
\EndOfBibitem
\bibitem[Yang \emph{et~al.}(2011)Yang, Hrymak, and Kamal]{Yang:2011}
D.~Yang, A.~N. Hrymak and M.~R. Kamal, \emph{Ind. Eng. Chem. Res.}, 2011,
  \textbf{50}, 11594--11600\relax
\mciteBstWouldAddEndPuncttrue
\mciteSetBstMidEndSepPunct{\mcitedefaultmidpunct}
{\mcitedefaultendpunct}{\mcitedefaultseppunct}\relax
\EndOfBibitem
\bibitem[Bauland \emph{et~al.}(2023)Bauland, Leocmach, Famelart, and
  Croguennec]{Bauland:2023}
J.~Bauland, M.~Leocmach, M.-H. Famelart and T.~Croguennec, \emph{Soft Matter},
  2023, \textbf{19}, 3562--3569\relax
\mciteBstWouldAddEndPuncttrue
\mciteSetBstMidEndSepPunct{\mcitedefaultmidpunct}
{\mcitedefaultendpunct}{\mcitedefaultseppunct}\relax
\EndOfBibitem
\bibitem[Martinelli \emph{et~al.}(2023)Martinelli, Caporaletti, Dallari,
  Sprung, Westermeier, Baldi, and Monaco]{Martinelli:2023}
A.~Martinelli, F.~Caporaletti, F.~Dallari, M.~Sprung, F.~Westermeier, G.~Baldi
  and G.~Monaco, \emph{Phys. Rev. X}, 2023, \textbf{13}, 041031\relax
\mciteBstWouldAddEndPuncttrue
\mciteSetBstMidEndSepPunct{\mcitedefaultmidpunct}
{\mcitedefaultendpunct}{\mcitedefaultseppunct}\relax
\EndOfBibitem
\end{mcitethebibliography}

\providecommand*{\mcitethebibliography}{\thebibliography}
\csname @ifundefined\endcsname{endmcitethebibliography}
{\let\endmcitethebibliography\endthebibliography}{}

\end{document}